\documentclass[12pt,preprint]{aastex}

\begin{document}


\title{A comprehensive study of GRB 070125, a most energetic gamma ray burst
}

\author{Poonam Chandra \altaffilmark{1,2}, 
S.~Bradley Cenko\altaffilmark{3},
Dale A.~Frail\altaffilmark{4},
Roger A.~Chevalier\altaffilmark{2},
Jean-~Pierre Macquart\altaffilmark{1,5},
Shri R.~Kulkarni\altaffilmark{5},
Douglas C.-J.Bock\altaffilmark{6},
Frank Bertoldi\altaffilmark{7},
Mansi Kasliwal\altaffilmark{5},
Derek B.~Fox\altaffilmark{8},
Paul A.~Price\altaffilmark{9},
Edo Berger\altaffilmark{10,11},
Alicia M.~Soderberg\altaffilmark{10,11,12},
Fiona A.~Harrison\altaffilmark{3},
Avishay Gal-Yam\altaffilmark{13},
Eran O.~Ofek\altaffilmark{5},
Arne Rau\altaffilmark{5},
Brian P.~Schmidt\altaffilmark{14},
P.~Brian Cameron\altaffilmark{5},
Lennox L.~Cowie\altaffilmark{15},
Antoinette Cowie\altaffilmark{15},
Katherine C.~Roth\altaffilmark{16},
Michael Dopita\altaffilmark{17},
Bruce Peterson\altaffilmark{17},
Bryan E.~Penprase\altaffilmark{18}
}
\altaffiltext{1}{Jansky Fellow, National Radio Astronomy Observatory}
\altaffiltext{2}{Department of Astronomy, University of Virginia, P.O. Box
        400325, Charlottesville, VA 22904}
\altaffiltext{3}{Space Radiation Laboratory, MS 220-47,
        California Institute of Technology, Pasadena, CA 91125}
\altaffiltext{4}{National Radio Astronomy Observatory, P.O. Box O,
Socorro, NM 87801}
\altaffiltext{5}{Division of Physics, Maths and Astronomy, 
        California Institute of Technology, Pasadena, CA 91125}
\altaffiltext{6}{Combined Array for Research in Millimeter-wave Astronomy,
P.O. Box 968, Big Pine, CA 93513}
\altaffiltext{7}{Argelander-~Institut f\"ur Astronomie, Auf dem H\"ugel 71,
D-53121 Bonn, Germany}
\altaffiltext{8}{Department of Astronomy and Astrophysics, Pennsylvania
        State University, 525 Davey Laboratory, University Park, PA 16802}
\altaffiltext{9}{Institute for Astronomy, University of Hawaii, 2680 Woodlawn
        Drive, Honolulu, HI 96822}
\altaffiltext{10}{Princeton University Observatory, Peyton Hall, Ivy Lane,
        Princeton, NJ 08544}
\altaffiltext{11}{Observatories of the Carnegie Institute of Washington,
        813 Santa Barbara Street, Pasadena, CA 91101}
\altaffiltext{12}{Hubble Fellow}
\altaffiltext{13}{Benoziyo Center for Astrophysics, The Weizmann
Institute of Science, Rehovot 76100, Israel}
\altaffiltext{14}{Research School of Astronomy and Astrophysics, Australian
        National University, Mt Stromlo Observatory, via Cotter Rd, Weston
        Creek, ACT 2611, Australia}
\altaffiltext{15}{Institute of Astronomy, University of Hawaii, 2680 Woodlawn Drive, Honolulu, HI 96822-1897}
\altaffiltext{16}{Gemini Observatory, 670 N. Aohoku Place, Hilo, HI 96720}
\altaffiltext{17}{Institute of Advanced Studies, The Australian National University, ACT2611, Australia}
\altaffiltext{18} {Department of Physics and Astronomy, Pomona College, 610 N. College Ave Claremont CA 91711}
\email{pc8s@virginia.edu}



\shorttitle{Comprehensive study of GRB 070125}
\shortauthors{Chandra \textit{et al.}}



\newcommand{\vdag}{(v)^\dagger}
\newcommand{\myemail}{pc8s@virginia.edu}
\newcommand{\Swift}{\textit{Swift}}
\newcommand{\event}{GRB 070125}


\begin{abstract}

We present a comprehensive multiwavelength 
analysis of the bright, long 
duration gamma-ray burst GRB 070125, comprised of observations 
in  $\gamma$-ray, X-ray, optical, millimeter and
centimeter wavebands. 
Simultaneous fits to the 
optical and X-ray light curves favor a break on 
day 3.78, which we interpret as the jet break from a collimated
outflow. Independent fits to optical and X-ray bands give similar 
results in the optical bands but shift the jet break to around day 10
in the X-ray light curve. 
We show that for the physical parameters derived for 
GRB 070125, inverse Compton scattering effects are important
throughout the afterglow evolution. 
While inverse Compton scattering does not affect 
radio and optical bands, it may be a promising candidate to
delay the jet break in the X-ray band. 
Radio light curves show rapid flux variations, which are interpreted
as due to interstellar scintillation, and are used to derive an 
upper limit of $2.4 \times 10^{17}$ cm on the radius
 of the fireball in the lateral expansion phase of the jet. 
Radio light curves and spectra suggest a high synchrotron
self absorption frequency indicative of 
the afterglow shock wave  moving in a dense medium.
Our broadband modeling favors a constant density profile 
for the circumburst medium over a wind-like profile ($R^{-2}$).
However, keeping in mind the uncertainty of the parameters, it is
difficult  to unambiguously distinguish between the two density profiles.
Our broadband fits suggest that  \event\ 
is a burst with high radiative efficiency ($> 60 \%$).  

\end{abstract}




\section{Introduction}
\label{sec:introduction}

To understand the inner workings of gamma-ray burst (GRB) central
engines, it is necessary to constrain their true energy release. If a 
redshift is known, the isotropic energy release in gamma-rays,
$E_{\mathrm{iso,}\gamma}$, is a readily measurable quantity. The
``energy crisis'' brought on by implied energy releases of $>10^{54}$
erg  \citep[e.g.,][]{kdo+99,ftm+99} was resolved when it was realized
that GRB blast waves are collimated with opening angles $\theta_j$ of
2-30$^\circ$ \citep{rho99,sph99,sgk+99b,hbf+99}. 
Thus the beaming-corrected
gamma-ray energy release, $E_\gamma$, is smaller than $E_{\mathrm{iso,}
\gamma}$ by a factor $\sim \theta_j^2/2$.

Whatever energy is not released in the prompt 
emission powers a relativistic
blast wave that plows into the circumburst medium (CBM). 
Measuring the kinetic
energy of this outflow, $E_{\mathrm{K}}$, is challenging because
only a fraction of this energy is radiated.  
Several methods have been used
in this endeavor \citep[see][]{bkp+03}. 
The simplest method, using only the X-ray afterglow light
curve, was suggested by \citet{kumar00}, \citet{fw01}, and \citet{bkf03}.  Provided that the 
X-rays are predominately synchrotron and the electrons radiate in the cooling
regime, this method yields the energy per unit solid angle and the
fraction of the shock energy carried by electrons $\epsilon_e$. 
For those
afterglows with high quality multiwavelength data sets, a fit of the 
data can be made using models that describe the dynamics of jet/circumburst
interaction and that calculate the expected synchrotron and inverse
Compton emission \citep[e.g.,][]{pk01,yhsf03}.  
Finally, for bright, long-lived afterglows,
 $E_{\mathrm{K}}$ can be measured more
robustly using the late-time radio light curve. Months or in some
cases years after the burst, the blast wave becomes sub-relativistic
(i.e. Sedov self-similar evolution) 
and the outflow is expected to be quasi-spherical 
\citep[e.g.,][]{fwk00,bkf04,fsk+05}.

Early studies, based on these methods, suggested that the energy
release of long-duration GRBs lies within a narrow range with a
mean of $\sim 10^{51}$ erg \citep[e.g.][]{fks+01,bkf03}. There are 
several problems with this simple picture, however, and the true situation
is likely more complicated.  For example, energy input into the blast wave 
now appears to extend beyond the prompt emission and into the afterglow phase. 
This could be due either to a long-lived central engine or slow-moving ejecta
that take a long time to reach the shock front \citep{zfd+06}.

The simple pictures of collimated explosions has been challenged by
\Swift\ detected GRBs.
Kocevski \& Butler (2007\nocite{kb07}) give an
excellent summary of the difficulties and their implications for
collimation-corrected energy release $E_{\gamma}$. 
If (for whatever reason) the \Swift\ sample of bursts has no
jet breaks then we are faced with a growing number of GRBs that
(by virtue of inferred isotropic energy release)
are hyper-energetic 
i.e.~well in excess of the canonical 10$^{51}$ erg 
\citep[e.g.,][]{br07,ckh+06,fck+06}. If 
accepted, these events would provide 
stringent tests of existing progenitor models. 

In this paper, we present multiwavelength observations of GRB\,070125.
Its bright afterglow has allowed us to follow the GRB until day
350 and obtain the most extensive radio data
in the \Swift\ era, 
coupled with well-sampled X-ray and optical
light curves indicative of a jet break.  Together with the well
characterized prompt emission extending beyond 1 MeV \citep{bhp+07}, 
GRB\,070125 is truly a rare event.  

In \S \ref{sec:observation} we provide details of observations
for GRB\,070125. In \S \ref{sec:result}
we describe our results, and find evidence for a jet break
in the optical and X-ray data. The radio spectra show evidence 
for evolution from an optically thick to an optically thin phase, while
the radio light curves show short timescale variability
that we ascribe to scintillation. 
We carry out a straightforward analytic
modeling and derive some physical parameters of the shock 
and the circumburst medium. Finally, we carry out a detailed 
model fit of the entire broadband dataset and summarize 
our results in \S \ref{sec:multiwave}.
We discuss the energetics, environment, 
relevance of reverse shock emission, efficiency of the
GRB and the inverse Compton scattering effects for \event\ in \S
\ref{sec:discussion}. 
The  main conclusions are listed in \S \ref{sec:conclusion}.

\section{Observations}
\label{sec:observation}

In this section, we present  multiwavelength observations of GRB\,070125.
We supplement our data set with the measurements reported in literature 
(mostly notices from the Gamma-ray Burst Circulars 
Network\footnote{http://gcn.gsfc.nasa.gov/gcn3\_archive.html}).

\subsection{Gamma-Ray Observations}
\label{sec:gammaray}

\begin{figure}
\begin{center}
\includegraphics[angle=0,width=1\textwidth]{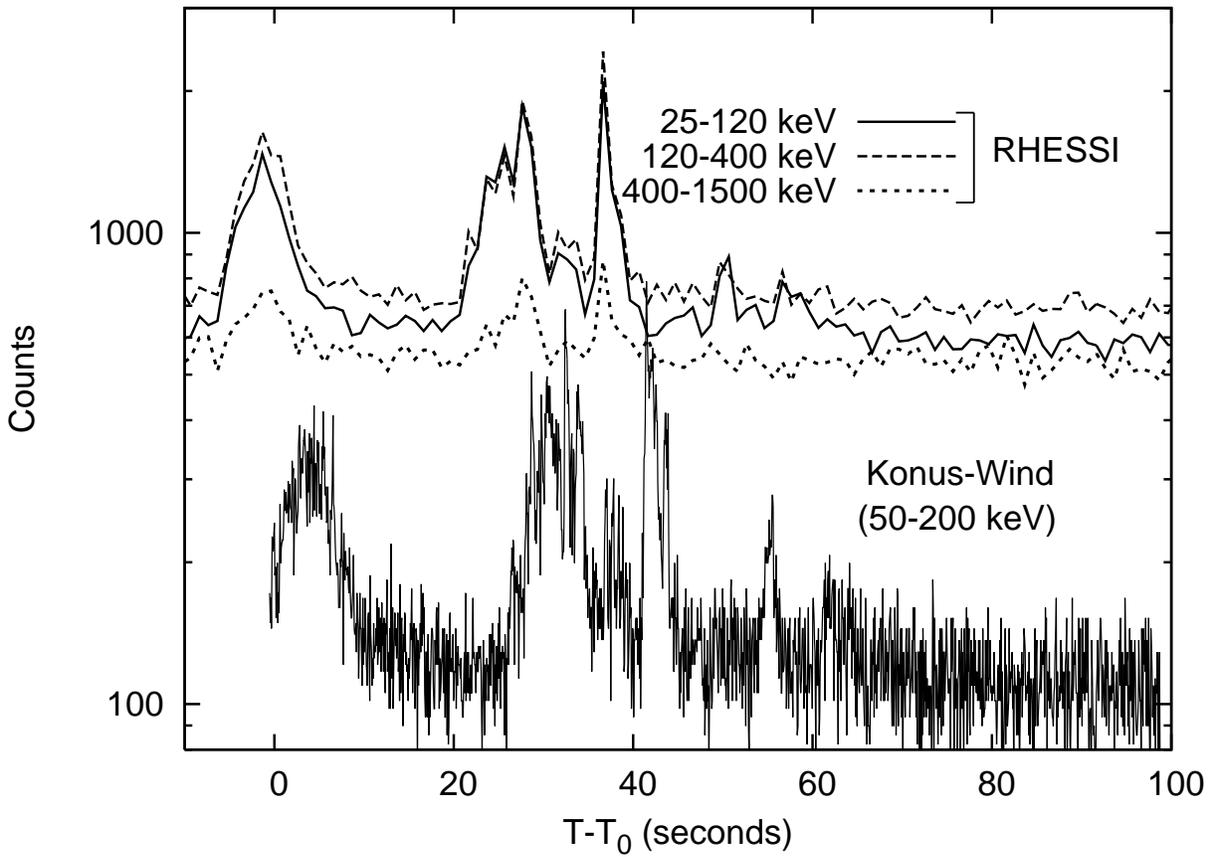}
\caption{Plot of RHESSI data and Konus-Wind data. Their peaks are
shifted by 8s because of the difference in light 
travel time between the two instruments. RHESSI recorded the burst explosion
time ($T_0$) to be
07:20:42 UT and Konus-Wind measured it to be 07:20:50.
}
\label{fig:high}
\end{center}
\end{figure}

GRB\,070125 was discovered by the Inter Planetary Network (IPN) of GRB detectors
at 07:20:45 UT 25 January 2007 \citep{hcm+07}.  
Mars Odyssey (HEND and GRS), Suzaku (WAM), INTEGRAL (SPI-ACS), RHESSI and
Konus-Wind, all observed this intense, $\sim 70$ s long, event
(Fig.~\ref{fig:high}). The burst was not in the field of view of
 {\it Swift}. Four minutes later the burst fell into the
Burst Alert Telescope (BAT) field of view \citep{hcm+07}. The 
\Swift\ BAT position was
consistent with the position of the GRB as triangulated
by IPN. 
The small $4'$ error circle from the BAT 
enabled follow-up observations to identify the bright 
 X-ray (\S \ref{sec:xray}) and optical
(\S \ref{sec:optical}) afterglow.

Because of the high-energy (up to 1 MeV) coverage provided by RHESSI, 
Konus-Wind, and Suzaku, tight constraints can be made on the prompt emission 
spectrum of GRB\,070125.  In particular, the gamma-ray fluence was much more 
accurately measured than a typical \Swift\ GRB (with coverage of 
only $\sim 300$\,keV).  A detailed analysis of the high-energy properties 
of GRB\,070125 has been performed by \citet{bhp+07}, 
who found that the 
spectrum is well fit by a Band model \citep{bmf+93} with a peak energy
$E_{\mathrm{p}} = 430$ keV and the 
 power-law photon indices (as defined in Band model)
 of $-1.14$ and $-2.11$ below and above
the peak energy, respectively.
 The resulting fluence in the 20 keV -- 10 MeV band was 
$1.7 \times 10^{-4}$ erg cm$^{-2}$.

\subsection{X-ray observations}
\label{sec:xray}

The X-ray Telescope (XRT; \citealt{bhn+05}) on board \Swift\ began observing
the field of \event\ 46 ks after the burst trigger \citep{GCN.6030}.  Due to
the relatively low source flux, observations were conducted exclusively 
in Photon Counting (PC) mode, and so the effects of photon pile-up were 
negligible.  During the first five orbits, 
the X-ray spectrum is well-fit by a power-law model with a photon index 
$\Gamma = 2.1 \pm 0.3$ \citep[$N(\gamma)\propto \gamma^{-\Gamma} 
d\gamma$, where $N$ 
is the photon flux density,][]{GCN.6030}. The 
inferred absorption is consistent with the
Galactic absorption value of $N_{\mathrm{H}} 
\approx 8 \times 10^{20}$\,cm$^{-2}$
\citep{GCN.6030}.
The XRT continued to monitor the X-ray afterglow of \event\ over the course
of the next two weeks, until the source faded below the XRT sensitivity
threshold.

Motivated by 
the proposed lack of X-ray jet breaks observed in
the \Swift\ era \citep{br07}, we obtained Director's
Discretionary time on the \textit{Chandra X-ray Observatory} to observe the
X-ray afterglow of \event\ at very late times.  We obtained a 30 ks exposure
using the Advanced CCD Imaging Spectrometer on board \citep{gbf+03} beginning at
21:28 UT 2007 March 5 ($\sim 40$\,d after the burst trigger).  No source
was detected at the position of the X-ray afterglow at this time.  Formally,
using a circular aperture with a one-arcsecond
 diameter, we detect $0.9 \pm 5.0$ 
photons in the energy range from $0.3-10$\,keV at the location of \event\
(a $5- \sigma$ upper limit).

The log of X-ray observations and the measured fluxes are
summarized in Table~\ref{tab:xray}.
The \Swift\ XRT light curve is obtained from the on-line 
repository\footnote{http://www.Swift.ac.uk/xrt\_curves} \citep{ebp+07}.
We converted counts to flux using spectral parameters derived from
the first five XRT orbits, ranging from $47 - 389$ ks.
There is no evidence for spectral evolution in the X-ray 
light curve, implying the derived 
conversion factor is applicable at all times
 \citep{rcm07}. 
We then converted the $0.3-10.0$ keV 
flux to a flux density 
($F_\nu \propto \nu^{-\beta}$, where $\beta=\Gamma-1=1.1$)
at $E = 1.486$\,keV ($\nu_{0} = 3.594 \times 10^{17}$\,Hz).  This value was 
so chosen that the flux was equally divided below and above this cutoff.  

\subsection{Optical observations}
\label{sec:optical}

In response to the IPN-\Swift\ localization, we began observing the field of
\event\ with the automated Palomar 60-inch telescope (P60; \citealt{cfm+06})
on the night of 2007 January 26.  Inspection of the first images
revealed a bright, stationary source at $\alpha = 07^{\mathrm{h}}
51^{\mathrm{m}} 17\farcs75$, $\delta =
 +31^{\circ} 09\arcmin  04\farcs2$ (J2000.0),
not present in the Sloan Digital Sky Survey (SDSS) images of the field 
\citep{aaa+07}.  This position was promptly reported as the afterglow of \event\
\citep{GCN.6028}, allowing several groups to obtain spectroscopy of the
afterglow while still quite bright ($R \sim 18$\,mag).

The optical afterglow of GRB\,070125 was observed by a variety of facilities
worldwide, as it turned out to be one of the brightest optical
afterglows ever detected for a GRB (Updike \textit{et al.}, in prep.)
We continued to monitor the afterglow of \event\ with the 
P60 in the Kron $R$ and 
Sloan $i^{\prime}$ filters for the following four nights, until the afterglow 
faded below our sensitivity limit.  All P60 data were reduced using our custom
software pipeline (see \citealt{cfm+06} for details) using IRAF\footnote{IRAF 
is distributed by the National Optical Astronomy Observatory, which is operated 
by the Association for Research in Astronomy, Inc., under cooperative agreement 
with the National Science Foundation.} routines.

In addition to our P60 monitoring, we obtained three epochs of late-time optical
photometry with the Gemini Multi-Object Spectrograph and Imager (GMOS;
\citealt{hab+03}) mounted on the 8-m Gemini North telescope.  All three epochs
consisted of either single $r^{\prime}$ or $i^{\prime}$ exposures obtained as
acquisition images for spectroscopic observations.  All GMOS images
were reduced using the IRAF \texttt{gemini} package.

Finally, we obtained a single epoch of simultaneous $g$- and $R$-band
imaging on the night of 2007 February 16 with the Low-Resolution Imaging
Spectrometer (LRIS: \citealt{occ+95}) mounted on the 10-m Keck Telescope.
Individual images were bias-subtracted and flat-fielded using
standard IRAF routines.  Co-addition was performed using
Swarp\footnote{http://terapix.iap.fr}.  Photometric calibration for all of our
optical imaging was performed relative to the SDSS, with empirical
filter transformations from \citet{jga06} applied where necessary.
Typical RMS photometric uncertainties were $\approx 0.05$ mag in all of our
images.

\citet{gfj+07} imaged the position of the GRB 070125 afterglow
with the LBC-blue CCD camera and 8.4-m SX mirror on the LBT on 2007 February
21.1 (UT) in r-band.  A faint source is detected at the position of the 
afterglow with brightness r=$26.3 \pm 0.3$ mag. Since the source may be 
contaminated by the host galaxy, this observation represents an upper-limit on 
the magnitude of the afterglow 26.8 days after the GRB.

To convert magnitudes to flux densities, we used zeropoint measurements from
\citet{fsi95}.  We have incorporated a modest amount of
Galactic extinction ($E(B-V) = 0.052$; \citealt{dl90,sfd98}) into these
results.  The results of our optical monitoring of \event, together with 
measurements reported by other observatories via the GCN, are shown in 
Table~\ref{tab:opt}.

\subsection{Millimeter \&\ Sub-Millimeter observations}
\label{sec:submm}
We obtained director's discretionary time for observations in the 1.2-mm band 
(250 GHz) at the Max-Planck Millimeter Bolometer Array (MAMBO), installed at
the IRAM 30 m telescope on Pico Veleta, Spain.  We used the MAMBO-2 version 
with 117 channels.  The bandwidth used was $210-290$ GHz half power. Our first 
observations took place on 2007 January 30 and we detected the afterglow of 
\event\ at a flux density of $3.14 \pm 0.59$ mJy.  We monitored the afterglow 
regularly until it had dropped 
below the instrumental sensitivity.

Observations were also obtained at 95 GHz using the Combined Array for
Research in Millimeter-wave Astronomy (CARMA)\footnote{Support for
CARMA construction was derived from the Gordon and Betty Moore
Foundation, the Kenneth T. and Eileen L. Norris Foundation, the
Associates of the California Institute of Technology, the states of
California, Illinois, and Maryland, and the National Science
Foundation. Ongoing CARMA development and operations are supported by
the National Science Foundation under a cooperative agreement, and by
the CARMA partner universities.}, in good weather on 2007 February 5,
9, 12, and 18.
Individual observations were between $6.7-8$ hours in length. 
System temperatures ranged from 200 to 400 K, scaled to outside the
atmosphere. A single linear polarization was received in a bandwidth
of $1.2-1.5$ GHz per sideband, depending on the antenna. CARMA was in
the C configuration, with baselines $26-370$ m.

The bright quasar 0748+240 was observed as a phase calibrator at
20-minute intervals.  A fainter source nearer GRB\,070125, 0741+312
(2\degr\ distant), was observed for 30 seconds after each observation
of the GRB. The flux density of this source when self-calibrated was
compared to the flux density when calibrated with 0748+240. The ratio
of these measurements provides an estimate of the atmospheric
decorrelation affecting the observations. The measured flux densities
of GRB 070125 were corrected by this factor, which ranged from 9 to
14\%. The absolute flux density scale was derived from observations of
Uranus on February 2 and 19, and transferred by reference to the
quasar 3C84, which was observed on those dates and near the beginning
of each observation of GRB 070125.

Images were made for each observation using the MIRIAD software package 
(\citealt*{stw95}).  
The quoted uncertainty in the
measurement is dominated by the sensitivity of the map, but also
includes an estimate for the flux density scale uncertainty of 15\%.
We detected the afterglow in the first three observations. The
flux density of the mean observation time 2007 February
 5, 0700 UT was 2.3 mJy
The log and flux densities of the MAMBO
and CARMA observations can be found in Table~\ref{tab:radio}.

\subsection{Centimeter band observations}
\label{sec:radio}

The earliest measurement of the radio flux density was taken from the Westerbork
Synthesis Radio Telescope (WSRT) in 5 GHz band by  
\citet{van07}), just 1.5 days after the 
burst. A 2$-\sigma$ upper limit on the 
5 GHz flux density of $F_{\nu} < 174$\,$\mu$Jy was obtained.
Shortly
thereafter, we triggered Very Large Array (VLA)
observations of the field.  Our first measurement at 
8.46\,GHz, four days after the explosion, resulted in a strong detection
\citep[$F_{\nu} = 360 \pm 42 $\,$\mu$Jy;][]{cf07b}.  
Encouraged by this detection, we triggered 
observations with the Giant Meterwave Radio Telescope (GMRT) in the 610 MHz 
band on 2007 January 31.  However, we did not detect the afterglow with the GMRT
to a 2$-\sigma$ limiting flux density of $F_\nu < 300$\,$\mu$Jy \citep{ccg07}.

We continued the followup observations of \event\ 
with the VLA from 2007 January 29th until 2007
February 2nd at 8.46 GHz and 4.86 GHz.
After 2007 February 5, we undertook 
observations in the 1.46, 4.86, 8.46, 14.94 and 22.5 GHz bands.  
We followed the GRB until day 342 since explosion.

Each observation at a single frequency was from 30 minutes to one 
hour duration.  
The bright radio quasar 
3C48 (0137+331) was used as a flux calibrator. We used phase calibrators
0745+317 and 0741+312 to track the
instrumental and/or atmospheric gain and phase variations,
as well as to monitor the
quality and the sensitivity of the data.
A bandwidth of  $2 \times 50$ MHz 
was used for all the 
observations. The data were analyzed using standard data reduction procedures 
within the  Astronomical Image Processing System (AIPS).
During initial observations,
 the VLA was in ``C$\rightarrow$D'' and ``D'' array configuration 
for early observations; because of the
short baselines, there were many sources in the field of view at lower 
frequencies. For such cases at 4.86 and 1.43 GHz, we did 3-D cleaning with 
several rounds of self-calibration inside AIPS.

On 2007 February 7, 8  and 14, we made longer observations for
durations of $\sim 5.5$, $\sim 4$, and $\sim 8$ hours, 
respectively, in order to search for variability 
due to Interstellar Scintillation (ISS) (see \S \ref{sec:scintillation}). 
We combined every 20 minutes of the data and imaged 
the afterglow field of view to 
measure short timescale scintillations.
The results of our radio monitoring of the afterglow of GRB\,070125 can
be found in Table~\ref{tab:radio}.

%

\section{Results}
\label{sec:result}

In this section, we carry out a  simple analysis with minimum model
assumptions to derive robust conclusions.
 We find strong evidence for a jet break in the 
combined fitting of the X-ray
and the optical light curves 
(\S \ref{sec:tjet}).  In \S \ref{sec:scintillation} 
we use the radio scintillation data to constrain the size of the emitting 
region. Together, the jet break time and the inferred
 radius allow us to constrain 
the prompt energy release (\S \ref{sec:energetics}) and the density of the 
circumburst medium (\S \ref{sec:density}).  In the next section
(\S \ref{sec:multiwave}), we undertake a detailed analysis utilizing the
full machinery of afterglow models.

\subsection{Break in the Light Curve}
\label{sec:tjet}

\begin{figure}
\begin{center}
\includegraphics[angle=0,width=0.70\textwidth]{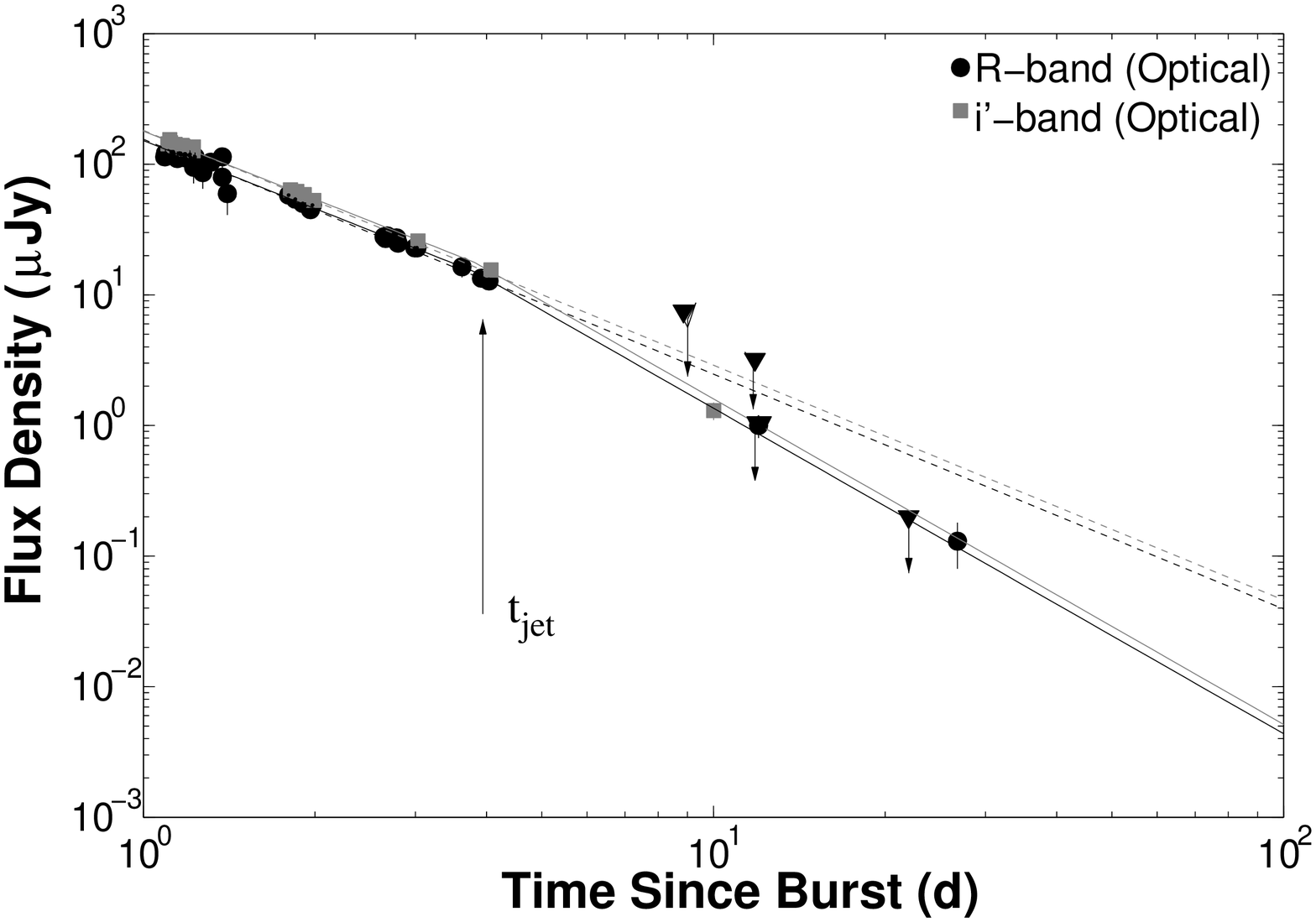}
\includegraphics[angle=0,width=0.70\textwidth]{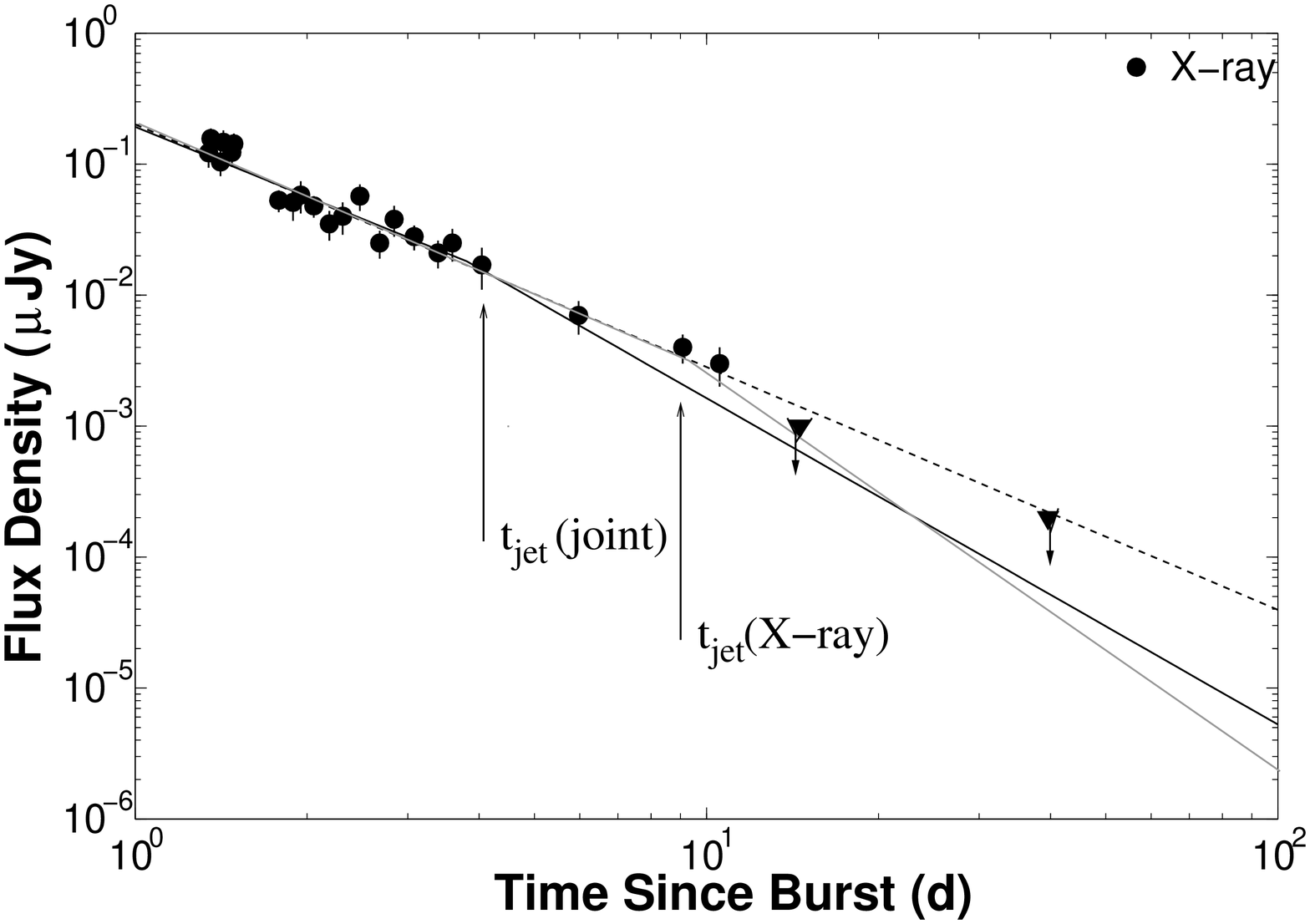}
\caption[]{\small
\textit{Top}: Optical ($R$- and $i^{\prime}$-band) light curves of \event\
as a result of joint optical and X-ray fits.
Best-fit single power-law models are shown with dashed lines, while the broken
power-law models are shown in solid lines.  It is clear that in 
the optical bands, a broken power-law (indicating a 
jet break) is strongly favored.
\textit{Bottom}: X-ray light curve of \event, 
the the joint fit to optical and X-ray data.  Again the single
power-law model is shown as a dashed line, while the broken power-law model
is shown as a solid line. Grey solid line indicates the independent fit to
X-ray data. The independent fit
is consistent with the joint fit in the optical bands but
shifts the jet break to $\sim$ $9-10$ days. We discuss this
in \S \ref{sec:tjet} and \S \ref{sec:discussion}.
}
\label{fig:jet}
\end{center}
\end{figure}

We performed a joint fit
on the $R$, $i^{\prime}$, and X-ray light curves of \event\ using both
a single power-law (spl) and a broken power-law (bpl) model.  
The results of these two fits are shown in
Figure~\ref{fig:jet} and Table~\ref{tab:jet}.  We also perform the spl and
the bpl fits letting the X-ray and optical 
pre-break indices vary independently. The results are consistent 
with the previous fits in which we constrained both the
indices to be the same. 
Altogether, the broken power-law 
model, indicating a jet break at $t_{\mathrm{j}} = 3.8$\,d, 
is strongly favored
($\chi^{2}_{\mathrm{r}}(\mathrm{bpl}) = 1.30$ for 97 
d.o.f.~vs.~$\chi^{2}_{\mathrm{r}}(\mathrm{spl}) = 2.23$ for 99 d.o.f.).

We note, however, that the distinction between the two models is significantly 
more pronounced in the optical ($\chi^{2}_{\mathrm{r}}(R\mathrm{-band, bpl}) = 
1.54$ for 52 d.o.f.~vs.~$\chi^{2}_{\mathrm{r}}
(R\mathrm{-band, spl}) = 2.22$ for 
52 d.o.f.; and 
$\chi^{2}_{\mathrm{r}}(i'\mathrm{-band, bpl}) = 
0.77$ for 25 d.o.f.~vs.~$\chi^{2}_{\mathrm{r}}
(i'\mathrm{-band, spl}) = 3.43$ for
52 d.o.f.) than the X-rays.  As first noted by \citet{GCN.6181}, without the
late-time \textit{Chandra} data, the X-ray light curve cannot distinguish 
between the spl and bpl models.  Formally, 
we find the single power-law model is 
actually favored in the X-rays ($\chi^{2}_{\mathrm{r}}(\mathrm{X-ray, spl}) = 
0.88$ for 22 d.o.f.~vs.~$\chi^{2}_{\mathrm{r}}(\mathrm{X-ray, bpl}) = 1.32$ for 20
d.o.f.). This could perhaps be due to the denser sampling at early times.

We also perform independent optical and X-ray fitting. The optical 
fits are consistent with our joint fits.
However, when we fit the X-ray light curve independently of the optical, the jet
break appears much later, $t_{\mathrm{j}} \approx 9$\,d (Fig.~\ref{fig:jet}).
While not formally required by our simple analysis, we consider the possibility
that this break may in fact be \textit{chromatic}. We discuss this further in
\S \ref{sec:discussion}. 

\subsection{Scintillation and fireball size}
\label{sec:scintillation}

\begin{figure}
\begin{center}
\includegraphics[angle=0,width=1.00\textwidth]{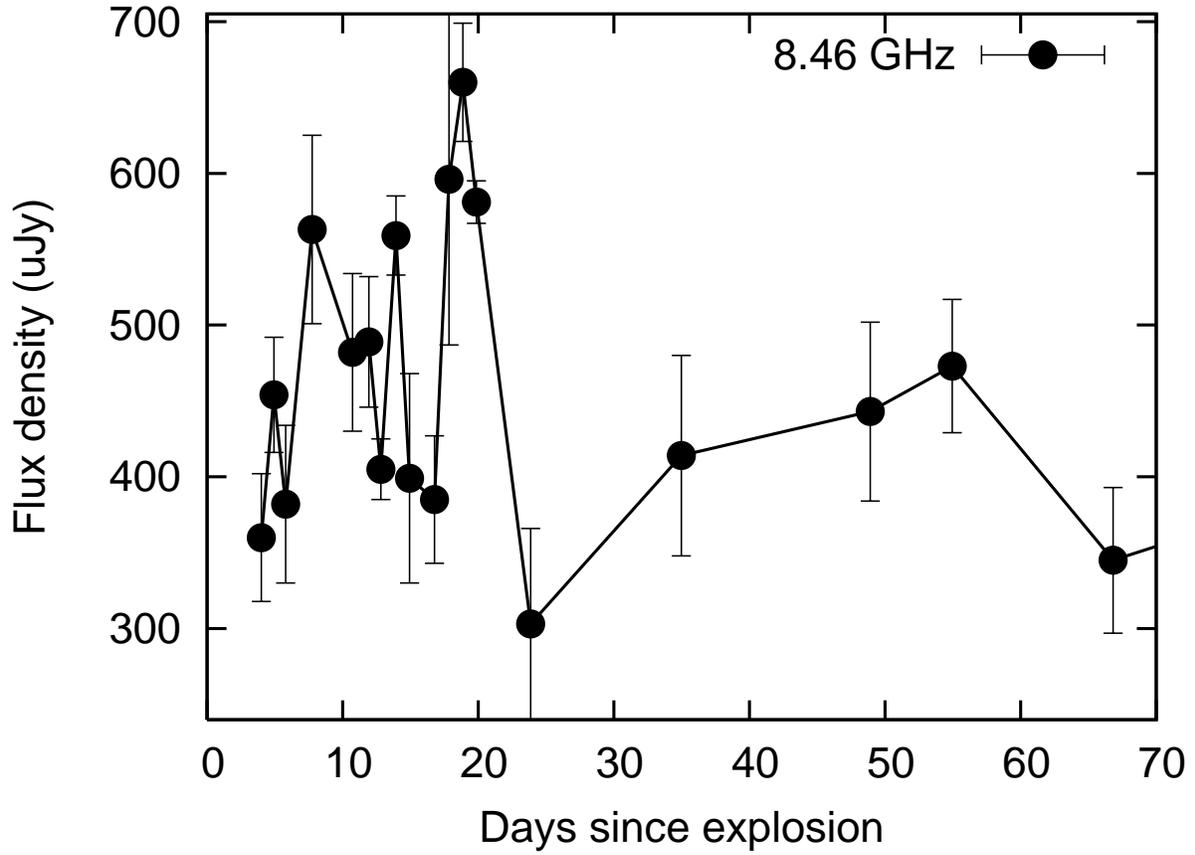}
\caption{Long term light curve at 8.46 GHz showing
evidence of interstellar scintillation. This scintillation
is refractive in nature and starts dominating once diffractive
scintillation quenches (see text for more details).
}
\label{fig:refscint}
\end{center}
\end{figure}

As can be seen in Fig. \ref{fig:refscint} and 
Table \ref{tab:radio}, there are significant day-to-day 
deviations in the low frequency radio
light curves. These modulations are likely due 
to scintillation caused by interstellar propagation effects.

\begin{figure}
\begin{center}
\includegraphics[angle=0,width=0.49\textwidth]{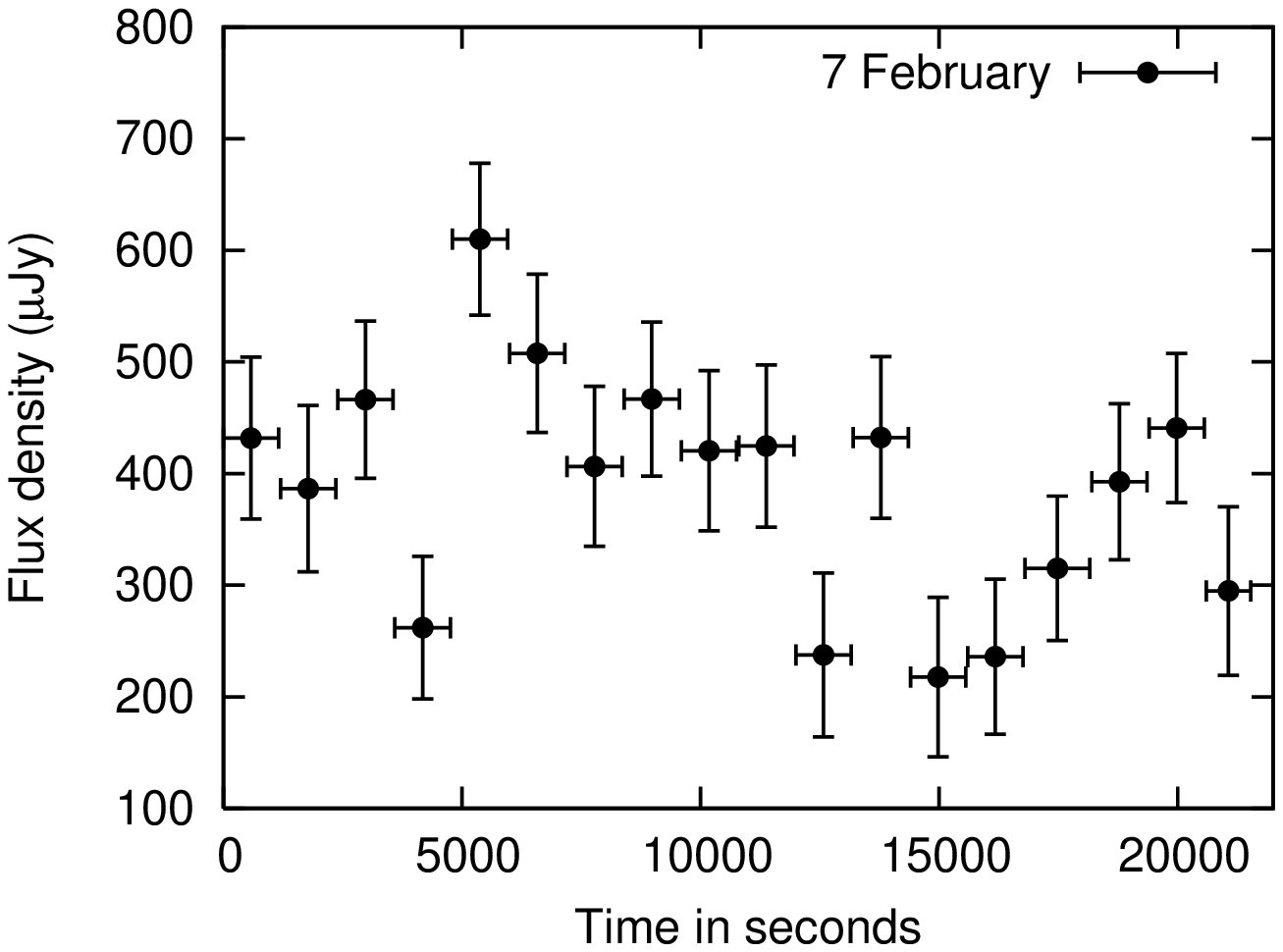}
\includegraphics[angle=0,width=0.49\textwidth]{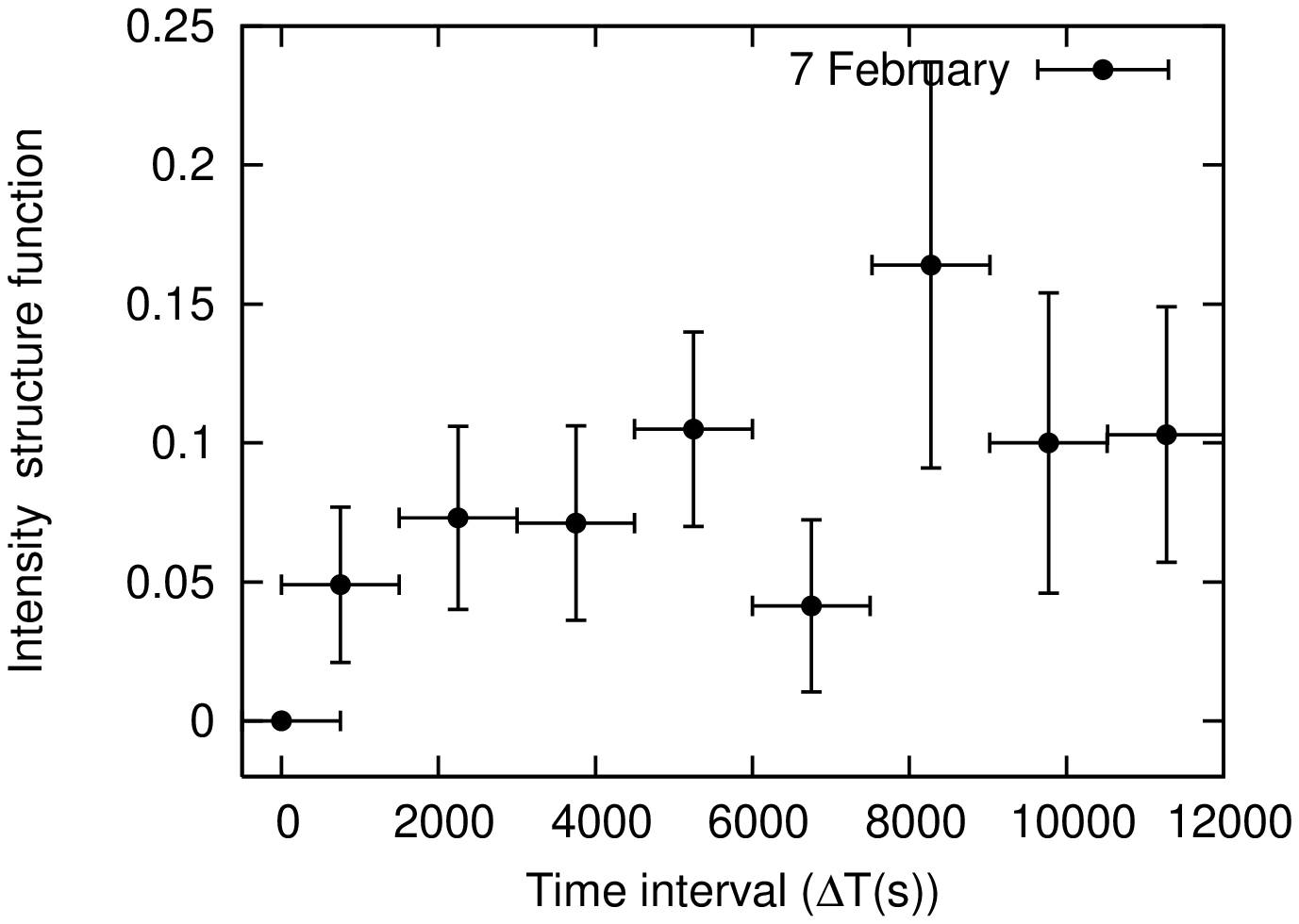}
\includegraphics[angle=0,width=0.49\textwidth]{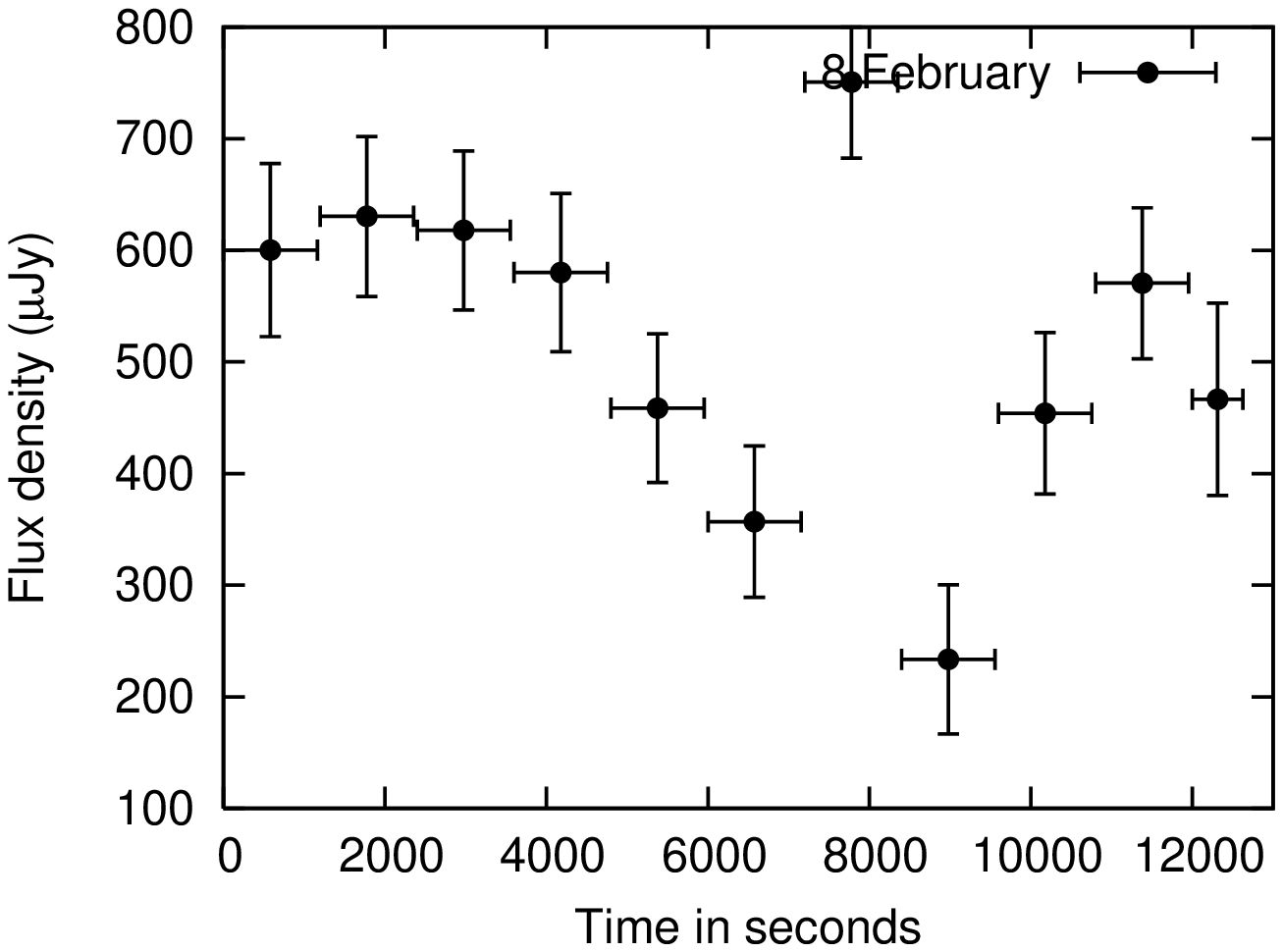}
\includegraphics[angle=0,width=0.49\textwidth]{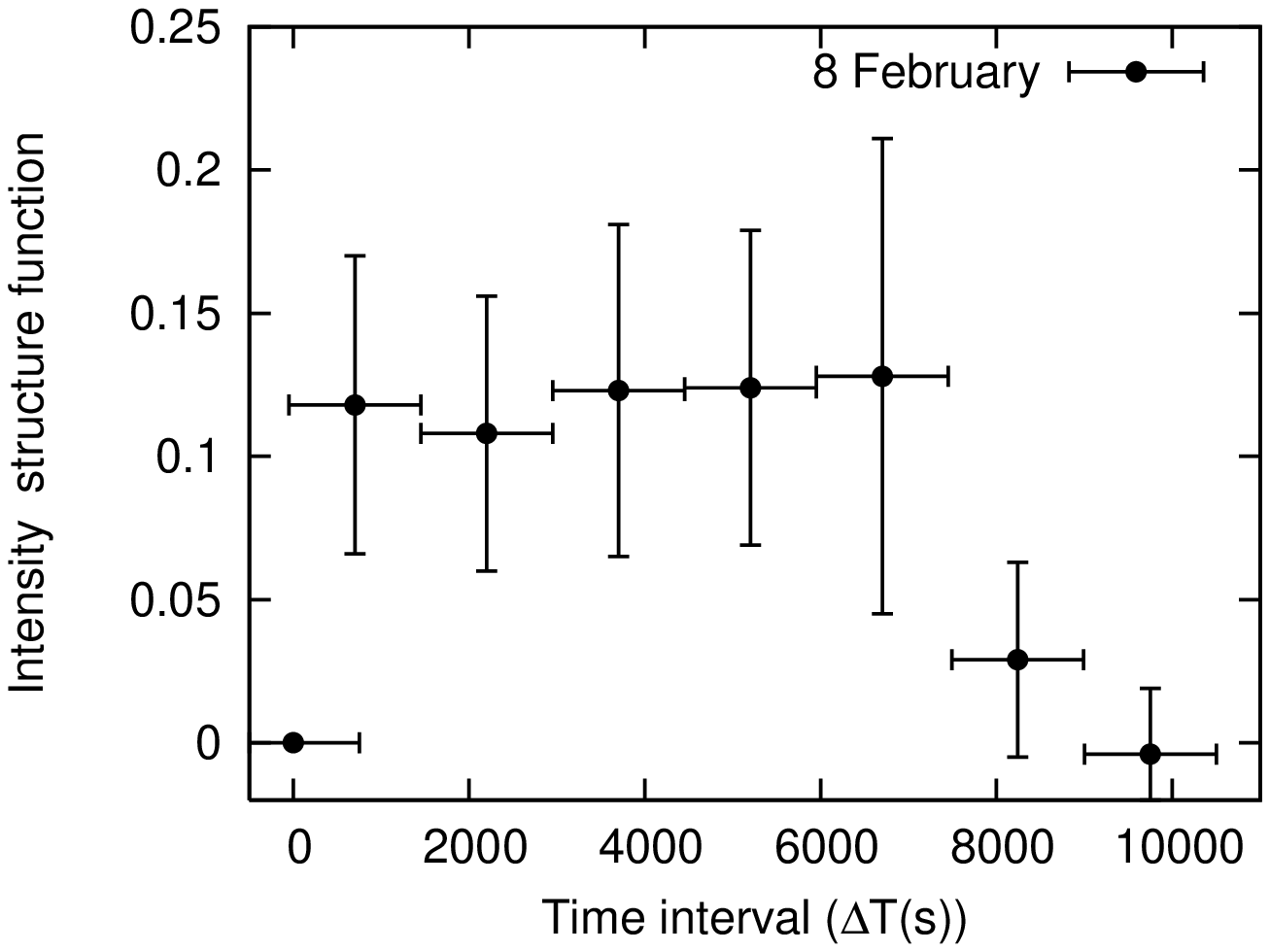}
\includegraphics[angle=0,width=0.49\textwidth]{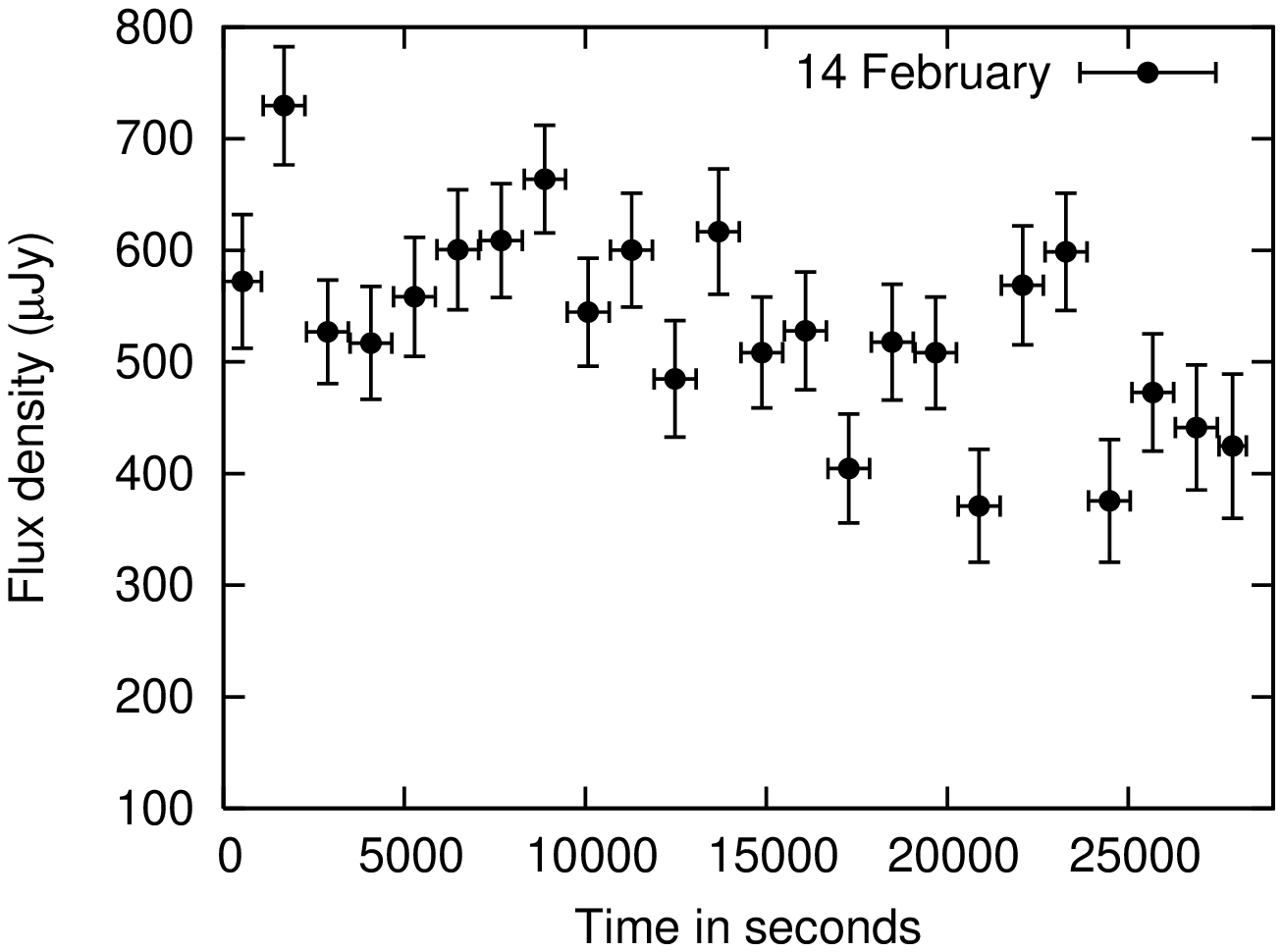}
\includegraphics[angle=0,width=0.49\textwidth]{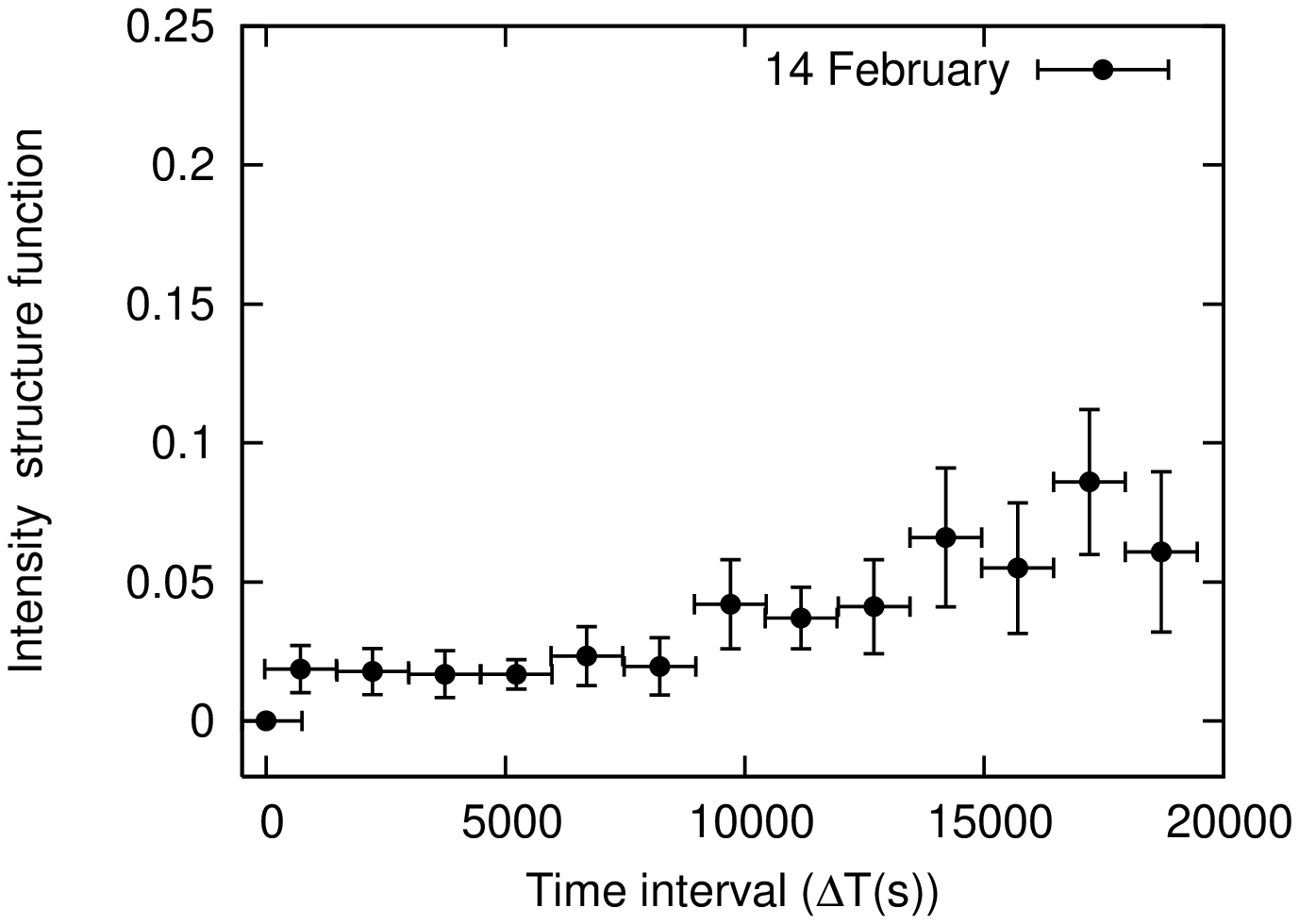}
\caption{Left panel:
Lightcurves of the short timescale
variations on 7, 8 and 20 February 2007.  The data 
were sampled at intervals
of 20 mins. Right panel: Structure functions of the
intensity fluctuations on 7, 8 and 14 February 2007.  
The bias of the thermal
noise fluctuations, which adds in quadrature to the source variations in
the structure function, has been subtracted.
}
\label{fig:scint}
\end{center}
\end{figure}

We obtained long-duration observations of the afterglow of GRB\,070125 at
8.5 GHz on three separate occasions: February 7 (5.5 hour duration), 
February 8
(4 hour duration), and  February 14 (8 hour duration).  The data were
split into 20\,minute blocks and imaged in order to extract information on
the fast variability of the source.  GRB\,070125 exhibits flux density
variations with a significance exceeding 99.8\% on each of the three
epochs (see Fig.~\ref{fig:scint}).  The reduced $\chi^2$ values for the 
hypothesis of no variability are 2.31 (with 17 degrees of freedom), 4.42 
(10 d.o.f.) and 2.84 (23 d.o.f.) for the three dates, respectively.  

We used
intensity structure functions to determine the variability timescale on each
date.  While the short baseline hindered a definitive determination, we 
tentatively identify breaks with $20-30$\% 
accuracy at $\Delta T \approx 6 \times 10^3, 7 \times 10^3$ 
and $9 \times 10^3$ s, in the three structure functions.  The breaks are 
marginally significant in the data from 7 February and 8 February, but not so 
prominent in the 14 February images. This can be explained by quenching of the 
scintillation as the source expands at late times.  These measurements are 
congruent with the timescales of the peaks and troughs apparent in the 
corresponding centimeter-wavelength lightcurves (Fig.~\ref{fig:refscint}).

The interpretation of the variability 
depends on whether the scintillation occurs in the weak or strong regime.  
Strong scintillations require that the so-called Fried parameter (coherence
length scale, $s_d$) be smaller than the Fresnel size ($r_F$).
Strong chromatic scintillations is possible only for sources
smaller than $\lambda/s_d$. These lead to the following two conditions
\citep{goo97}:

\begin{equation}
\nu<13.4 \left(\frac{\rm SM}{10^{-3.19} \rm m^ {-20/3}\, kpc} \right)^{6/17}
\left(\frac{D_{\rm scr}}{\rm kpc}  \right)^{5/17}\;\; \rm GHz \equiv 
\nu_{ss}
\label{scint1}
\end{equation}

\begin{equation}
\theta_d=6.5 \left(\frac{\nu}{8.46 \; \rm GHz} \right)^{-11/5}
\left(\frac{\rm SM}{10^{-3.19} \rm m^ {-20/3}\, kpc} \right)^{3/5}\;\;
\mu \rm as \;\; 
\ll \sqrt{\frac{c}{2 \pi \nu D_{{\rm scr}}}}\, ,
\label{scint2}
\end{equation}
where SM is the scattering measure and $D_{\rm scr}$ is the effective
distance to the scattering material, which is essentially the 
Galactic ionized medium \citep{cl02}. 
To  determine 
this, we first estimate the scattering distance and the scattering measure 
using the formulation of \citet{cl02}.  Given the Galactic coordinates of 
\event, $(l,b) = (189.4,25.6)$, the expected SM is 
$10^{-3.19}\,$m$^{-20/3}\,$kpc and the effective distance to the scattering 
material is $D_{\rm scr}=0.84\,$kpc. 
For these parameters, the critical frequency obtained from
Eq. \ref{scint1} is 12.7 GHz. These parameters also satisfy the condition of
Eq. \ref{scint2}. Hence, 
the GRB is likely to be in the strong scintillation regime.
However, given the paucity of our knowledge of the distribution of 
scattering material off the Galactic plane, 
these estimates should be taken with caution.

Strong scintillation can be diffractive as well refractive in nature.  
Diffractive scintillation of a point-like source is characterized by 
variations with a modulation index ($m_p$) close to unity, where the modulation 
index is a measure of how much the modulated flux varies around its 
intrinsic flux density level. 
 The size limit for a source to exhibit diffractive 
scintillation ($\theta_1$) is \citep{goo97}:
\begin{equation}
\theta_1 = 1.2 \left(\frac{\nu}{8.46 \, \rm GHz} \right)^{6/5}
\left(\frac{D_{\rm scr}}{\rm kpc}  \right)^{-1}
\left(\frac{\rm SM}{10^{-3.19} \rm m^ {-20/3}\, kpc} \right)^{-3/5}\;
\mu \rm as.
\end{equation}
For actual size of the source, $\theta_s$,
$\theta_s \le \theta_1$.
Diffractive scintillation occurs when $\theta_d>\theta_1$, which is indeed 
the case for \event\ using the parameters discussed above.

We now attempt to limit the size of the emitting region using the formulation
of diffractive scintillation \citep{wal98,wal01,cl02}, but with more robust
estimates of $D_{\rm scr}$ and  SM. The timescale of diffractive scintillation 
can be expressed as:
\begin{eqnarray}
t_{\rm diff}= 5950 \left(\frac{\nu}{8.46 \, \rm GHz} \right)^{6/5}
\left(\frac{\rm SM}{10^{-3.19} \rm m^ {-20/3}\, kpc} \right)^{-3/5}
\left( \frac{v_{\rm ISS}}{30\, \rm km \, s^{-1}} \right)^{-1}\,\rm  s,
\end{eqnarray}
where $v_{\rm ISS}$ is the speed of the scattering material transverse to the 
line of sight.  The decorrelation bandwidth ($\Delta \nu_{\rm dc}$) for 
diffractive scintillation is then \citep{goo97}:
\begin{eqnarray}
\Delta \nu_{\rm dc} = 1.55 \left(\frac{\nu}{8.46 \, \rm GHz} \right)^{22/5}
\left(\frac{D_{\rm scr}}{\rm kpc}  \right)^{-1}
\left(\frac{SM}{10^{-3.19} \rm m^ {-20/3}\, kpc} \right)^{-6/5}\, {\rm GHz}.
\label{eq:nu}
\end{eqnarray}

Using the variability timescale of $\Delta T \sim 7 \times 10^{3}$\,s measured
on 8 February, along with $v_{\rm ISS} =30\,$km\,s$^{-1}$, we infer a scattering
measure of $SM=0.76 \times 10^{-3.19}\, \rm m^{-20/3}\, {\rm kpc}$.  In 
Eq.~\ref{eq:nu} we take the decorrelation bandwidth to be roughly half of the
frequency of observation, i.e.~$\Delta \nu_{\rm dc} \approx (1/2)\times \nu$,
because at the boundary of strong and weak scintillation $\Delta \nu_{\rm dc} 
\approx \nu$ \citep{goo97}.
This yields a distance to the scattering screen of $D_{\rm scr}=0.51$ kpc.

Knowing these two parameters, we can put an upper limit on the 
angular size of the emitting region of $\theta_{\rm src} =2.8\pm0.5\, 
\mu\rm as$. Here the error in angular size corresponds to 20\% error in
the determination of $\Delta\,T$. 
At $z = 1.547$ \citep{cfp+07}, the angular size translates\footnote{All the 
calculations were done with $H_0=71$, $\Omega_m=0.27$ and $\Omega_{vac}=0.73$.}
 to a linear size of $(4.7 \pm 0.8) \times 
10^{17}$\,cm or a radius of $2.4 \times 10^{17}$\,cm. 

Variability due to refractive scintillation is
expected at late stages, even after the source has expanded sufficiently.
Refractive scintillations 
are expected to be broadband in nature  
and start dominating once
 diffractive scintillations are quenched.  
The source exhibited refractive scintillation during the two months
subsequent to 2007 February 8, as can be seen from the first
70\,days of the 8.46\,GHz data plotted in Figure~\ref{fig:refscint}.

In a standard GRB  afterglow model,  the jet
starts to spread sideways after the jet break
\citep{pir99, pir05, mes06}. During this stage, the fireball 
size remain constant. Once the
jet has become spherical, it reaches the
non-relativistic regime. In this regime,
the equations of motion follow self-similar
Sedov-Taylor solutions.
Thus, this estimate
represents the size of the fireball in the post jet break regime until the
expansion becomes sub-relativistic, which occurs between 
days 30 and 50.

\subsection{Energetics}
\label{sec:energetics}

The isotropic gamma ray energy of a GRB can be written as:
$$E_{\rm \gamma, iso}=3 \times 10^{51}\, {\rm erg}\,
\left(\frac{2}{1+z}\right)
\left(\frac{d_L}{7.12\,{\rm Gpc}}\right)^2 \left(
\frac{f_\gamma}{10^{-6}}\right),$$
where $d_L$ is the luminosity distance of the GRB and $f_\gamma$ 
is total fluence. 
The fluence in the 20 keV -- 10 MeV energy range is
$1.74 \times 10^{-4}$ erg cm$^{-2}$ \citep{gam+07,bhp+07}, yielding 
$E_{\rm \gamma, iso}= 1.06 \times 10^{54}$ erg.  This is one of the largest
isotropic energy releases (top 1\%) 
ever reported for a GRB \citep{a06}.

To determine the true prompt energy release, this isotropic value needs
to be corrected for collimation.  The combined fit to the optical and X-ray 
data gave a jet break at $t_{\mathrm{j}} \approx 3.78$ day (\S \ref{sec:tjet}).
The collimation correction depends on the density profile
of the circumburst medium.  
We derive corrections for a uniform density (ISM; $n=$ constant) 
medium and a wind-like (wind; $n=Ar^{-2}; 
\; A=3\times 10^{35}A_\star\;\rm cm^{-1}$) medium \citep[][and references
therein]{wdhl05}.  
The collimation angle for a radiative afterglow can be written as 
\citep{spn98,fks+01,lc03}:
\begin{eqnarray}
\theta_j({\rm ISM})=0.20\left(\frac{t_j}{1\, \rm day}\right)^{3/7}
\left(\frac{2}{1+z} \right)^{3/7}
\left(\frac{\Gamma_0}{200} \right)^{1/7}
\left(\frac{E_{52}}{n} \right)^{-1/7} \\ 
\theta_j({\rm Wind})=0.50\left(\frac{t_j}{1\, \rm day}\right)^{1/3}
\left(\frac{2}{1+z} \right)^{1/3}
\left(\frac{\Gamma_0}{200} \right)^{1/3}
\left(\frac{E_{52}}{A_\star} \right)^{-1/3}.
\end{eqnarray}
where $t_j$ is the break in the optical light curve in days, and 
$E_{52}$ is the isotropic kinetic energy of the 
fireball in units of $10^{52}$ erg. The term $\Gamma_0$ is the 
initial Lorentz factor of the fireball.

Similarly, the size of the spherical fireball can be expressed in the ISM and
wind media as \citep{spn98,lc03}:
\begin{eqnarray}
R({\rm ISM})=1.3 \times 10^{17} \left(\frac{E_{52}}{n}\right)^{2/7}
 \left(\frac{\Gamma_{0}}{200}\right)^{-2/7}
 \left(\frac{1+z}{2}\right)^{-1/7} t_{\rm days}^{1/7}\;\; \rm cm\\ 
R({\rm Wind})=0.17 \times 10^{17} \left(\frac{E_{52}}{A_\star}\right)^{2/3}
 \left(\frac{\Gamma_{0}}{200}\right)^{-2/3}
 \left(\frac{1+z}{2}\right)^{-1/3} t_{\rm days}^{1/3}\;\; \rm cm,
\end{eqnarray}
which after jet break in \event\ translates to:
\begin{eqnarray}
&&R({\rm ISM})=1.3 \times 10^{17} \left(\frac{E_{52}}{n}\right)^{1/3}
  \left(\frac{\Gamma_{0}}{200}\right)^{-1/3}
 \left(\frac{\theta_j}{0.2}\right)^{1/3} \;\; \rm cm 
\label{eq:r_jet1} \\
&&R({\rm Wind})=0.17 \times 10^{17} \left(\frac{E_{52}}{A_\star}\right)
  \left(\frac{\Gamma_{0}}{200}\right)^{-1}
 \left(\frac{\theta_j}{0.5}\right) \;\; \rm cm.
\label{eq:r_jet}
\end{eqnarray}
Using the emission radius derived from scintillation studies ($R \le2.4 \times 
10^{17}\,\rm cm$), we find an opening angle of $\theta \le 0.25$ rad 
($14^{\circ}$) for the ISM model and $\theta \le 0.23 $ rad 
($13^{\circ}$)
for the wind model.  The Lorentz factor of the 
    shocked ejecta at the time of
jet break is $\gamma (t_{\rm jet}) \cong 1/\theta = 4$ in the ISM model
and $5$ in the wind model.

\subsection{Circumburst density}
\label{sec:density}

From Eq.~(\ref{eq:r_jet1}) and (\ref{eq:r_jet}), 
the circumstellar density can be written in terms of 
the kinetic energy of the afterglow as $n \ge E_{52}/0.11$ cm$^{-3}$ and 
$A_\star \ge E_{52}/1.54$ for the ISM and the wind models, respectively.
Let $\eta_\gamma$ be the efficiency factor for converting the fireball
energy into the radiation energy, i.e.~$\eta_\gamma=E_\gamma/(E_\gamma+E_K)$.
For an empirical value of $\eta_\gamma=0.35$ \citep{fyb+03}, 
the number density of \event\ is $n\approx 50 \, \rm cm^{-3}$ in the ISM model 
and $A_\star \approx 2.5$ in the wind model.  This value is quite high, even
for GRBs, and indicates that the afterglow of GRB\,070125 is expanding into 
a  dense medium.

\begin{figure}
\begin{center}
\includegraphics[angle=0, width=0.49\textwidth]{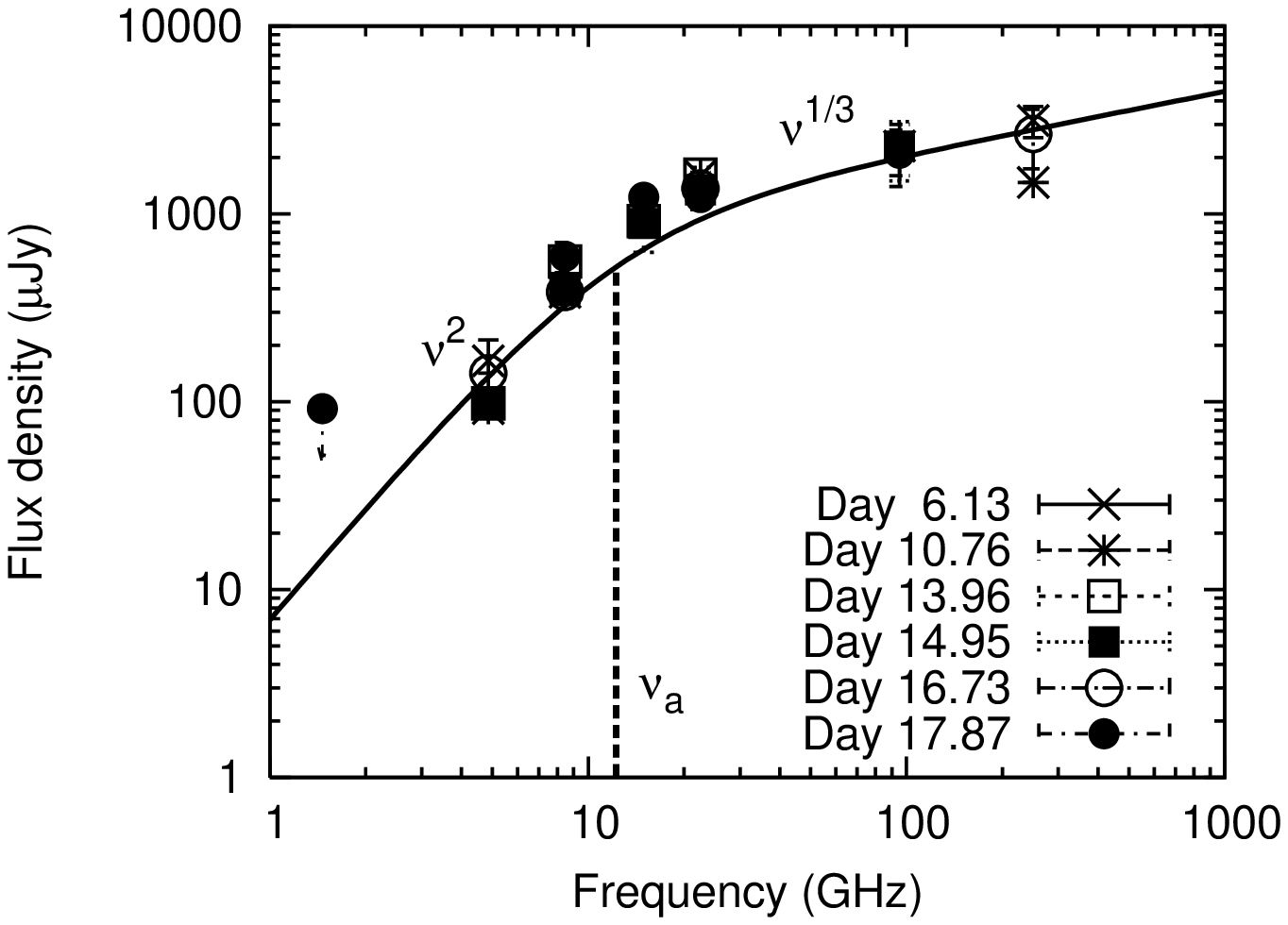}
\includegraphics[angle=0, width=0.49\textwidth]{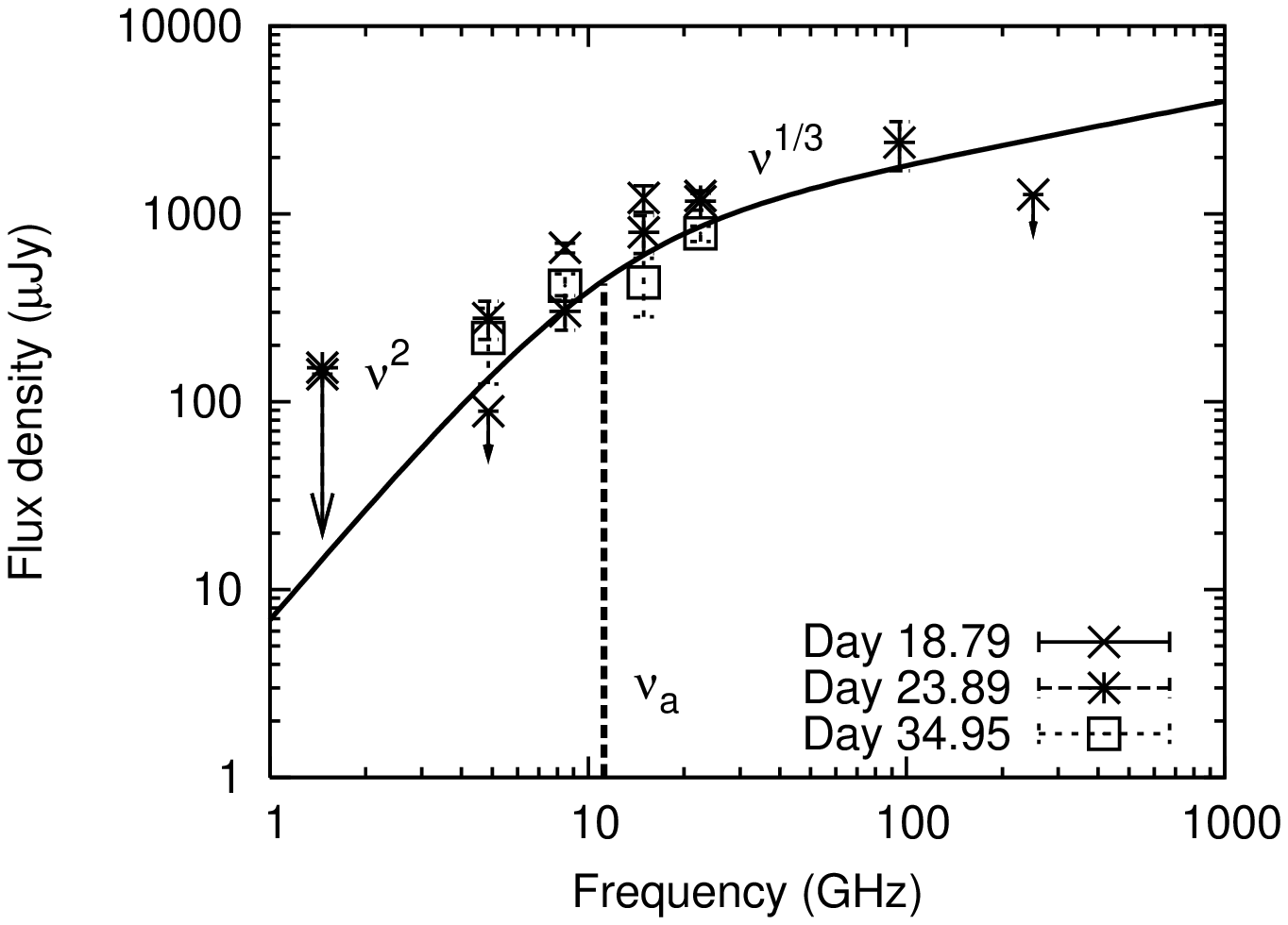}
\includegraphics[angle=0, width=0.49\textwidth]{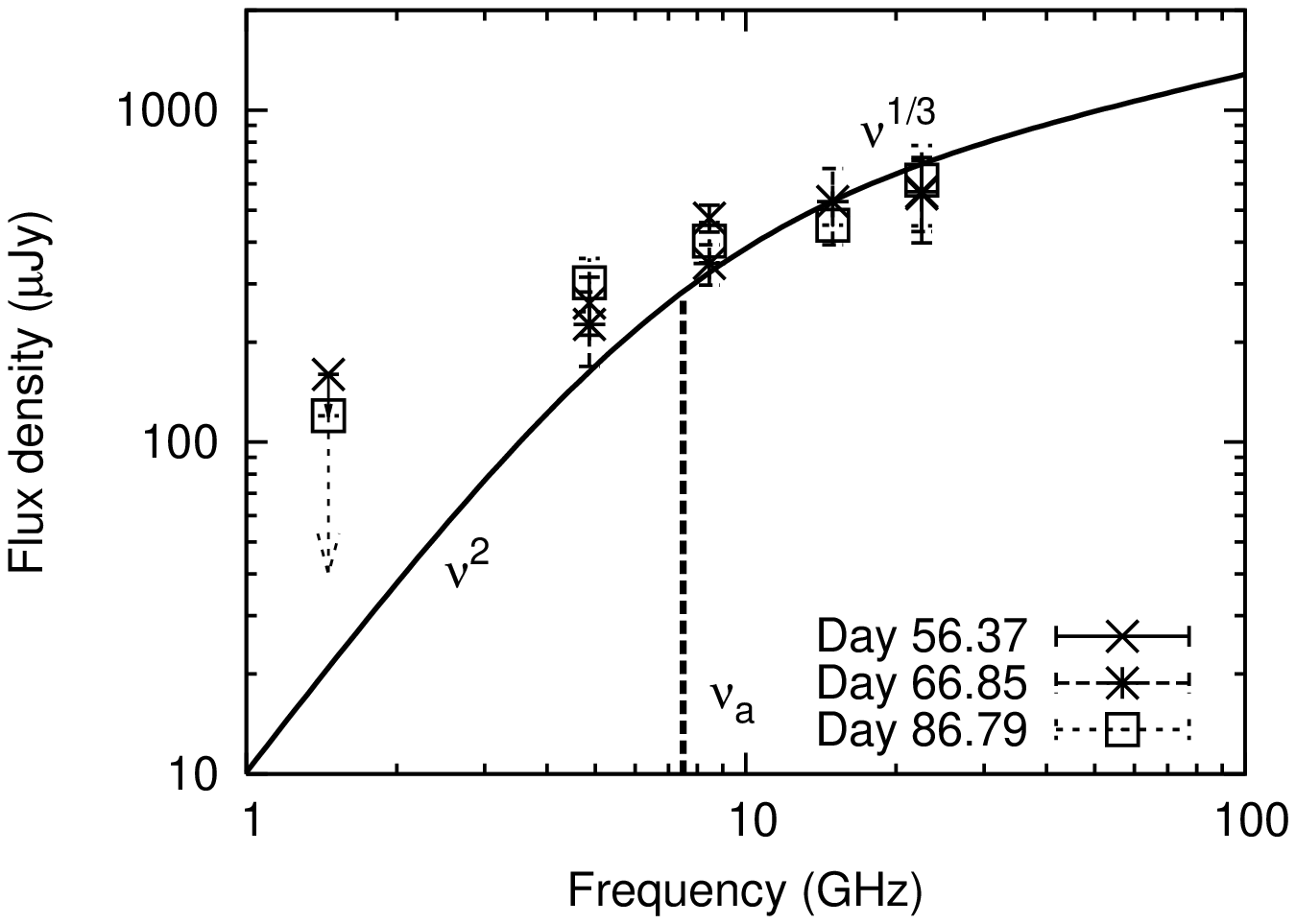}
\includegraphics[angle=0, width=0.49\textwidth]{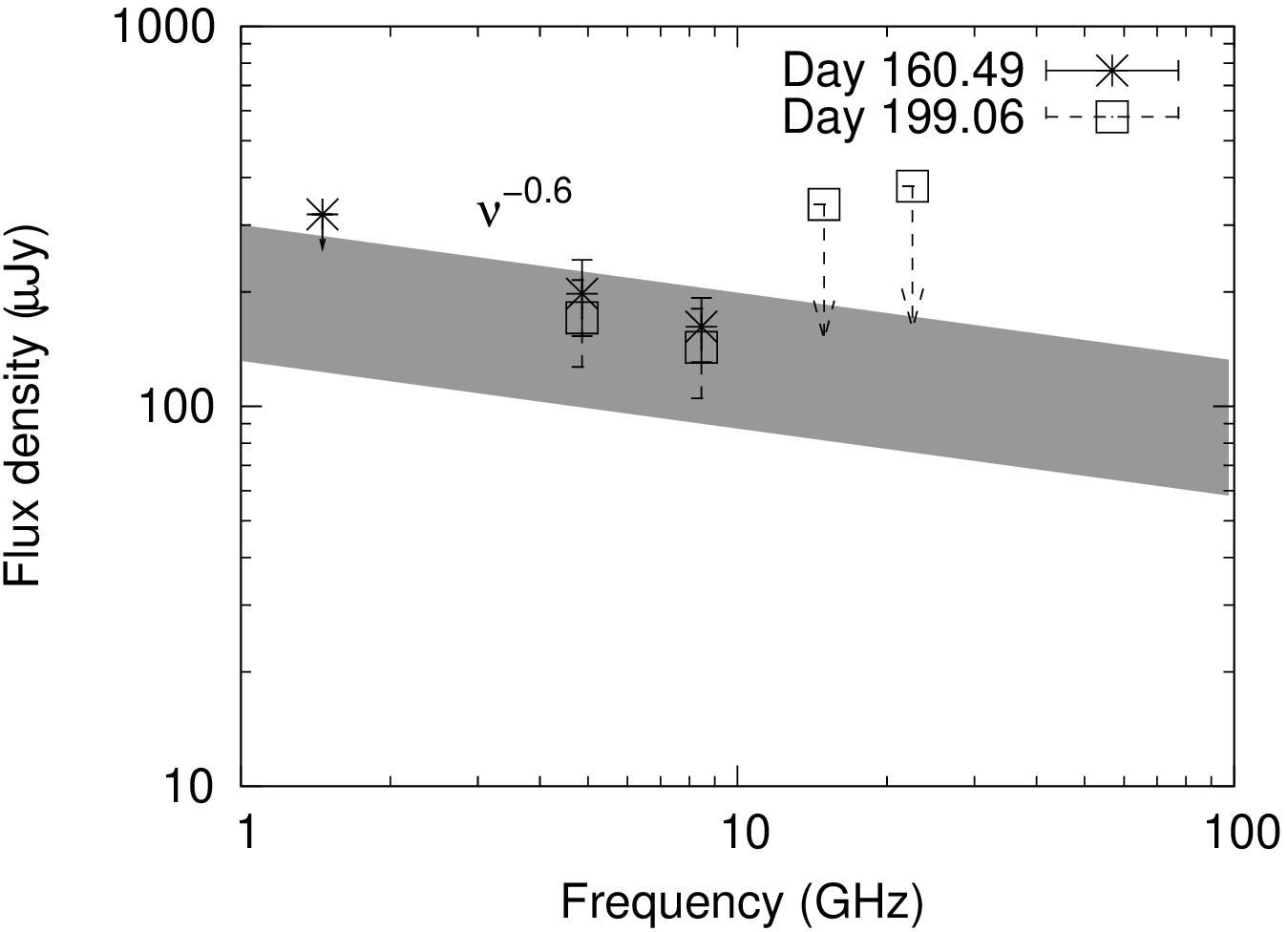}
\caption{Radio spectra of GRB 070125. 
We combined spectra between days
$6-17$, $18-34$,  $56-86$ and $160-200$ and extracted the
best fit $\nu_a$. 
Between day $160-200$ the spectrum is in the optically thin phase, 
$\nu_a$ is below the observed frequencies. The grey area indicates the
uncertainity region for the powerlaw index slope.
}
\label{fig:radio_spectra}
\end{center}
\end{figure}

A  natural consequence of a high circumburst density is a high
synchrotron self-absorption frequency.
To this end, we plot broadband radio spectra (Figure~\ref{fig:radio_spectra}).
To improve the signal-to-noise ratio, we binned the spectra into 4 
groups ($t = 6 -17$ d, $18 -35$ d, $56 -86$ d, and $160 - 200$ d).  The 
division was initially done on the basis of similarly-looking spectra, but 
later in this section we justify it by
demonstrating that the spectral evolution has a weak 
time dependence.  
As can be seen, 
there is a clear turnover in the spectra in the first three epochs.  Moreover, 
there is some indication that the turnover frequency evolves to lower 
frequencies with time. At  the last epoch, no turnover frequency is discernible 
and the spectrum is inverted.

This spectral behavior has been seen in many previously well-studied 
GRB afterglows.  We interpret this behavior in terms of the evolution of the 
afterglow from an optically thick to optically thin phase, parametrized by a 
single unknown, the synchrotron self-absorption frequency $\nu_a$.  We can 
measure the value of $\nu_a$ from the radio spectra using a very simple 
formulation given below:
\begin{equation}
 F_\nu= \cases{
F_{\rm max}\left(\displaystyle \frac{\nu}{\nu_a}\right)^2,
        & $\nu < \nu_a$, \cr
F_{\rm max}\left(\displaystyle \frac{\nu}{\nu_a}\right)^{\frac{1}{3}},
        & $\nu > \nu_a$. \cr
}
\label{eqn:nua}
\end{equation}
The above relation is a broken power-law with a break at $\nu_a$.  We used the 
following smooth approximation of eq.~(\ref{eqn:nua}):
\begin{equation}
F_\nu=F_{\rm max}\left(\frac{\nu}{\nu_a}\right)^2
\left[1+\left(\frac{\nu}{\nu_a}\right)^{2-\frac{1}{3}}\right]^{-1}
\end{equation}
We fit this function to the first three radio spectra and obtain the
following values of synchrotron self-absorption frequency: \\
\noindent $\nu_a = 12.25^{+1.70}_{-0.92}$ GHz in the range $6-17$ days,\\
\noindent $\nu_a = 11.22^{+1.00}_{-0.61}$ GHz in the range $18-35$ days,\\
\noindent $\nu_a = 7.49^{+0.36}_{-0.28}$ GHz in the range $56-86$ days.\\

Our approximation of a constant $\nu_a$ within each epoch is justified because
of the slow evolution of $\nu_a$.  The best fit time dependence to the values 
given above is $\nu_a \propto t^{-0.24\pm0.05}$, which agrees well with the
time dependence predicted in the fireball model 
($\nu_a \propto t^{-0.2}$, \citet{fyb+03,mes06}).  The spectrum in the final
epoch was well into the optically thin phase so we could not determine $\nu_a$
at this epoch. 

We also plot the multiwaveband spectra 
on day 10.7 and day 23.4 (Fig. \ref{fig:analytic}), the 
two epochs at which we had observations in all the bands 
simultaneously.  
This shows radio
data points to be on the optically thick part of the spectra.
The afterglow peaks at 3-mm band and the optical and X-ray data 
fall in the optically thin regime.
This gives a rough estimate of the
break frequencies, $\nu_m$, corresponding to the
minimum electron Lorentz factor, and $\nu_c$, the
cooling frequency. We also plot $\nu F_{\nu}$, a measure of energy, 
against $\nu$. The curve peaks at the cooling frequency
$\nu_c$
(Fig. \ref{fig:analytic}). 


\section{Broadband Modeling}
\label{sec:multiwave}

In the previous section, we used simple analytical techniques to estimate
four fundamental physical properties of the explosion: the opening angle 
($\theta \approx 0.23$ -- 0.25 radian), the size of the emitting region 
($R \approx (2.4  \times 10^{17}$\,cm), the collimation-corrected prompt
energy release ($E_{\gamma} \approx 3 \times 10^{52}$\,erg), 
and the circumburst
density ($n \approx 50$\,cm$^{-3}$; $A_{*} \approx 2.5$).  Here
we combine \textit{all} our observations of the \event\ afterglow in an
attempt to derive a comprehensive model of the entire afterglow evolution.
Our modeling software assumes a standard 
synchrotron forward shock formulation, 
with possible contributions from inverse Compton (IC) emission 
and radiative losses
also included. This model also includes scintillation uncertainties,
and hence gives realistic estimates of various parameters.
Further details can be found in 
\citet{2004PhDT.......429Y} and \citet{yhsf03}.

As in \S \ref{sec:tjet}, we ignore all data before $t = 1$\,d due to the 
possibility of late-time energy injection.  Based on our results in 
\S \ref{sec:scintillation}, we have incorporated scintillation effects into
the radio regime.
We used
an LMC-like extinction model for the optical data \citep{pei92}; 
however, the
extremely small host contribution to the extinction makes the effects of 
differing extinction laws negligible.

\subsection{Wind Model}

Results of the best fit wind model parameters are tabulated in Table
\ref{tab:models}. In terms of $\chi^2$ and the model-fit statistic\footnote{
$-{\rm ln}(P)=0.5 \times (\chi^2+2 \Sigma\, ln (\sigma_i))+$constant, here $P$
is the probability function and $\sigma_i$ is standard deviation}, the
wind model does a slightly better job than the ISM model. However,
the resulting best-fit parameters for the wind model
are either unphysical 
(i.e.~$\epsilon_{\mathrm{e}}$, the fractional energy imparted to electrons in
the shock, approaching unity) or quite different from the values we derived in
\S \ref{sec:result}.
The extremely small isotropic 
afterglow kinetic energy ($E_{52} \approx 0.3$)
compared to the $\gamma$-ray isotropic energy is 
also troubling (the same is
true in the ISM case, but to a lesser extent).  
We also notice that the wind 
model is less stable to small changes in the parameter space. 

We fit the wind model with and without the inverse 
Compton effects. The model
without inverse Compton effect gives even more unphysical values
with many of the parameters asymptotically reaching very high values.
The magnetic
field fraction required reaches 100\% in this model.

Apparently allowing $\epsilon_e$ to be a free parameter is 
problematic. Microphysics evolution has been considered by \citet{yhsf03,
2004PhDT.......429Y}. 
 It makes everything unconstrained.
We fixed $\epsilon_e$ to be 0.4 and 
obtained a good fit (see Table \ref{tab:models}). 
This exercise demonstrates that 
microphysical parameters are not constrained by our observations, at
least for wind model.

\subsection{ISM Model}

The best-fit results for the ISM model are also provided in Table 
\ref{tab:models}.  The ISM model gives values of 
various parameters closer to 
the ones obtained from our simple analytical models.  
The jet break time, 
density scale, and collimation angle are in good 
agreement with our previous
results. 

The energy quoted in Table \ref{tab:models} is the 
isotropic blastwave energy 
at the time when $\nu_c=\nu_m$, i.e.~at the time of 
the transition from fast 
cooling to slow cooling ($t \sim 8$\,d in our model).  
This isotropic kinetic energy is
much smaller than the isotropic gamma-ray energy obtained from the 
Konus-Wind/RHESSI fluence (\S \ref{sec:energetics}).  
This may indicate that 
either there are high radiative losses at early times or 
the prompt emission
is rather efficient with an extremely high value of $\eta_\gamma$.

\subsection{Broadband Model Results and Interpretation}

\begin{figure}
\begin{center}
\includegraphics[angle=0,width=0.49\textwidth]{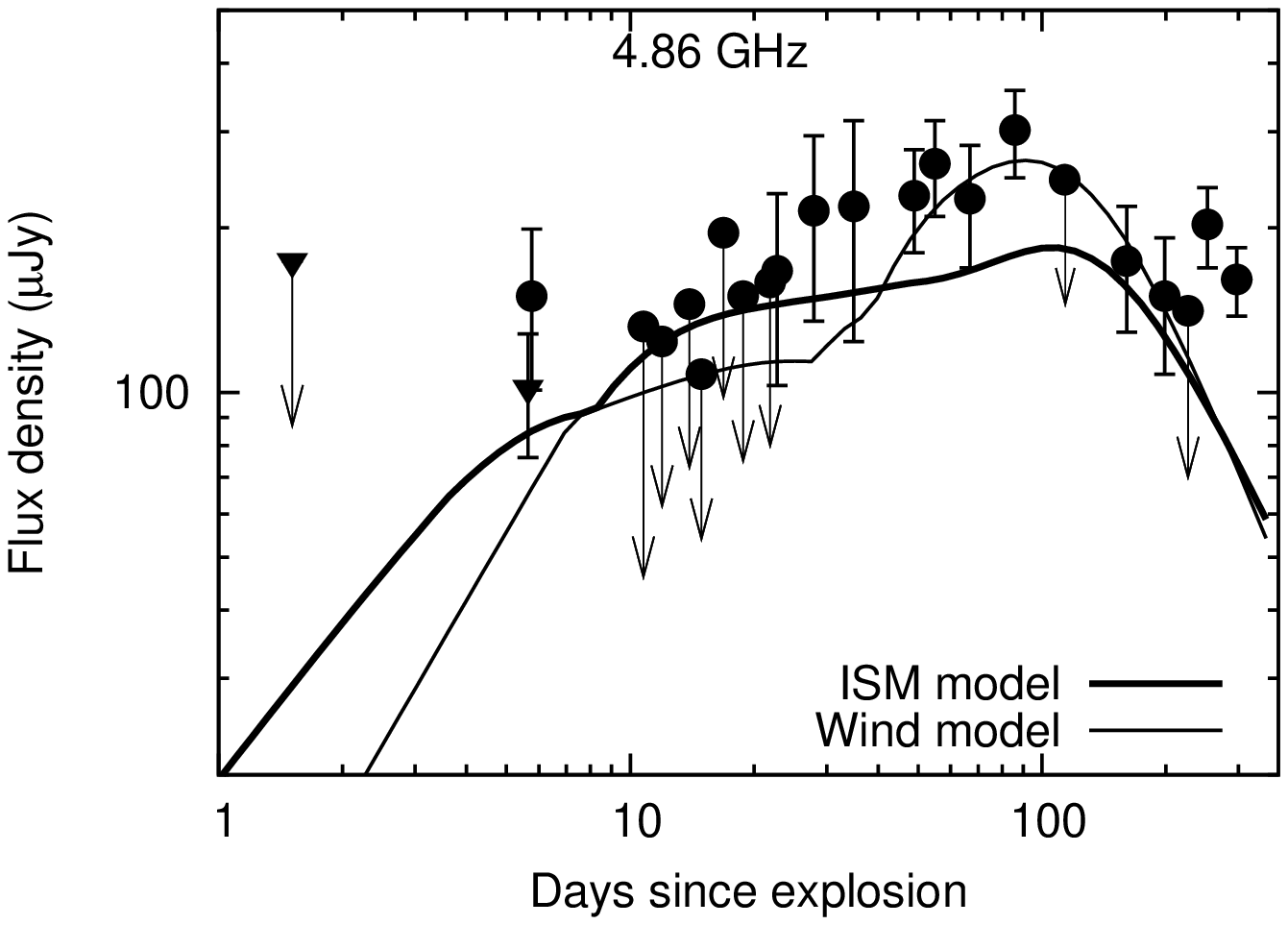}
\includegraphics[angle=0,width=0.49\textwidth]{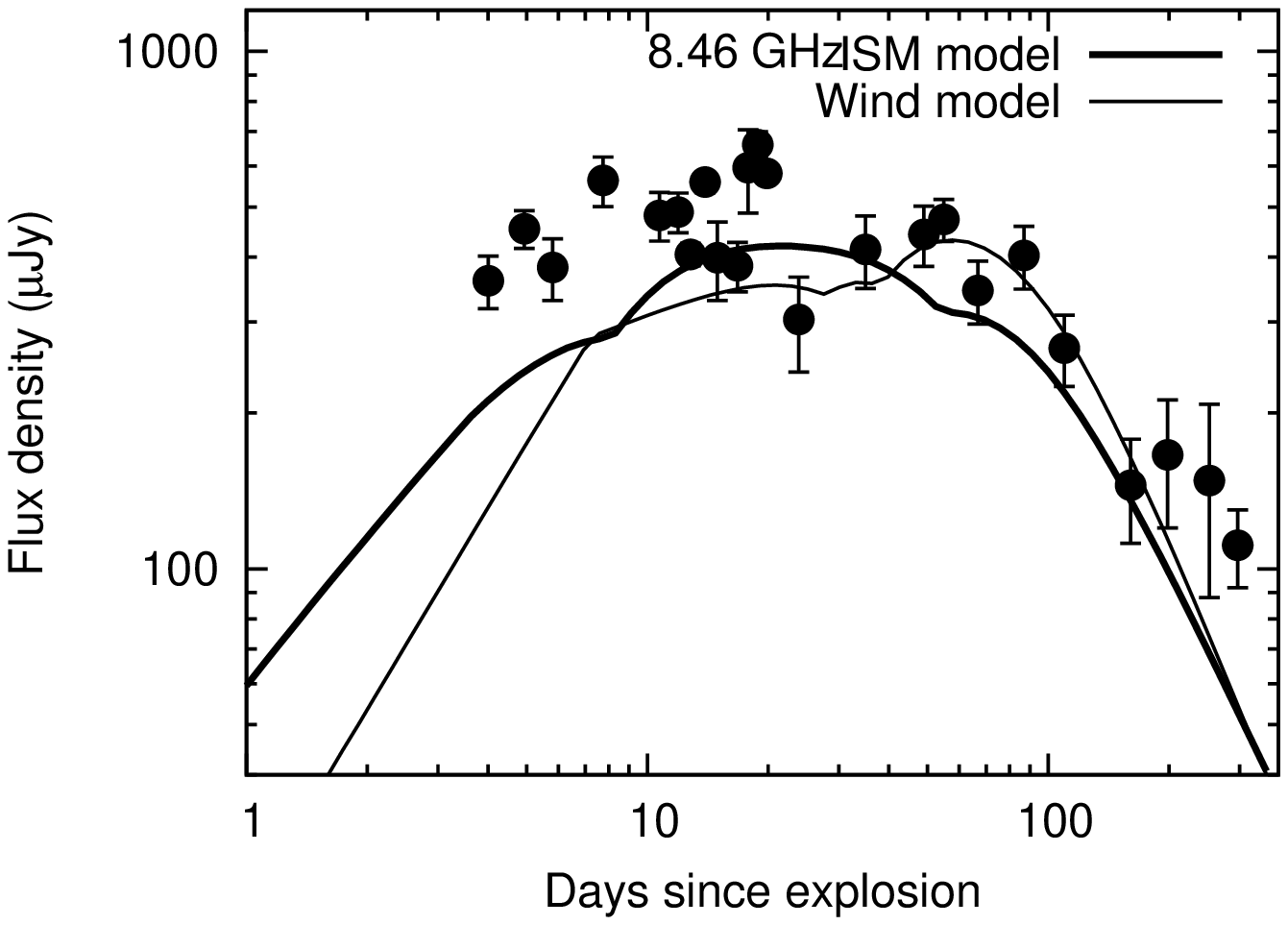}
\includegraphics[angle=0,width=0.49\textwidth]{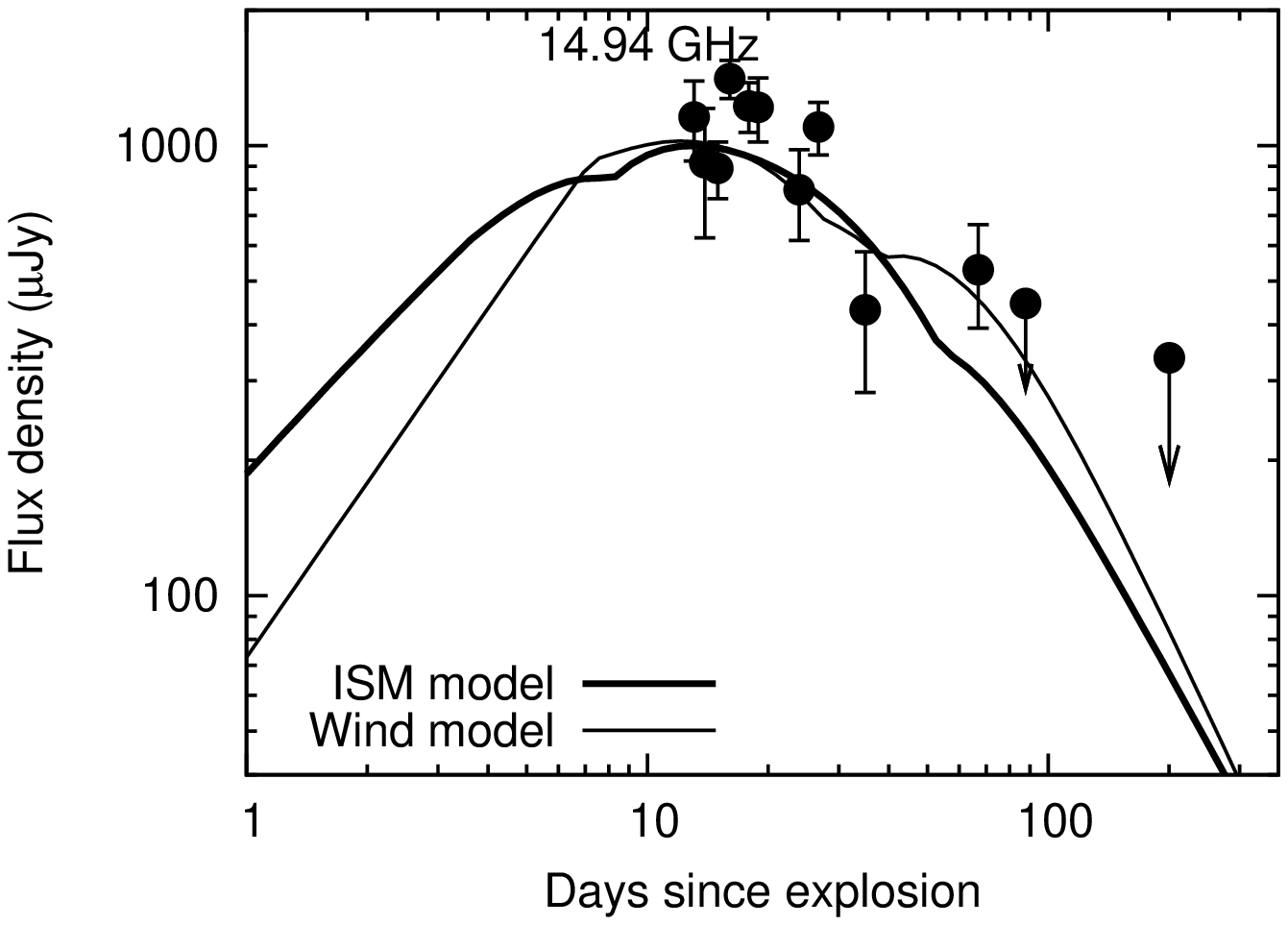}
\includegraphics[angle=0,width=0.49\textwidth]{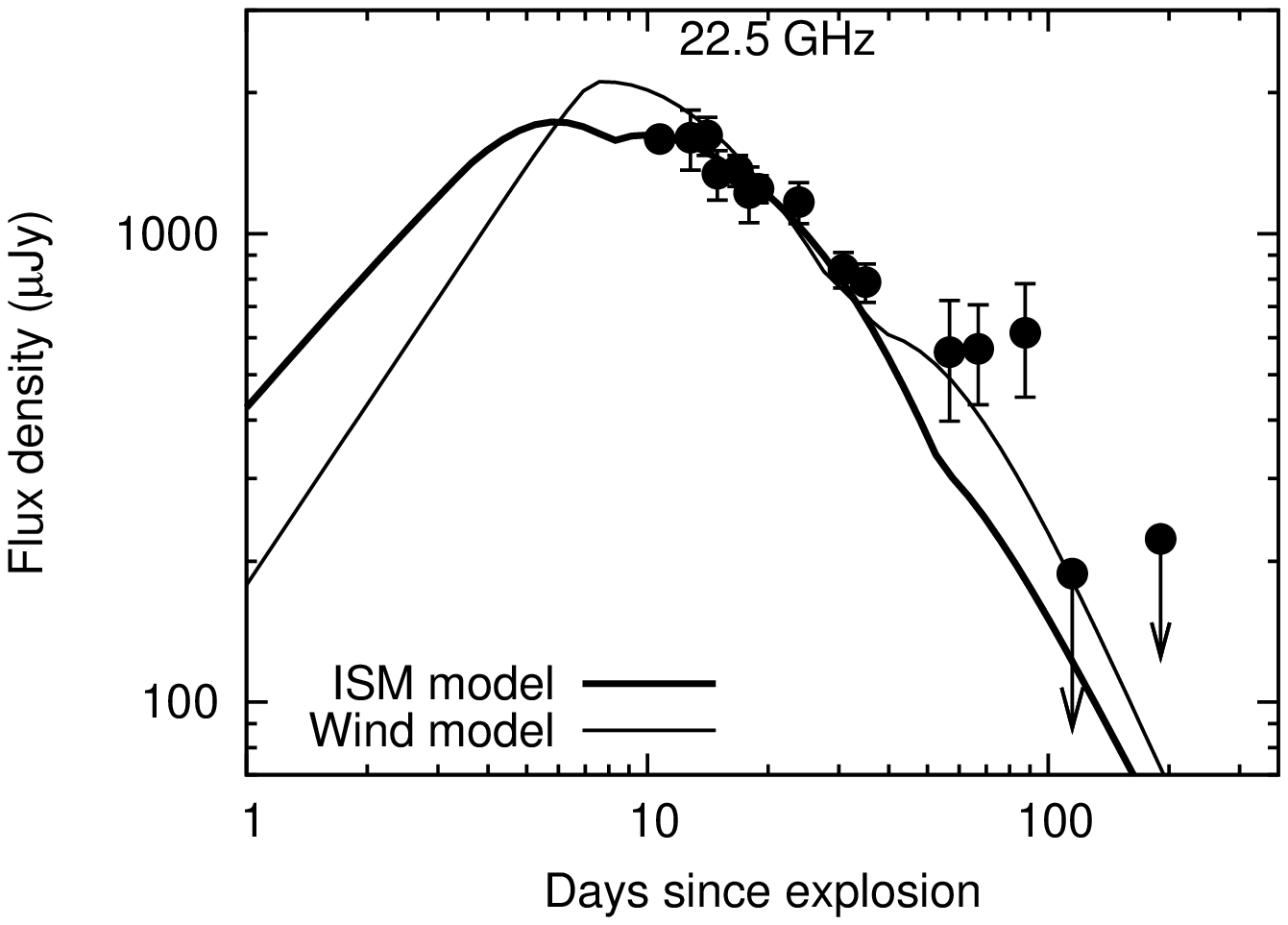}
\includegraphics[angle=0,width=0.49\textwidth]{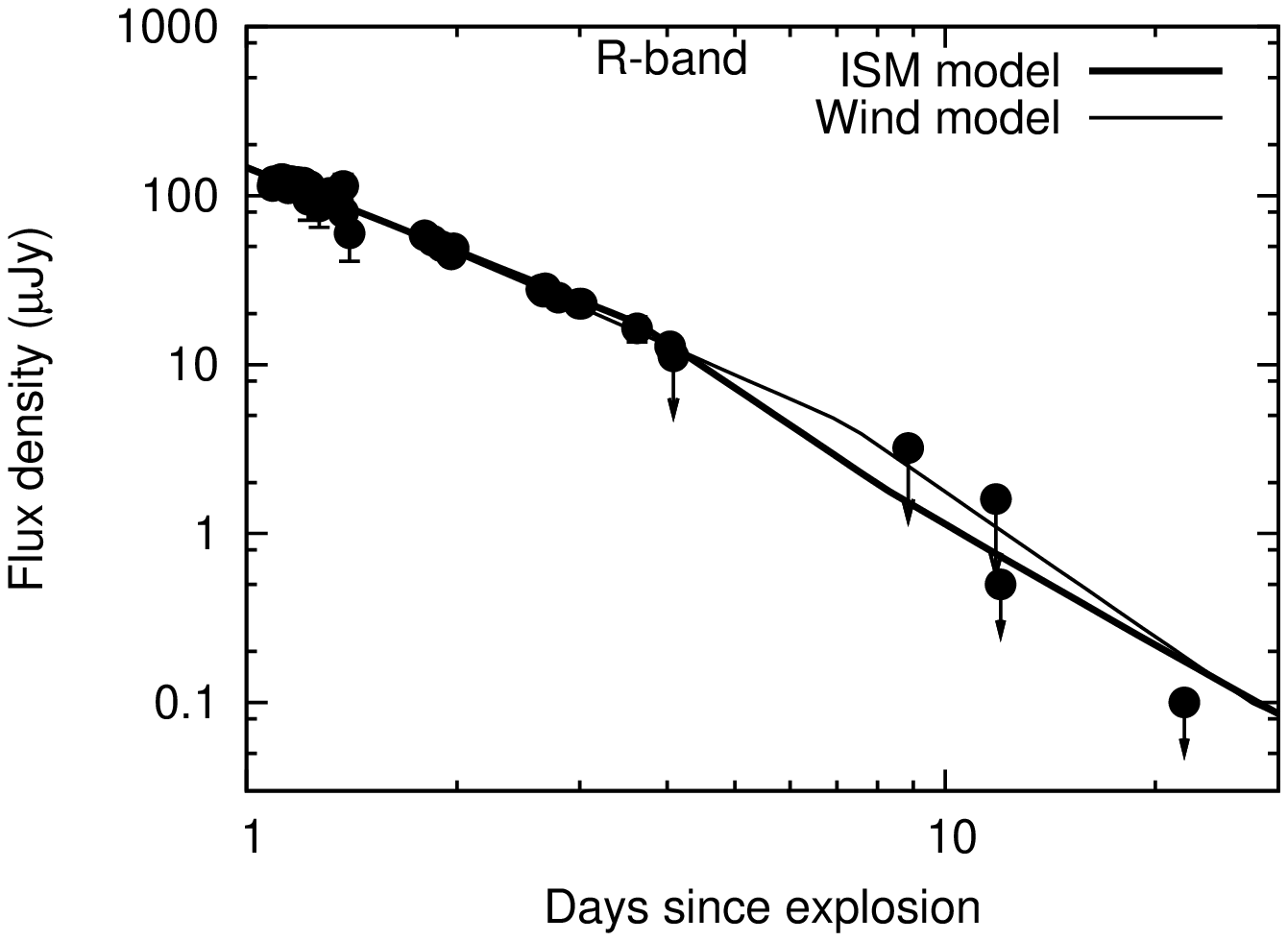}
\includegraphics[angle=0,width=0.49\textwidth]{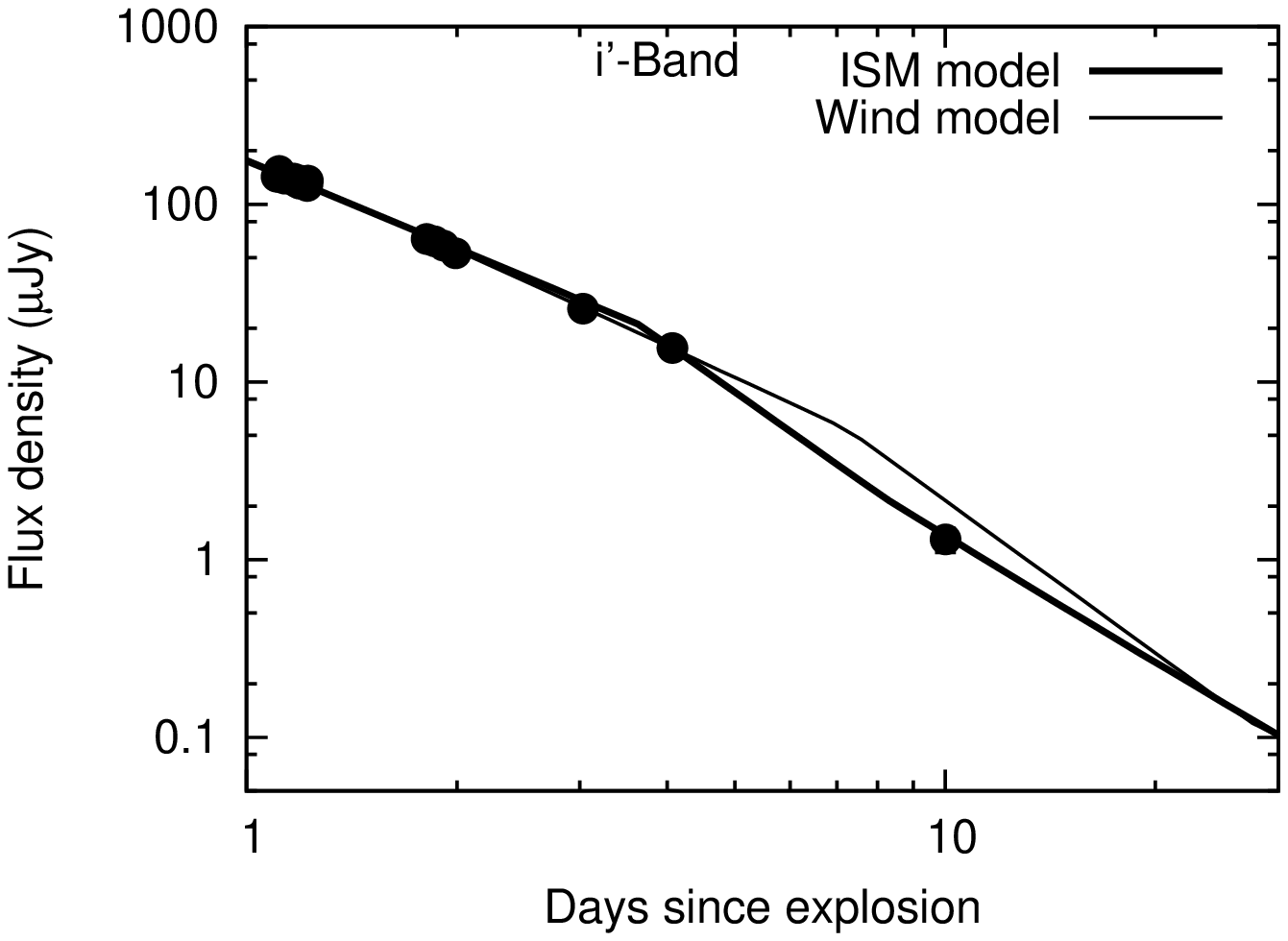}
\caption{Broadband modeling plots for the ISM model
as well as the Wind model.
We show light curves in  8.46 GHz, 4.86 GHz, 
$R$ band, and $i'$ band dataset. 
The darker line is for the ISM model and the Wind model is plotted with a thin line.
The fits in the radio bands are not good, probably because of 
scintillation effects.}
\label{fig:radio}
\end{center}
\end{figure}

\begin{figure}
\begin{center}
\includegraphics[angle=0,width=0.49\textwidth]{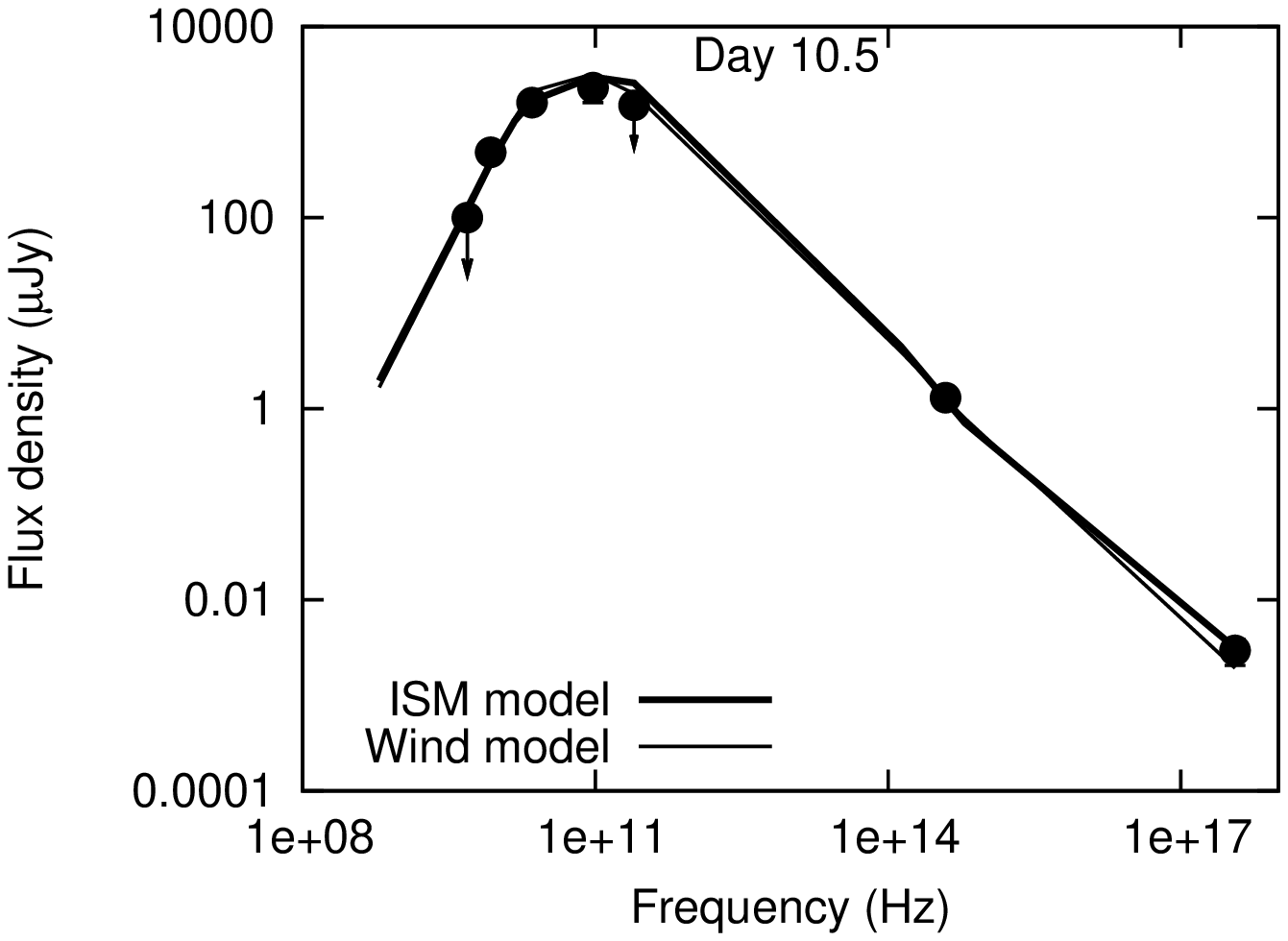}
\includegraphics[angle=0,width=0.49\textwidth]{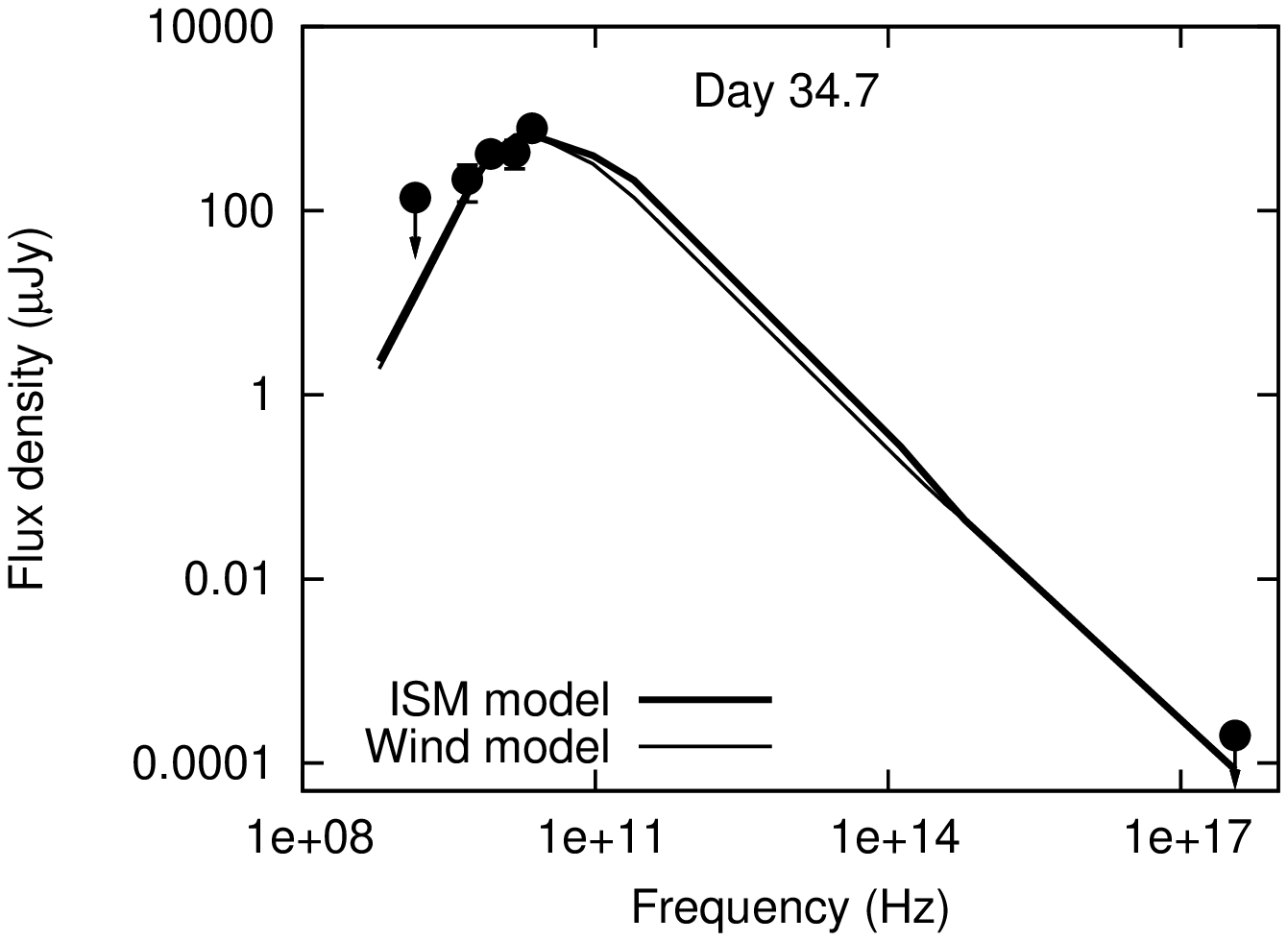}
\includegraphics[angle=0,width=0.49\textwidth]{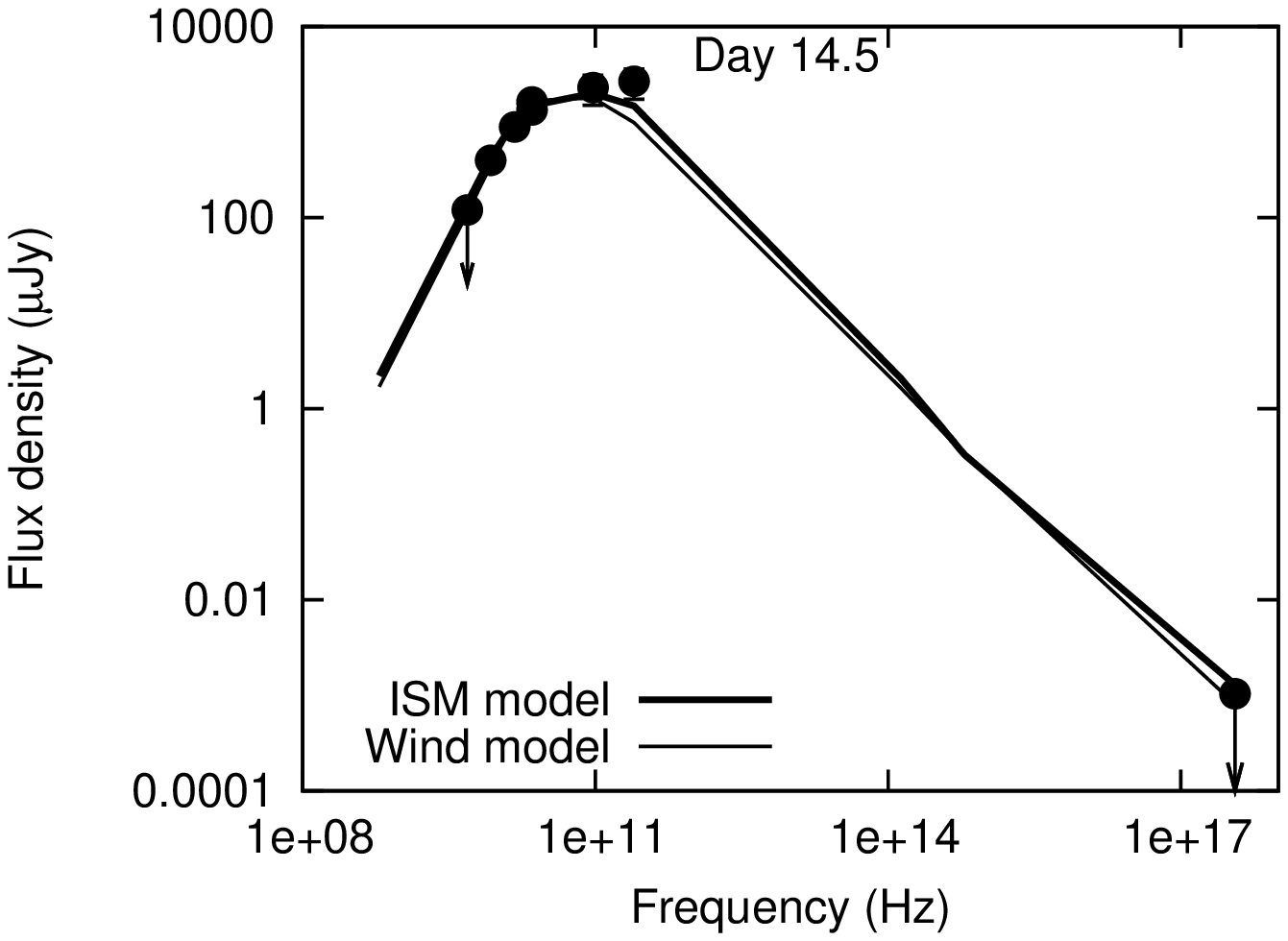}
\includegraphics[angle=0,width=0.49\textwidth]{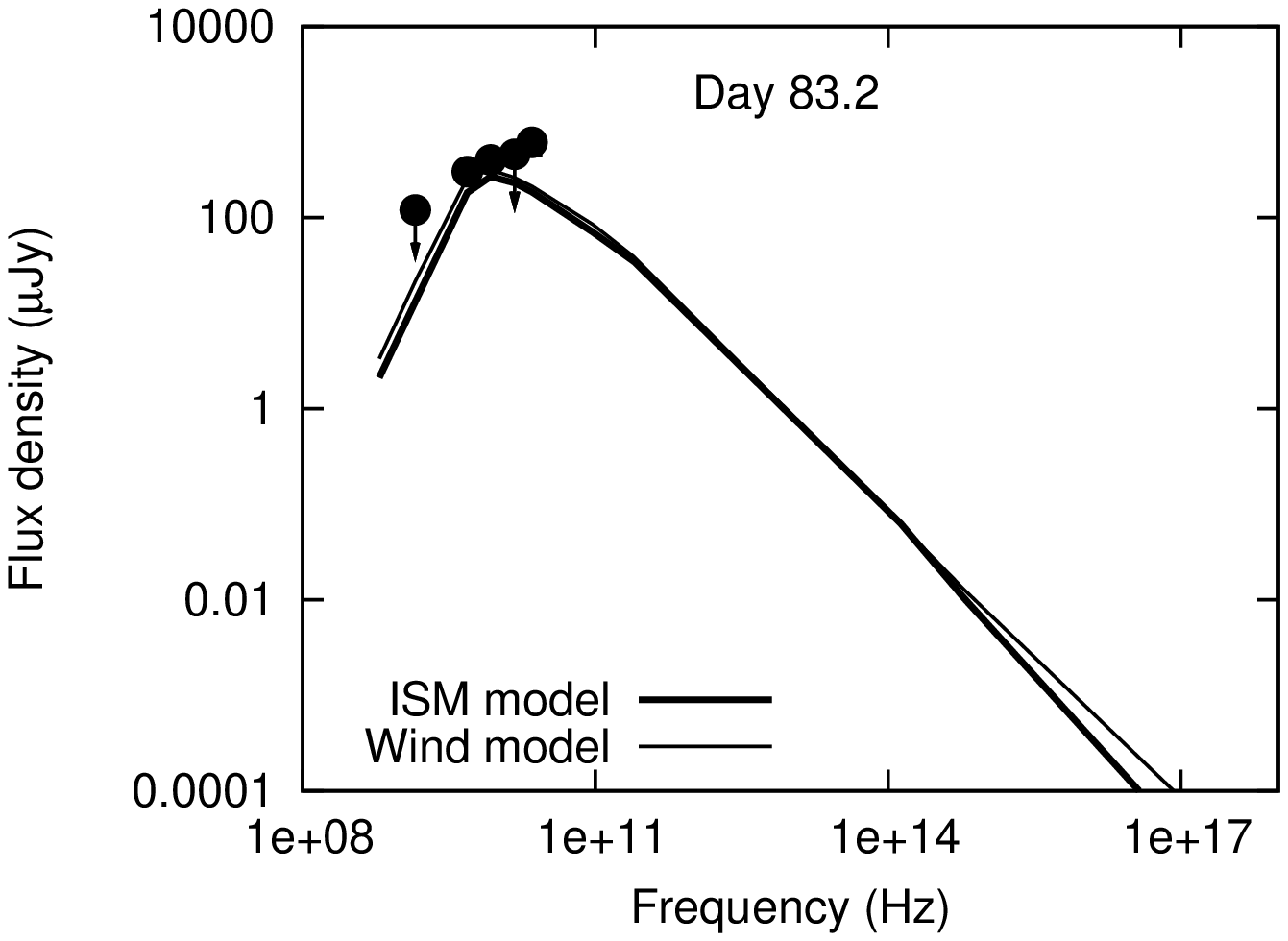}
\includegraphics[angle=0,width=0.49\textwidth]{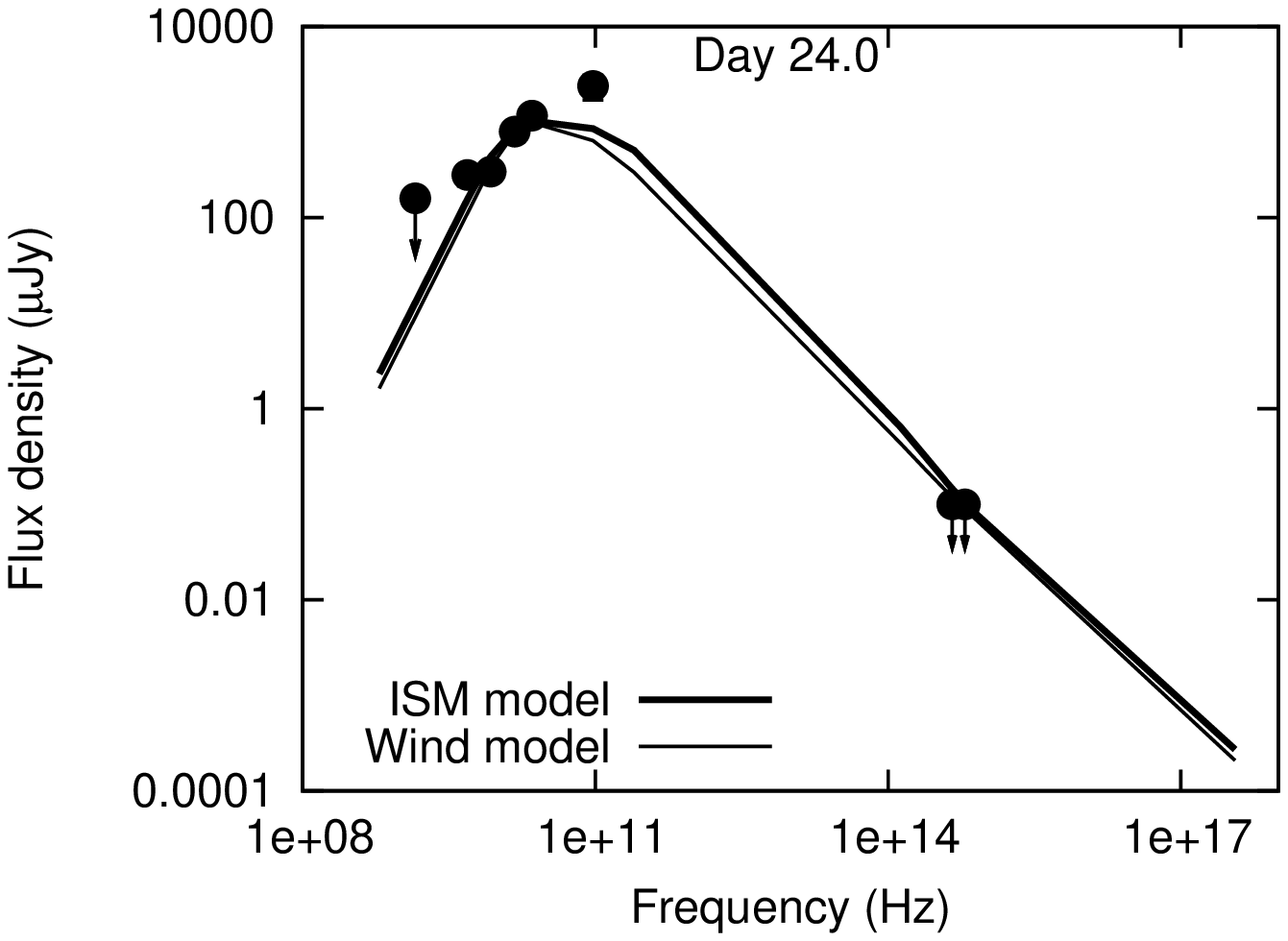}
\includegraphics[angle=0,width=0.49\textwidth]{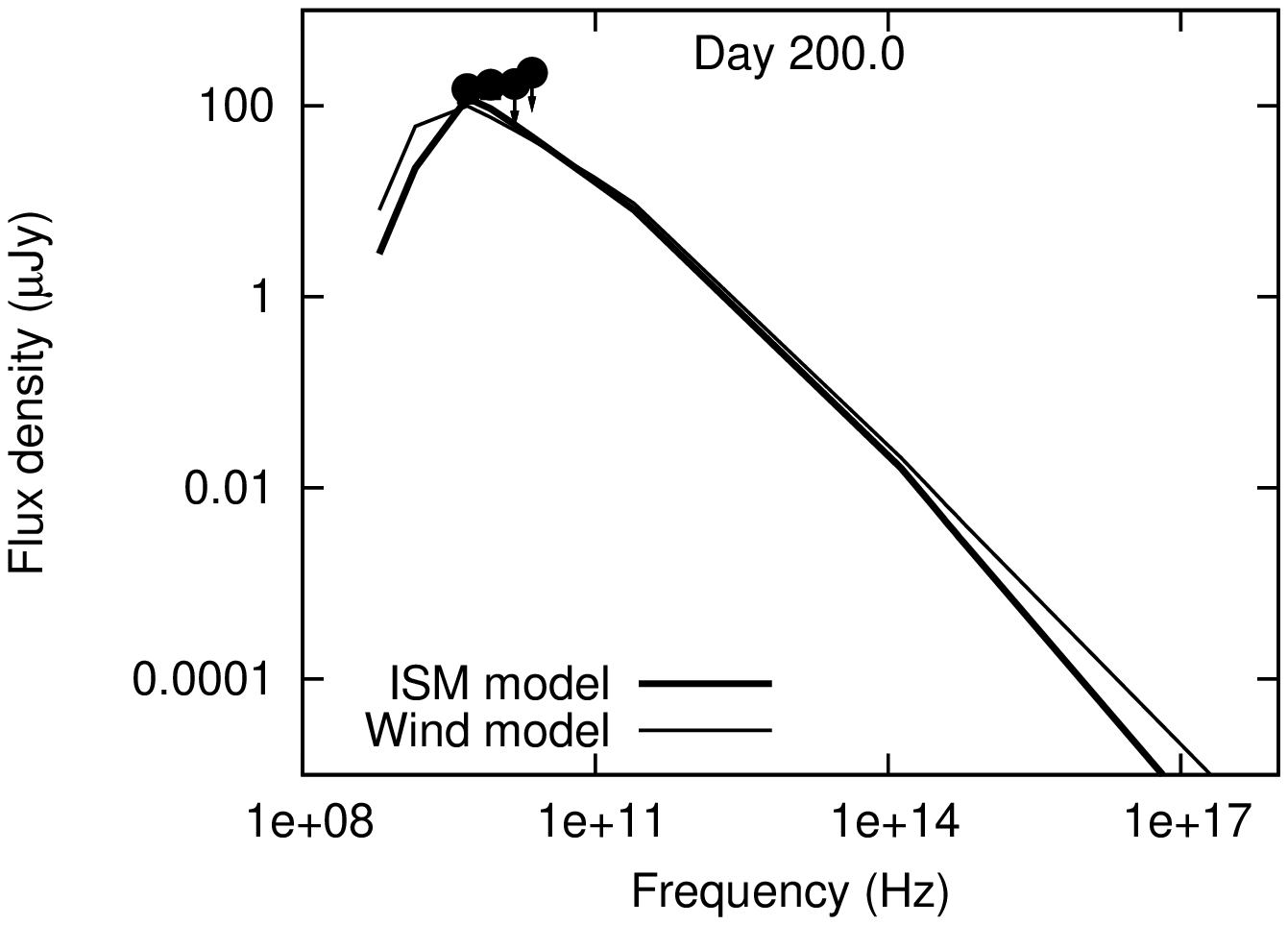}
\caption{Broad band spectra at various days. The thick solid line is the 
ISM model and the thin solid line represents the Wind model. The peak flux density is
well constrained due to submm observations. Both the models give equally 
good fits.
}
\label{fig:BBspec}
\end{center}
\end{figure}

\begin{figure}
\begin{center}
\includegraphics[angle=0,width=0.80\textwidth]{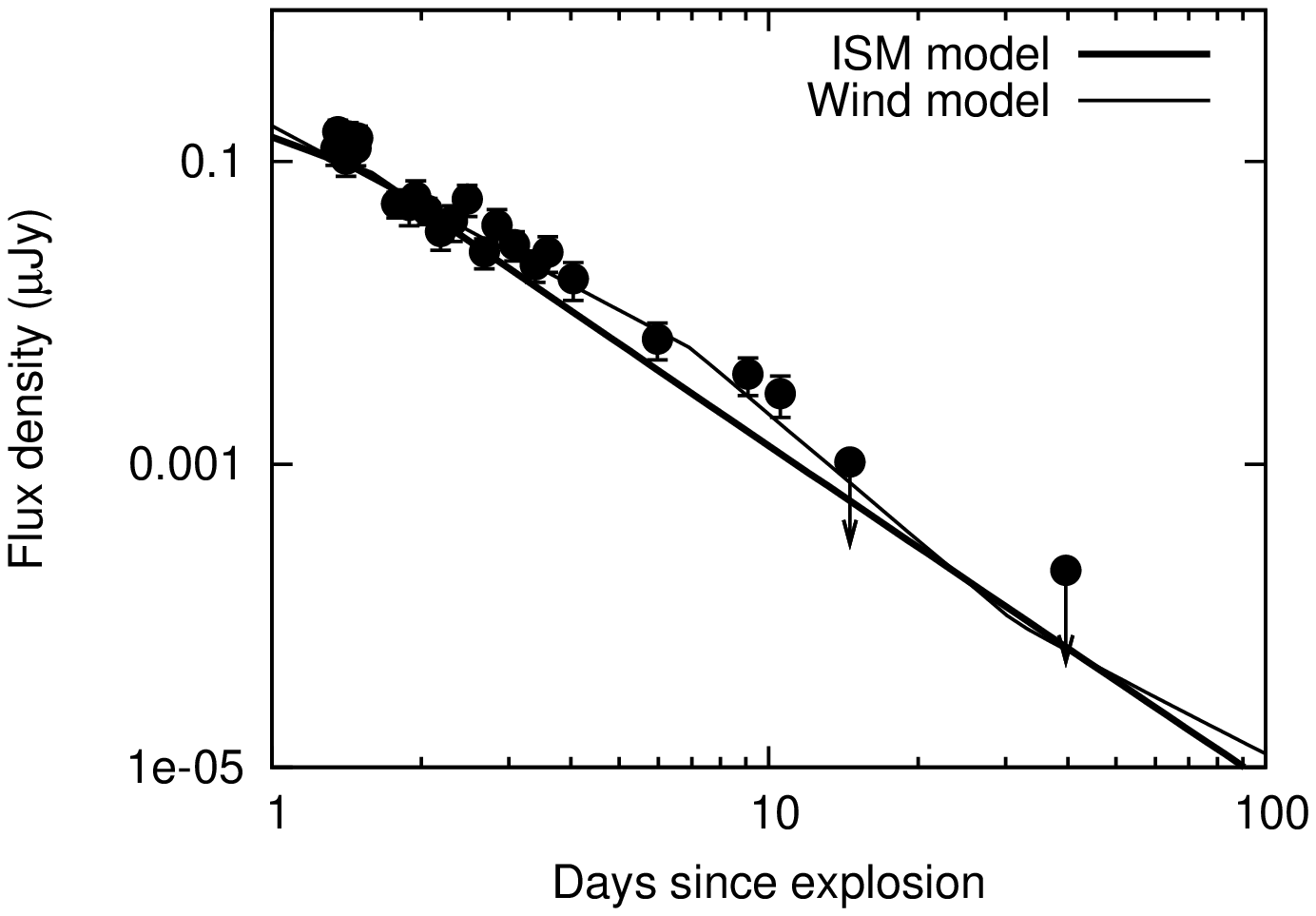}
\includegraphics[angle=0,width=0.80\textwidth]{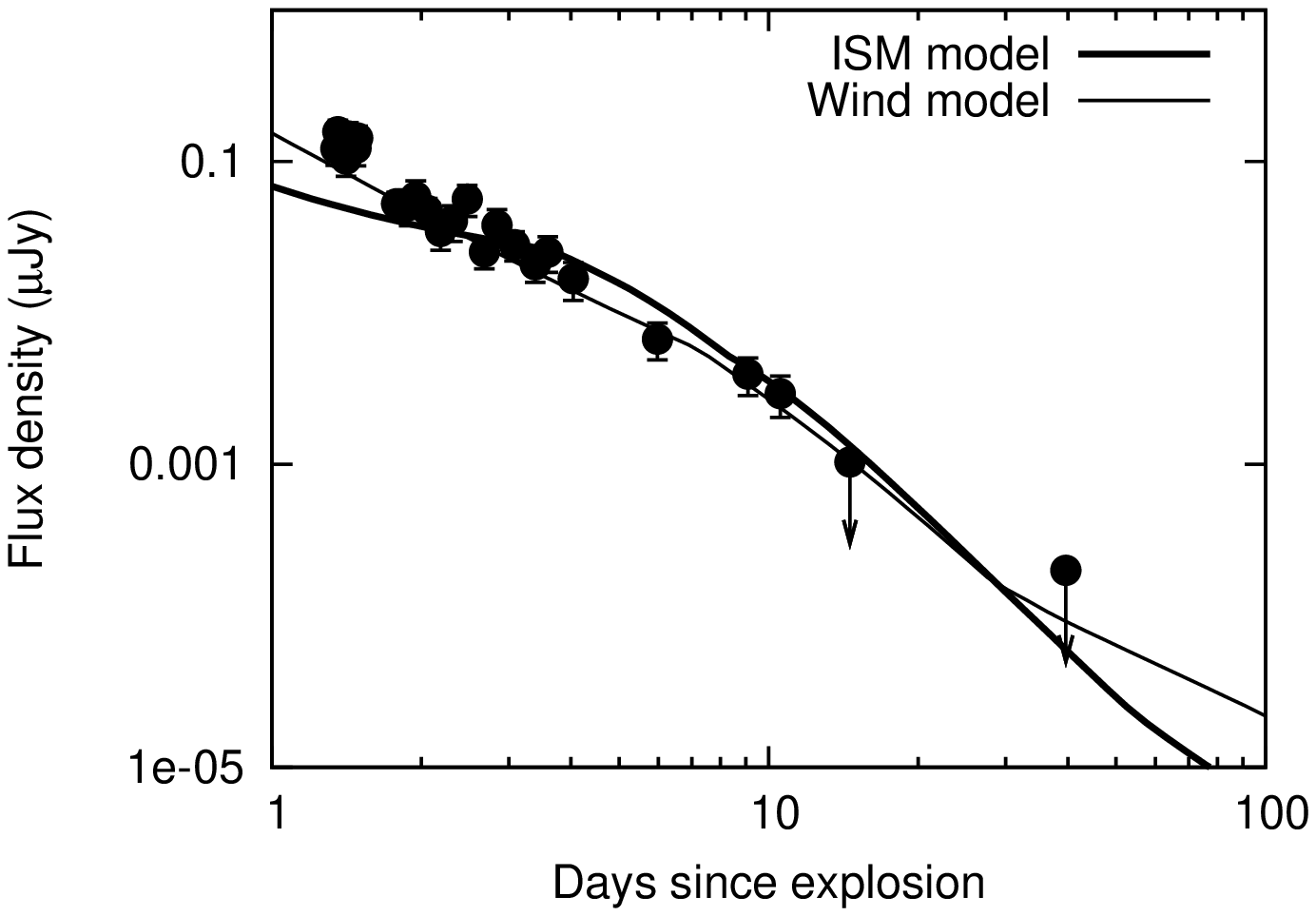}
\caption{X-ray plot only with the pure synchrotron model. 
The lower figure includes inverse Compton effects.
The IC effects start to dominate after 2.5 days.}
\label{fig:xray}
\end{center}
\end{figure}

We plot the results of our broadband modeling in Figures \ref{fig:radio}, 
\ref{fig:BBspec}, and \ref{fig:xray}.
It is difficult to differentiate between the wind and the ISM models
purely on the basis of these plots.  
Both models represent the optical data 
fairly well at early times  (Fig.~\ref{fig:radio}).  At late times ($t > 
4$\,d), both models over-predict the $R$-band flux, though with
significantly less discrepancy
in the constant density medium.  IC effects are negligible in
optical bands.

In the radio bands, none of the models
fit particularly well, especially at early times ($<15$ days). This is likely
caused partly by the
 diffractive scintillation (\S \ref{sec:scintillation}). IC scattering
has no influence in this band.
The most puzzling behavior is revealed in the 4.8 GHz band.  An early detection at
$t \approx 4$\,d, both in our data and the WSRT \cite{van07}, is followed
by almost two weeks of non-detections. Summing all these non-detections, we
can put very strict limits on the 4.8 GHz flux at this time: $f_{\nu} < 71$
$\mu$Jy.  The reason for this drop in flux is unclear, for it cannot
be due to scintillation.  We consider the possibility that the
early detection is due to the reverse shock 
emission, in \S \ref{sec:discussion}.

In Figure \ref{fig:BBspec}, we plot the broadband spectra from radio to X-ray 
at various times of the afterglow evolution. They are represented well with
our models. Both the wind and ISM environments seem to do an equally good job.

We plot the 
X-ray light curve of GRB\,070125 in Figure~\ref{fig:xray}. This is the 
only band affected by   IC emission.
In the upper panel of Figure~\ref{fig:xray}, we plot 
the model assuming only
 synchrotron radiation, while the lower panel incorporates  IC 
scattering as well.  The wind      
model in a pure synchrotron fit has very unphysical parameters 
(Table~\ref{tab:models}); hence we do not consider it further. In the ISM model,
the observations at $4 \lesssim t \lesssim 10$\,d do not fit the data very well
without any IC emission.  The fit improves significantly for both environments 
when we incorporate IC effects. 
The IC component 
seems to raise the X-ray flux roughly almost the same time 
the jet break becomes visible in
the optical bands.  This could well explain the jet break 
at a later stage in the X-rays.
We will discuss this further in \S \ref{sec:discussion}.


\section{Discussion}
\label{sec:discussion}
With our comprehensive broadband models in hand, we now turn to some of the
questions raised  in  the previous sections.

\subsection{Is Inverse Compton scattering delaying the jet-break?}

Here we examine the  IC scattering effect on the 
GRB afterglow lightcurve in the X-ray band. We adapt an approach in which we
use only the synchrotron  model for the GRb afterglow and derive various 
parameters such as $E,\;p,\;\epsilon_e,\;\epsilon_B,$, density etc
from the broadband data fitting.
In this approach, we force the broadband jet break to
be fixed on 
the day of the optical jet break from our analytical fits i.e. on day 3.7.
We then use these parameters to derive the light curve purely due to 
IC effect.
For reasons noted earlier we confine our discussion to the
ISM model. 
Here, we assume that the spectrum due to IC scattering has the same shape 
as that of the synchrotron model. Hence, IC spectrum in the X-ray band is
\begin{equation}
F_\nu^{\rm IC} = \cases{
 F_{\rm max}^{\rm IC} \displaystyle \left(\frac{\nu}{\nu_c^{\rm IC}}\right)^{-1/2},
& $\nu_c^{\rm IC} < \nu < \nu_m^{\rm IC}$, \cr
F_{\rm max}^{\rm IC} \displaystyle \left(\frac{\nu}{\nu_m^{\rm IC}}\right)^{-p/2}
\displaystyle \left(\frac{\nu_m^{\rm IC}}{\nu_c^{\rm IC}}\right)^{-1/2},
& $\nu>\nu_m^{\rm IC}$, \cr
}
\label{eq:IC}
\end{equation}
Here $F_{\rm max}^{\rm IC}$ is IC peak flux,
$\nu_c^{\rm IC}$ is IC cooling frequency and $\nu>\nu_m^{\rm IC}$
is IC peak frequency.

The best fit parameters from the synchrotron broadband model fit are:
$E_{52}=2.98,\;\theta_j=0.23\;{\rm rad},\;p=2.27,\;n=15.7\,{\rm cm^{-3}},
\;\epsilon_e=0.275,\;
\epsilon_B=0.274,\;t_c=7.8{\rm d}\;{\rm and}\;t_j=3.7{\rm d}$. 
Here $t_c$ is the transition time from the fast cooling to the
slow cooling state. $E_{52}$ is the isotropic-equivalent
 kinetic energ at $t=t_c$. IC scattering delays the
cooling time by a fraction $(1+\epsilon_e/\epsilon_B)^2$
\citep{se01}, i.e. the cooling time in 
presence of the IC effect is $t_c^{\rm IC}=31$d. Thus, for the time span 
of our observations, the afterglow  remains in the fast
cooling state and we will use formulation in this regime.

Using the formulation described in \citet{se01} and \citet{wl98}, 
we estimate the time ($t_{\rm IC}$) when
IC scattering starts becoming important
at 1.486 keV (the X-ray light curve
frequency).
Using  above best fit parameter values, we obtain $t_{\rm IC}=2.8$ day.
The IC light curve will satisfy the pre-jet break condition
in the timerange of $t_{\rm IC} \le t \le t_{\rm jet}$, i.e.
between day $2.8 - 3.7$.
The light curve will follow post-jet break formulation
 from day 3.7 onwards. 

The flux density, size, Lorentz factor and cooling frequencies
derived on day 2.8 are:
$R =2.86 \times 10^{17}{\rm cm},\; F_{\rm max}= 24.2\,{\rm mJy},\; 
\nu_m=4.09 \times 10^{12}{\rm Hz}\;{\rm and}\; 
 \nu_c=5.17 \times 10^{11}
{\rm Hz}$. 
Hence, the  derived IC parameters on day 2.8 are:
$F_{\rm max}^{\rm IC}= 0.024\,\mu{\rm Jy},\; 
\nu_m^{\rm IC}=2\gamma_m^2\nu_m=2.42 \times 10^{18}{\rm Hz},\;{\rm and}\;
\nu_c^{\rm IC}=2\gamma_c^2\nu_c=23.9 \times 10^{16}
{\rm Hz}$. Here $\gamma_m$ and $\gamma_c$ parameters are
defined in \citet{se01}.
The time dependences of various parameters in the pre-jet break epoch are:
$R \propto t^{1/4},\; \Gamma \propto t^{-3/8},\;
F_{\rm max} \propto t^0,\;\nu_m \propto t^{-3/2},
\; \nu_c \propto t^{-1/2},\;F_{\rm max}^{\rm IC} \propto t^{1/4},
\; \nu_m^{\rm IC} \propto t^{-9/4},\;{\rm and}\;\nu_c^{\rm IC} \propto 
t^{-1/4}$. 
Our frequency of observation ($\nu=3.59 \times 10^{17}$ Hz)
satisfies
$\nu_c^{\rm IC} < \nu < \nu_m^{\rm IC}$ condition at $t=2.8$d.
Therefore, using Eq. \ref{eq:IC}, the 1.5 keV light curve
for IC scattering between day $2.8-3.7$ becomes  
\begin{equation}
F_\nu^{\rm IC}(t) = 0.0079\; \left(\frac{t}{2.8
{\rm d}}\right)^{1/8}\;\mu{\rm Jy}.
\label{iclc1}
\end{equation}

From day 3.7 onwards, we use the post-jet break formulation
and derive the IC light curve. 
The time dependences
of various parameters in the post-jet break regime are:   
$R \propto t^{0},\; \Gamma \propto t^{-1/2},\;
F_{\rm max} \propto t^{-1},\;\nu_m \propto t^{-2},
\; \nu_c \propto t^{0},\;F_{\rm max}^{\rm IC} \propto t^{-1},
\; \nu_m^{\rm IC} \propto t^{-3},\;{\rm and}\;\nu_c^{\rm IC} \propto 
t^{1}$. 
On the jet break day, we find $\nu_m^{\rm IC}=1.29 \times 10^{18}$ Hz and 
$\nu_c^{\rm IC}=3.6 \times 10^{16} $ Hz, which still satisfies
$\nu_c^{\rm IC} < \nu < \nu_m^{\rm IC}$ condition. 
Therefore, we derive the light curve from day 3.7 onwards using Eq. \ref{eq:IC}
as
\begin{equation} 
F_\nu^{\rm IC}(t) = 0.0082\; \left(\frac{t}{3.7{\rm d}}\right)^{-1/2}\;\mu{\rm Jy}.
\label{iclc2}
\end{equation}
The IC frequency $\nu_m^{\rm IC}$ reaches the 
X-ray observation frequency of $\nu=3.594
\times 10^{17}$ Hz on day 5.7. Hence, the above light curve is valid
until day 5.7. 

After day 5.7, the IC flux density in the $\nu>\nu_m^{\rm IC}$
regime (Eq. \ref{eq:IC}), which gives the  light curve in this 
regime as follows:\\ 
\begin{equation}
F_\nu^{\rm IC}(t)=0.0066\; 
\left(\frac{t}{5.7{\rm d}}\right)^{-2.4}\;\mu{\rm Jy}.
\label{iclc3}
\end{equation} 

\begin{figure}
\begin{center}
\includegraphics[angle=0,width=1\textwidth]{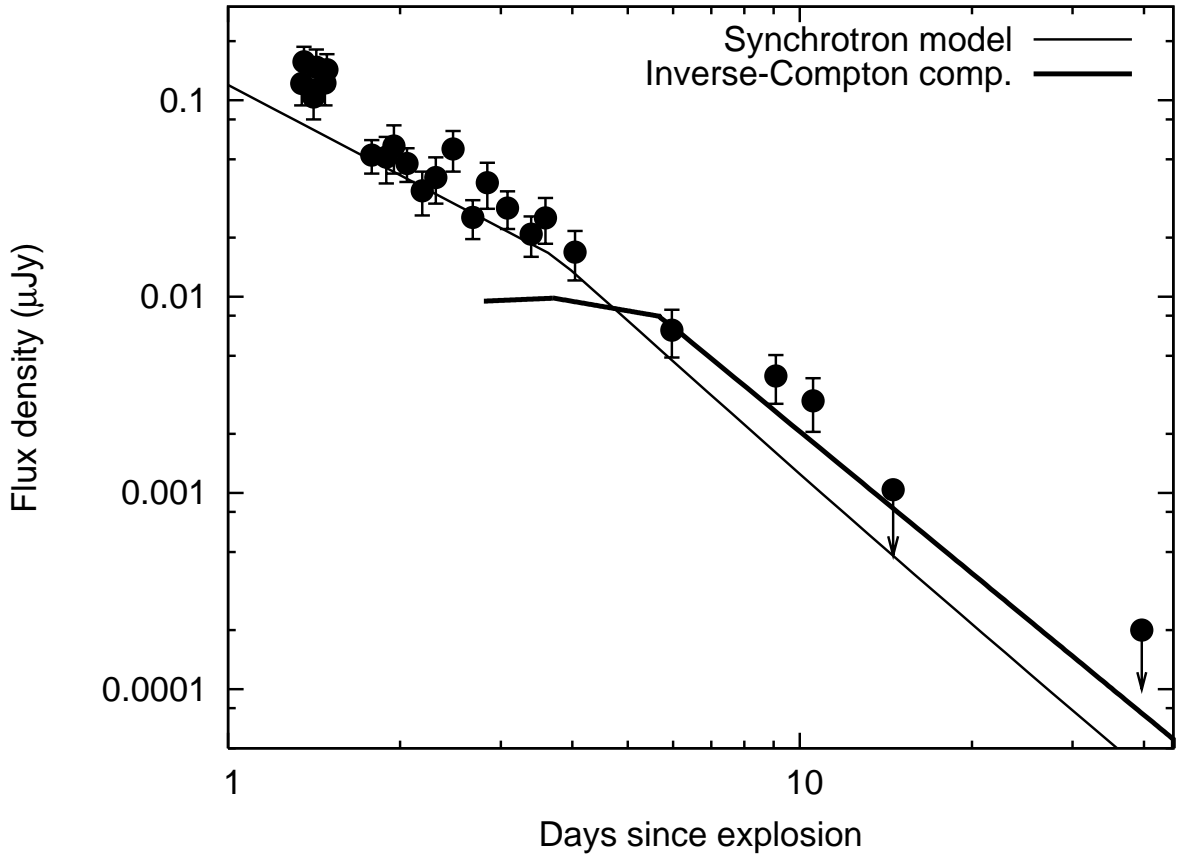}
\caption{
Contribution of IC in the synchrotron model. The thin line represents 
the broadband model with the synchrotron component only. The thick line represents
the IC light curve. }
\label{fig:IC}
\end{center}
\end{figure}

Combining Eqs. (\ref{iclc1}), (\ref{iclc2}) and (\ref{iclc3}), 
we plot the total IC light curve in Fig. \ref{fig:IC}.  The figure
clearly shows that IC scattering flattens the light curve and  
delays the jet break to later time. 
What happens after day $9-10$ is
unknown due to the lack of X-ray detection of the afterglow. 

While we have strong evidence that IC effects delay the X-ray jet break in
GRB\,070125, we would like to know if this effect could be 
seen in other GRBs
as well.  In the pre-\Swift\ era, 
the X-ray data were not sufficiently well sampled 
to search for jet breaks, and so collimation corrections were almost
exclusively calculated in the optical bands.  In the \Swift\ era, with a 
plethora of well-sampled XRT light curves, we may be missing the jet break due 
to IC effects in many GRBs.  The radio afterglow is not useful in determining
 the jet 
break, since GRBs most likely scintillate in radio bands at such early times.
This makes the optical the unique bandpass in 
which the real jet 
break can be determined unambiguously.

The inverse Compton effect is most prominent in a dense medium. Our radio 
observations have already shown that GRB\,070125 resides in a very dense medium.
It has been shown by \citet{wl00} that IC is important in relativistic ejecta 
and even in non-relativistic ejecta in high density.  
Harrison et al. (2001) found good evidence for IC production of X-rays in
GRB 000926 and derived a high circumburst density, $30$ cm$^{-3}$, comparable
to the density we find for \event.
\citet{cp06} have shown 
that the late time flattening in the X-ray light curve of XRF\,050406 can be 
explained as an effect of IC scattering.  For inverse Compton 
scattering to play an
important role, 
the electron energy density fraction ($\epsilon_e$) must be larger 
than the magnetic energy density fraction ($\epsilon_B$).

Recently, 
\citet{pan07} has discussed chromatic breaks occurring due to
scattering of the forward-shock synchrotron emission by 
a relativistic outflow located behind the leading
 blast-wave. This model may have X-ray jet breaks showing up at later
times than the optical breaks. However, this model requires a long-lived
central engine, which may not be the case with most of the GRBs.  

\subsection{An Emerging Class of Hyper-Energetic ($E > 10^{52}$\,erg) GRBs?}
\label{sec:hyper}

With GRB\,070125, we now have found three \Swift\ events with total energy
release in excess of $10^{52}$\,erg: GRB\,050904 \citep{fck+06} and GRB\,050820A
\citep{ckh+06}.  While both GRB\,050904 and GRB\,050820A appear to have
exploded in an dense circum-burst medium, the lack of a bright radio
afterglow from GRB\,050820A indicates a more typical environment.  Moreover,
with the exception of the total energy release, other parameters derived
from broadband modeling are in line with previous studies of less
energetic GRBs \citep{pk01,yhsf03}.  It seems likely, therefore, that some 
factor intrinsic to the progenitor system is responsible for the large energy
release.

At first blush, it seems surprising that \Swift\ has detected three of the most
energetic GRBs ever.  With its increase high-energy sensitivity, \Swift\
should preferentially select GRBs at the low end of the fluence distribution.
We note, however, that a strong selection bias exists.  As first noted by 
\citet{kb07}, in many \Swift\ X-ray light curves, the last XRT measurement
is not sufficient to rule out a collimation-corrected prompt energy
release of $\sim 10^{51}$\,erg.  Similarly, in the optical bandpass, 
\citet{dgp+07} have shown that at least some jet breaks occur at late times 
beyond the sensitivity of medium aperture facilities.  

While a detailed discussion of the relative rates of hyper-energetic events
is still premature, it is clear at this point that, at the very least, the
prompt $\gamma$-ray energy distribution is significantly broader than
previously believed \citep{kb07}.  Coupled with the recent controversy 
surrounding the validity of the many high-energy correlations
(e.g.~\citealt{bkb+07,wog+07}), we 
believe the future utility of GRBs as cosmological probes is significantly
lessened.

Even more importantly, however, hyper-energetic GRBs have important consequences
for progenitor models.  Sustained engine activity has been seen now in many 
GRBs \citep{brf+05}.  This poses a problem for the collapsar model, as the 
duration of the central engine should not significantly exceed the accretion
time scale onto the remnant black hole \citep{w93}.  Late-time engine activity 
is naturally accommodates by models in which the central object is a magnetar 
\citep{u92}.  The existence of hyper-energetic GRBs, however, is a direct
and severe challenge to the magnetar model.  

With the current rate of hyper-energetic events ($\sim 1$\,yr$^{-1}$), coupled
with the difficulty in measuring late jet breaks for more typical \Swift\
events, future prospects look grim.  However, the impending launch of 
\textit{GLAST} offers a new hope in the study of GRB energetics.  
Much like blazars, those GRBs capable of producing GeV photons detectable by the
Large Area Telescope should be the most energetic and narrowly beamed events.
Together, synergistic \textit{GLAST} and \Swift\ observations in the coming
years should be able to shed light on the opening angles and energy release of 
a large sample of GRBs. 

\subsection{The Unusual Environment of GRB\,070125}
\label{sec:environ}

Our observations presented here, particularly the bright, self-absorbed radio
afterglow, indicate GRB\,070125 exploded in a dense circum-burst medium.
In a separate work, however, \citet{cfp+07} have reported spectroscopic
observations indicating an environment almost completely devoid of
absorbing material.  \citet{cfp+07} furthermore find no evidence of an
underlying host galaxy to deep ($R > 25$\,mag) limits.  We briefly reiterate
here the resolution of this apparent paradox.

The key to understanding the environment of GRB\,070125 is to note that the
broadband afterglow emission and the super-posed spectroscopic absorption
features derive from distinct physical regions.  Afterglow emission is
caused by electrons in the circum-burst medium accelerated by the outgoing
blastwave (e.g.~\citealt{p05}).  These electrons reside relatively close to
the explosion center, typically at radii $r \le 1$\,pc.

We have strong evidence, however, that the absorption features seen super-posed
on GRB afterglow spectra derive from material at significantly larger distances
from the explosion site: $r \approx 1$\,kpc, or within the host galaxy ISM.  
Evidence in support of this large distance comes primarily from two lines of
argument.  First, the presence of \ion{Mg}{1} indicates a large distance
from the explosion site, as the first ionization energy of Mg falls below 1 Ryd,
and thus any Mg near the GRB would be ionized to at least \ion{Mg}{2}
\citep{pcd+07}.  Second, \citet{vls+07} have reported the detection of 
variability in the fine structure levels of \ion{Fe}{2} from UV pumping for 
GRB\,060418.  By modeling the variability over time, they were able to measure 
the GRB-absorber distance: $d = 1.7 \pm 0.2$\,kpc.  

While this explains the apparent density paradox, we are still left to explain
how a dense circum-burst medium could be embedded in such a tenuous ISM.
Taking a clue from the lack of an underlying host detection, we suggested
GRB\,070125 may have exploded in a dense stellar cluster enriched by galaxy
interactions.  Star formation in such extreme environments can be seen
in the local universe (e.g.~Tadpole galaxy; \citealt{jpf+06}), and under our
hierarchical picture of galaxy formation, such interactions should have 
occurred more frequently at $z > 1$.  The report of the detection of a
faint source at the afterglow location at late times \citep{dgp+07} may call 
this interpretation into question, although it is unclear whether this emission
is attributable to the fading afterglow or an underlying host (or some
combination thereof).  Regardless, future high-resolution imaging 
(i.e.~\textit{HST}) seems worthwhile to pin down the environment of this 
truly unique event.

\subsection{Early radio emission by the reverse shock?}

\event\ was not detected at $t \sim 1.5$ d with the WSRT, and then
was detected 
by both the VLA and the WSRT around day 5.  It remained below
detection level for the 
next 15 days, before rebrightening on day 22.  We explore the possibility that 
the flux from the GRB at $t \sim 5$\,d could be emission from a reverse shock.
We first consider the possibility that the non-detection was caused by 
modulations due to scintillation. 

The modulation in flux density due to the refractive scintillation can decrease 
the flux density at the most up to $\Delta f_{\nu} \sim 100\, \mu$Jy. To
check for this possibility, we combined all the eight observations taken during the 15 day non-detection phase.  This vastly improved the signal-to-noise
ratio.  The flux density we obtained at the GRB position is
$70\pm25$ $\mu$Jy,
much lower than scintillation can explain. 

According to the internal-external shock model for GRBs, the prompt emission
is produced by internal shocks within a relativistic outflow, while the 
afterglow is produced by external shocks with the interstellar medium.  The
reverse shock
has a much
lower temperature than that of the forward shock so it radiates at considerably 
lower frequencies.  In this scenario using the ISM model, 
the reverse shock emission peaks in the optical 
band at \citep{np05}:
$$t_o= {\rm max}(\Delta/c, t_{\rm dec}).$$
For \event\, we calculate $t_o \approx 30$ sec. 
This corresponds to the time of peak in the radio band to be:
$$t_{\rm radio}=\frac{\nu_a^r t_o}{\nu_{\rm radio}} $$
Here $\nu_a^r $ is the synchrotron self-absorption frequency, which can be
written as \citep{np05}:
$$\nu_a^r(t_0)=6 \times 10^{12}\; {\rm Hz} [(1+z)^{-(p+6)/8}
\epsilon_{e,-1}^{p-1} \epsilon_{B,-2}^{(p+2)/4}
(nE_{54})^{(p+6)/8} t_{o,2}^{-(3p+10)/8}]^{2/(p+4)}.$$
For values obtained from our multiwavelength analysis, we calculate
$\nu_a^r=3\times 10^{13}$ Hz and $t_{\rm radio}\approx$ 2.25 days. 
Hence, the reverse shock emission peaks in the radio bands around $t \sim 2$\,d.
We now estimate the peak radio flux density for the reverse shock.  This can 
be written as:
$$\frac{F_{\rm radio}^r}{F_o^r} \left(\frac{t_{\rm radio}}{t_o}\right)^{(
p-1)/2+1.3}=\left(\frac{\nu_{\rm opt}}{\nu_{\rm radio}}\right)^{(p-1)/2},$$
where $F_o$ is the peak optical flux density expressed as \citep{np05}:
$$F_o^r \sim 16.6\; {\rm mJy}\; (1+z)^{-(4+p)/8}
n^{(p+2)/8} E_{52}^{(p+8)/8}
\left(\frac{\epsilon_e}{0.1}
\right)^{p-1} \left(\frac{\epsilon_B}{0.01}\right)^{(p+1)/4}
\left(\frac{t_o}{100\, \rm s}\right)^{-3p/8} D_{28}^{-2}$$
For our best fit parameters, this value is $\sim 1$ mJy, which a gives
the peak radio flux density 
to be $ \sim 1\;\mu $Jy, which is 
two orders on magnitude than the observed one.
 This shows that
a reverse shock is not strong enough to explain the detection on
day 5. We cannot explain this strange behavior.

\subsection{GRB with high radiative efficiency?}
\label{sec:bb}

One of the major concerns for \event\ is the difference between the 
isotropic $\gamma$-ray energy obtained from the high-energy fluence and the  
isotropic-equivalent blastwave energy obtained from our best fit model on
the day when 
$\nu_c=\nu_m$. The isotropic-equivalent kinetic energy is an order of 
magnitude smaller than the isotropic $\gamma$-ray energy ($
10^{54}$ erg, \S \ref{sec:energetics}).  However, at very 
early times, the afterglow is in the fast cooling regime, where it undergoes 
significant loss of the energy because it is highly radiative. The fireball 
may lose as much as 80\% of its energy during this phase \citep{hys+01,cps98}. 
If we incorporate radiative corrections, we may derive a more accurate 
estimate of the actual isotropic blastwave energy.  From \citet{wdhl05, sar97}
we find:
$$E(t)=E_0\left(\frac{t}{t_0}\right)^{-\frac{17 \epsilon_e}{12}} $$
for the ISM model and 
$$E(t)=E_0\left(\frac{t}{t_0}\right)^{-\frac{3\epsilon_e}{2}} $$
for the wind model.
For the  broadband modeling parameters, the isotropic kinetic energy one 
hour after the explosion is $(5.04\pm0.87) \times 10^{53}$\, and
$(3.88\pm3.11)\times 10^{55}$ erg for the ISM and the 
wind models, respectively.
Here, the energy for the wind model is rather unphysical.
The efficiency $\eta_\gamma$ for the ISM model is 0.67. 

\subsection{Wind model vs ISM model}

In terms of reduced $\chi^2$ for best fit, wind model is slightly better.
However, the wind model requires an electron energy density fraction
close to 1, which is very unlikely. The best fit parameters in the wind model
are rather unphysical, with less constrained boundaries. However,
fixing the electron energy fraction to be 0.4 also gives 
reasonable fits. Evidence 
favoring the ISM model comes from the fireball size estimation from the 
scintillation data. 
However, the uncertainties in the
diffraction scintillation time estimates may bring large uncertainties 
in the size estimates. 
Based on the energetics
arguments stated above, the ISM model is favored over the wind model,
but we can by no means definitely dismiss the latter.


\section{Conclusion}
\label{sec:conclusion}

\begin{figure}
\begin{center}
\includegraphics[angle=0,width=0.49\textwidth]{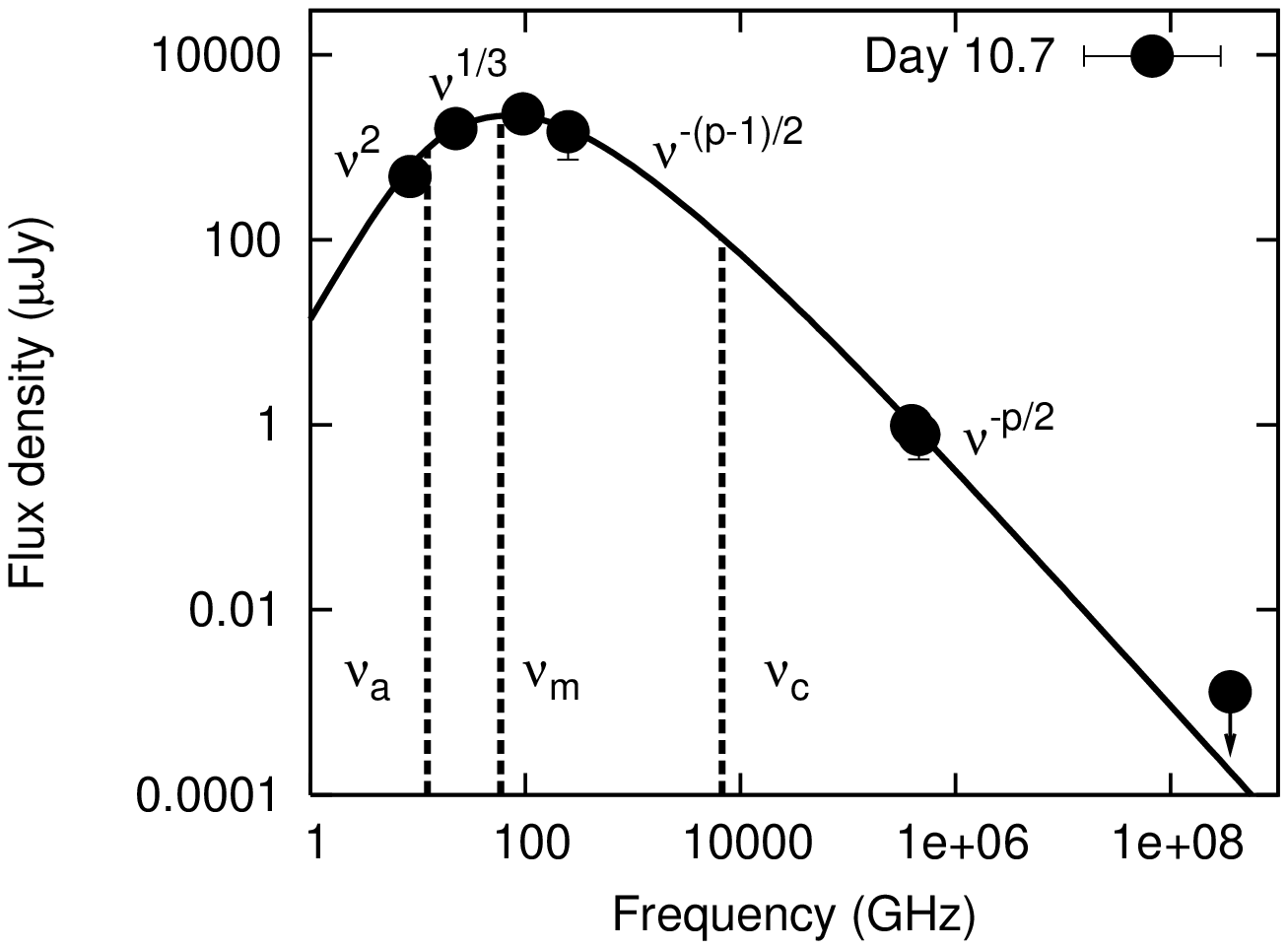}
\includegraphics[angle=0,width=0.49\textwidth]{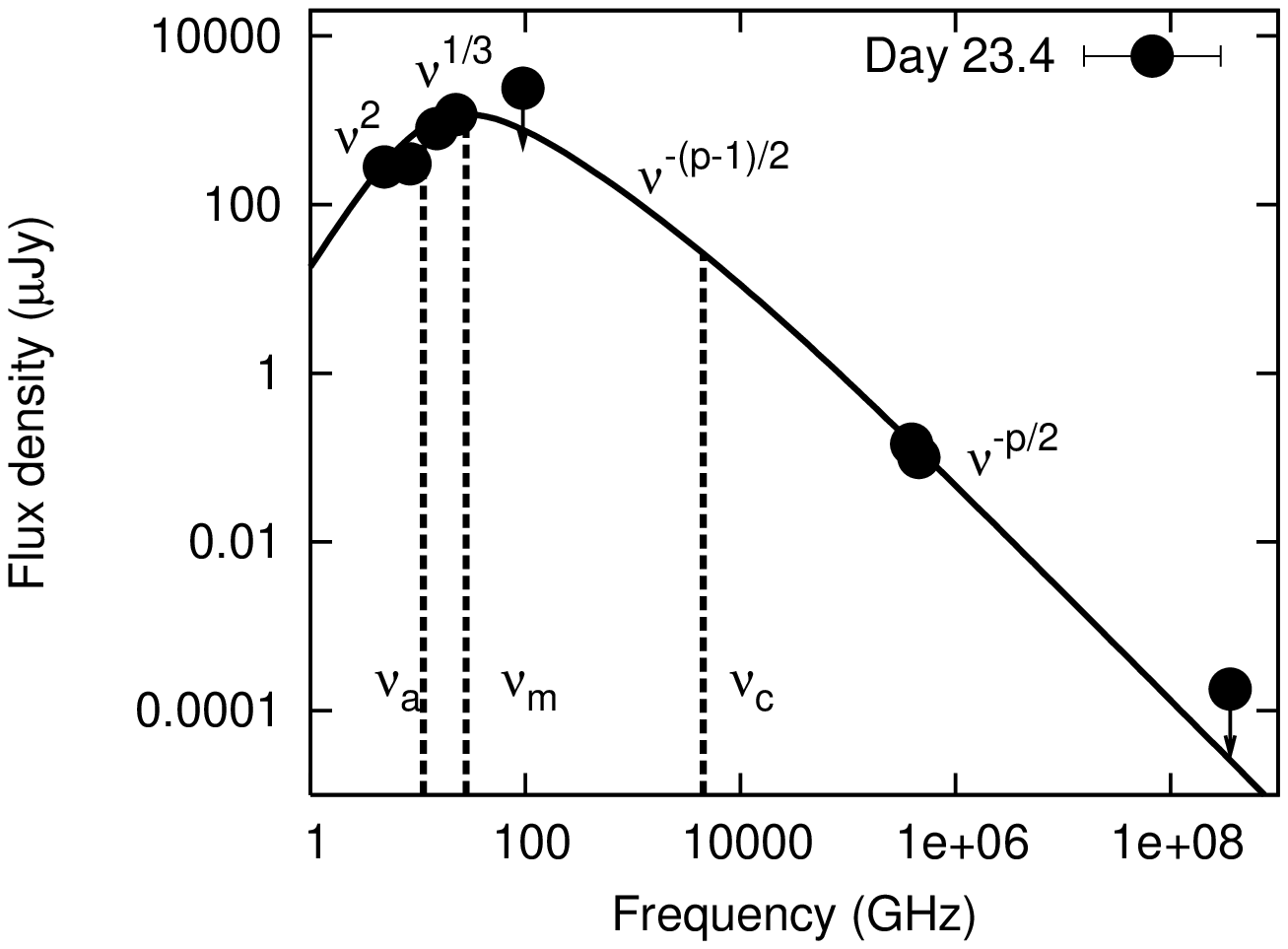}
\includegraphics[angle=0,width=0.49\textwidth]{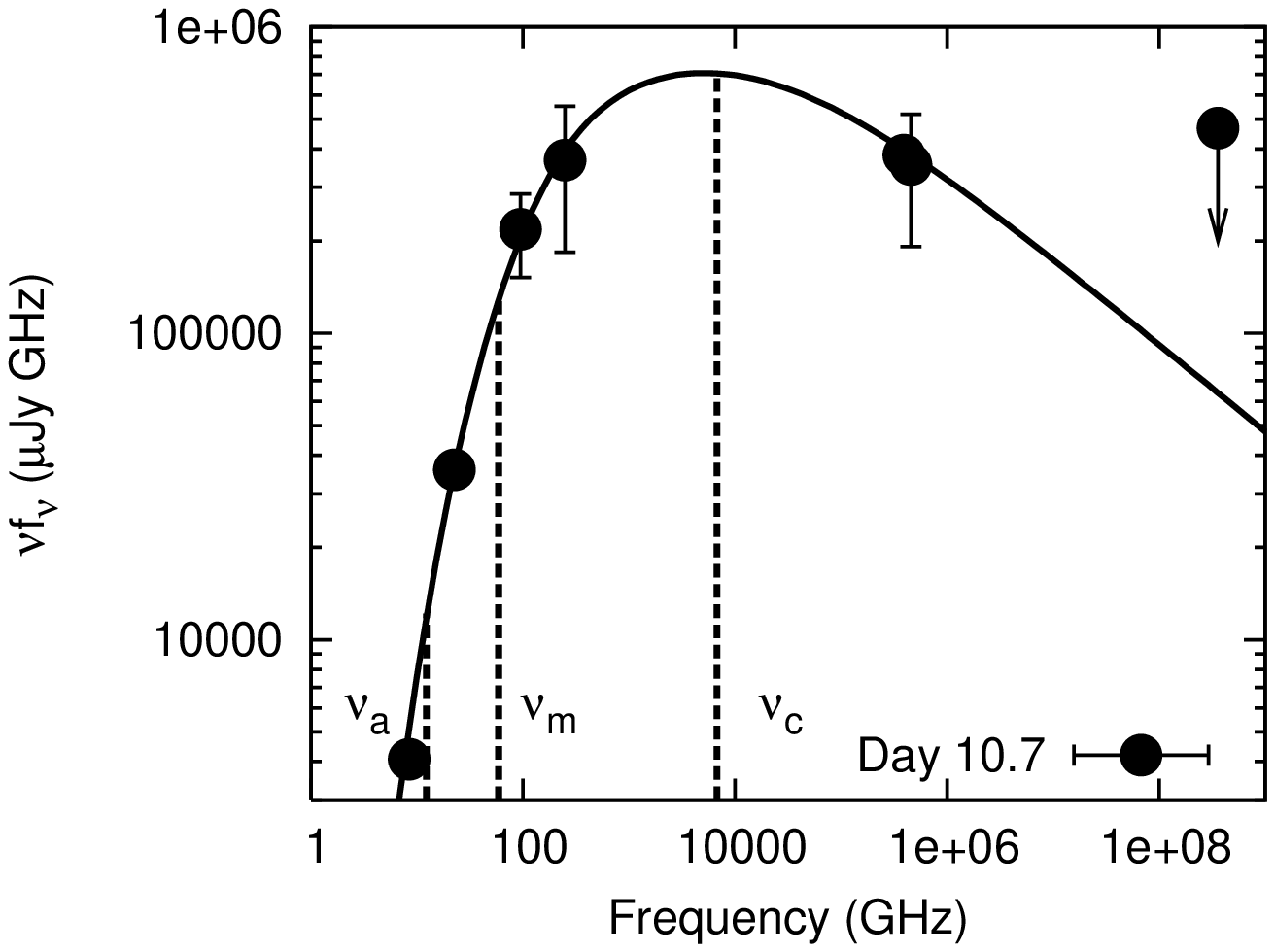}
\includegraphics[angle=0,width=0.49\textwidth]{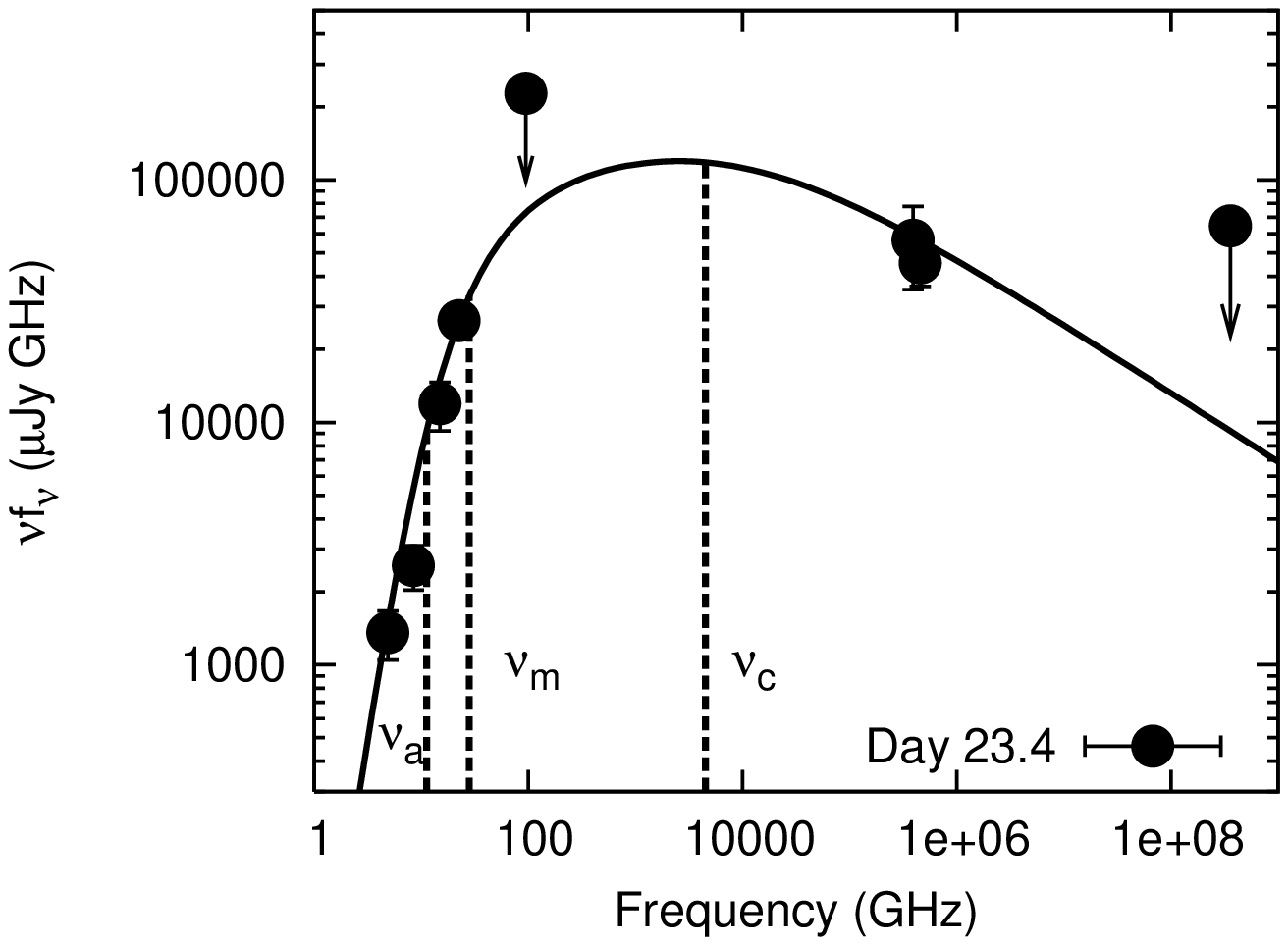}
\caption{Spectra on day 10.7 and day 23.4. We fit a
smooth  function for the GRB afterglow model and derive the
$F_{\rm max}$, $\nu_a$, $\nu_m$, and $\nu_c$ parameters.
The plots at the bottom are 
$\nu F_\nu$ plots. For slow cooling the flux should peak at 
$\nu_m$ and the total power ($\nu F_\nu$) should peak at $\nu_c$.
}
\label{fig:analytic}
\end{center}
\end{figure}

\event\ is one of the brightest GRBs ever detected, 
both in terms of its
prompt high-energy fluence and its optical and radio 
afterglows. The isotropic equivalent energy for the GRB is
$10^{54}$ erg. 
This  is  the most extensively followed GRB 
in multiwavebands in the \Swift\ era.
The richness of the data allowed us to derive many important 
properties of the GRB and place useful constraints on many
parameters.
\event\ was one of the few GRBs with sub-mm observations at various 
epochs. Our 95 GHz and 250 GHz observations with CARMA and 
IRAM, respectively, gave a robust determination of 
the peak flux density and $\nu_m$, and 
constrained the power ($\nu f_\nu$) (Fig. \ref{fig:analytic}).

Simultaneous fitting to optical and X-ray data favors a jet break
at day 3.78 than the single powerlaw model.
The  evidence for the jet break is 
indisputable in the optical $R$ and $i'$
bands with pre-break and post-break slopes being
1.73 and 2.49 respectively. However, the jet break is not very prominent 
in the X-ray band.
When we do the independent fit to optical and X-ray bands, the 
optical best fit is consistent with our joint fit. However,
the jet break in the X-ray band is shifted to day $\sim 10$. 
Using the best fit parameters of the model, we show that the 
inverse Compton effects will dominate throughout our observations 
with pronounced effects in X-ray frequencies. These effects delay the
jet break in the X-ray band.

We had long observations of the GRB at three epochs in the 8 GHz band.
These data gave evidence for diffractive scintillations,
which gave an upper limit on the size of the fireball after the jet break, 
until the Sedov-Taylor phase started. 
This estimate of the fireball
size is consistent with the one obtained from the broadband modeling 
in a constant density medium.

We obtained synchrotron self absorption frequency estimates
at various epochs from the VLA radio data. The evolution of the
synchrotron self absorption frequency is $t^{-0.24 \pm0.05}$,
which is consistent with the one expected ($t^{-0.2}$)
in the standard afterglow model. Synchrotron self absorption
frequency estimates indicate that the GRB afterglow is moving in a
dense medium. 

Our model fits could not distinguish between the ISM density profile and
the wind-like density profile. The $\chi^2$ fits were marginally better
for the wind model but it needed an unphysically high electron 
    energy fraction ($\sim 1$). 
    When we fixed the electron energy density fraction to 0.4
    in the wind model, it did give decent fits and 
    physical parameter values. However, the parameter values in
    the ISM model are more robust and change little with little
    change in the input values, unlike the wind model
    which is rather unstable. 
In both the ISM and the wind models, the radiative efficiency of the
GRB is very high ($> 60\%$).

We suggest that IC scattering is a potential candidate in 
flattening  the light curve and 
delay the jet break  in other \Swift\ events and 
explain the absence of a jet break
in X-ray light curves of some of the \Swift\ bursts.
IC effects are more prominent in high density medium. 
Frequent radio measurements are necessary to 
measure the circumburst density of the medium.
Hence, in absence of good radio data, one
cannot determine the importance of IC scattering. 
GRB 070125 is unique because this has the richest radio data 
in \Swift\ era, having closely spaced X-ray light curve.

Even though \event\ has rich multiwaveband data, we could not
nail down some of the lingering issues, such as 
Wind vs ISM density
profile. 
One reason for this is that much of the evolution is in the
jet break phase, when the ISM and Wind models have similar properties.
Very dense samples of sensitive radio, X-ray, 
$\gamma$-ray and optical
data from the very beginning until the GRB fades below the detection limit are needed.
In the future, a combination of \Swift\, GLAST, ALMA, EVLA, 
and various optical telescopes will
provide this opportunity.

\acknowledgments
PC thanks the VLA staff for making radio observations,
without which this work was not possible.
PC is a Jansky fellow at National Radio Astronomy 
Observatory. The National Radio Astronomy Observatory is a 
facility of the National
Science Foundation operated under cooperative agreement by Associated
Universities, Inc. 
We thank Sarah Yost for providing us the GRB broadband modeling 
code and helping out in running the code.
PC used the fussy calculator (http://fussy.googlecode.com) for error
propagation calculations and wish to thank its author, 
Sanjay Bhantagar.
RAC was supported in part by NASA grant NNG06GJ33G.

\bibliographystyle{apj}

\begin{thebibliography}{77}
\expandafter\ifx\csname natexlab\endcsname\relax\def\natexlab#1{#1}\fi

\bibitem[{{Adelman-McCarthy} {et~al.}(2007)}]{aaa+07}
{Adelman-McCarthy}, J.~K. {et~al.} 2007, arXiv:astro-ph/0707.3413

\bibitem[{{Amati}(2006)}]{a06}
{Amati}, L. 2006, preprint (astro-ph/0601553)

\bibitem[{{Band} {et~al.}(1993){Band}, {Matteson}, {Ford}, {Schaefer},
  {Palmer}, {Teegarden}, {Cline}, {Briggs}, {Paciesas}, {Pendleton}, {Fishman},
  {Kouveliotou}, {Meegan}, {Wilson}, \& {Lestrade}}]{bmf+93}
{Band}, D., {Matteson}, J., {Ford}, L., {Schaefer}, B., {Palmer}, D.,
  {Teegarden}, B., {Cline}, T., {Briggs}, M., {Paciesas}, W., {Pendleton}, G.,
  {Fishman}, G., {Kouveliotou}, C., {Meegan}, C., {Wilson}, R., \& {Lestrade},
  P. 1993, \apj, 413, 281

\bibitem[{{Bellm} {et~al.}(2007){Bellm}, {Hurley}, {Pal'shin}, {Yamaoka},
  {Bandstra}, {Boggs}, {Hong}, {Kodaka}, {Kozyrev}, {Litvak}, {Mitrofanov},
  {Nakagawa}, {Ohno}, {Onda}, {Sanin}, {Sugita}, {Tashiro}, {Tretyakov},
  {Urata}, \& {Wigger}}]{bhp+07}
{Bellm}, E.~C., {Hurley}, K., {Pal'shin}, V., {Yamaoka}, K., {Bandstra}, M.~E.,
  {Boggs}, S.~E., {Hong}, S., {Kodaka}, N., {Kozyrev}, A.~S., {Litvak}, M.~L.,
  {Mitrofanov}, I.~G., {Nakagawa}, Y.~E., {Ohno}, M., {Onda}, K., {Sanin},
  A.~B., {Sugita}, S., {Tashiro}, M., {Tretyakov}, V.~I., {Urata}, Y., \&
  {Wigger}, C. 2007, ArXiv e-prints, 710

\bibitem[{{Berger} {et~al.}(2003{\natexlab{a}}){Berger}, {Kulkarni}, \&
  {Frail}}]{bkf03}
{Berger}, E., {Kulkarni}, S.~R., \& {Frail}, D.~A. 2003{\natexlab{a}}, \apj,
  590, 379

\bibitem[{{Berger} {et~al.}(2004){Berger}, {Kulkarni}, \& {Frail}}]{bkf04}
---. 2004, \apj, 612, 966

\bibitem[{{Berger} {et~al.}(2003{\natexlab{b}}){Berger}, {Kulkarni}, {Pooley},
  {Frail}, {McIntyre}, {Wark}, {Sari}, {Soderberg}, {Fox}, {Yost}, \&
  {Price}}]{bkp+03}
{Berger}, E., {Kulkarni}, S.~R., {Pooley}, G., {Frail}, D.~A., {McIntyre}, V.,
  {Wark}, R.~M., {Sari}, R., {Soderberg}, A.~M., {Fox}, D.~W., {Yost}, S., \&
  {Price}, P.~A. 2003{\natexlab{b}}, \nat, 426, 154

\bibitem[{{Burrows} {et~al.}(2005{\natexlab{a}}){Burrows}, {Hill}, {Nousek},
  {Kennea}, {Wells}, {Osborne}, {Abbey}, {Beardmore}, {Mukerjee}, {Short},
  {Chincarini}, {Campana}, {Citterio}, {Moretti}, {Pagani}, {Tagliaferri},
  {Giommi}, {Capalbi}, {Tamburelli}, {Angelini}, {Cusumano}, {Br{\"a}uninger},
  {Burkert}, \& {Hartner}}]{bhn+05}
{Burrows}, D.~N., {Hill}, J.~E., {Nousek}, J.~A., {Kennea}, J.~A., {Wells}, A.,
  {Osborne}, J.~P., {Abbey}, A.~F., {Beardmore}, A., {Mukerjee}, K., {Short},
  A.~D.~T., {Chincarini}, G., {Campana}, S., {Citterio}, O., {Moretti}, A.,
  {Pagani}, C., {Tagliaferri}, G., {Giommi}, P., {Capalbi}, M., {Tamburelli},
  F., {Angelini}, L., {Cusumano}, G., {Br{\"a}uninger}, H.~W., {Burkert}, W.,
  \& {Hartner}, G.~D. 2005{\natexlab{a}}, Space Science Reviews, 120, 165

\bibitem[{{Burrows} \& {Racusin}(2007{\natexlab{a}})}]{GCN.6181}
{Burrows}, D.~N. \& {Racusin}, J. 2007{\natexlab{a}}, GRB Coordinates Network,
  6181, 1

\bibitem[{{Burrows} \& {Racusin}(2007{\natexlab{b}})}]{br07}
---. 2007{\natexlab{b}}, ArXiv Astrophysics e-prints, astro-ph/0702633

\bibitem[{{Burrows} {et~al.}(2005{\natexlab{b}}){Burrows}, {Romano}, {Falcone},
  {Kobayashi}, {Zhang}, {Moretti}, {O'Brien}, {Goad}, {Campana}, {Page},
  {Angelini}, {Barthelmy}, {Beardmore}, {Capalbi}, {Chincarini}, {Cummings},
  {Cusumano}, {Fox}, {Giommi}, {Hill}, {Kennea}, {Krimm}, {Mangano},
  {Marshall}, {M{\'e}sz{\'a}ros}, {Morris}, {Nousek}, {Osborne}, {Pagani},
  {Perri}, {Tagliaferri}, {Wells}, {Woosley}, \& {Gehrels}}]{brf+05}
{Burrows}, D.~N., {Romano}, P., {Falcone}, A., {Kobayashi}, S., {Zhang}, B.,
  {Moretti}, A., {O'Brien}, P.~T., {Goad}, M.~R., {Campana}, S., {Page}, K.~L.,
  {Angelini}, L., {Barthelmy}, S., {Beardmore}, A.~P., {Capalbi}, M.,
  {Chincarini}, G., {Cummings}, J., {Cusumano}, G., {Fox}, D., {Giommi}, P.,
  {Hill}, J.~E., {Kennea}, J.~A., {Krimm}, H., {Mangano}, V., {Marshall}, F.,
  {M{\'e}sz{\'a}ros}, P., {Morris}, D.~C., {Nousek}, J.~A., {Osborne}, J.~P.,
  {Pagani}, C., {Perri}, M., {Tagliaferri}, G., {Wells}, A.~A., {Woosley}, S.,
  \& {Gehrels}, N. 2005{\natexlab{b}}, Science, 309, 1833


\bibitem[Butler et al.(2007)]{bkb+07} Butler, N.~R., Kocevski, 
D., Bloom, J.~S., \& Curtis, J.~L.\ 2007, \apj, 671, 656 

\bibitem[{{Cenko} \& {Fox}(2007)}]{GCN.6028}
{Cenko}, S.~B. \& {Fox}, D.~B. 2007, GRB Coordinates Network, 6028, 1

\bibitem[{{Cenko} {et~al.}(2006{\natexlab{a}}){Cenko}, {Fox}, {Moon},
  {Harrison}, {Kulkarni}, {Henning}, {Guzman}, {Bonati}, {Smith}, {Thicksten},
  {Doyle}, {Petrie}, {Gal-Yam}, {Soderberg}, {Anagnostou}, \& {Laity}}]{cfm+06}
{Cenko}, S.~B., {Fox}, D.~B., {Moon}, D.-S., {Harrison}, F.~A., {Kulkarni},
  S.~R., {Henning}, J.~R., {Guzman}, C.~D., {Bonati}, M., {Smith}, R.~M.,
  {Thicksten}, R.~P., {Doyle}, M.~W., {Petrie}, H.~L., {Gal-Yam}, A.,
  {Soderberg}, A.~M., {Anagnostou}, N.~L., \& {Laity}, A.~C.
  2006{\natexlab{a}}, \pasp, 118, 1396

\bibitem[{{Cenko} {et~al.}(2006{\natexlab{b}}){Cenko}, {Kasliwal}, {Harrison},
  {Pal'shin}, {Frail}, {Cameron}, {Berger}, {Fox}, {Gal-Yam}, {Kulkarni},
  {Moon}, {Nakar}, {Ofek}, {Penprase}, {Price}, {Sari}, {Schmidt}, {Soderberg},
  {Aptekar}, {Frederiks}, {Golenetskii}, {Burrows}, {Chevalier}, {Gehrels},
  {McCarthy}, {Nousek}, \& {Piran}}]{ckh+06}
{Cenko}, S.~B., {Kasliwal}, M., {Harrison}, F.~A., {Pal'shin}, V., {Frail},
  D.~A., {Cameron}, P.~B., {Berger}, E., {Fox}, D.~B., {Gal-Yam}, A.,
  {Kulkarni}, S.~R., {Moon}, D.-S., {Nakar}, E., {Ofek}, E.~O., {Penprase},
  B.~E., {Price}, P.~A., {Sari}, R., {Schmidt}, B.~P., {Soderberg}, A.~M.,
  {Aptekar}, R., {Frederiks}, D., {Golenetskii}, S., {Burrows}, D.~N.,
  {Chevalier}, R.~A., {Gehrels}, N., {McCarthy}, P.~J., {Nousek}, J.~A., \&
  {Piran}, T. 2006{\natexlab{b}}, \apj, 652, 490


\bibitem[Cenko et al.(2008)]{cfp+07} Cenko, S.~B., et al.\ 
2008, \apj, 677, 441 

\bibitem[{{Chandra} {et~al.}(2007){Chandra}, {Chandra}, \& {Gupta}}]{ccg07}
{Chandra}, P., {Chandra}, I., \& {Gupta}, N. 2007, GRB Coordinates Network,
  6102, 1

\bibitem[{{Chandra} \& {Frail}(2007)}]{cf07b}
{Chandra}, P. \& {Frail}, D.~A. 2007, GRB Coordinates Network, 6061, 1

\bibitem[{{Cohen} {et~al.}(1998){Cohen}, {Piran}, \& {Sari}}]{cps98}
{Cohen}, E., {Piran}, T., \& {Sari}, R. 1998, \apj, 509, 717

\bibitem[{{Cordes} \& {Lazio}(2002)}]{cl02}
{Cordes}, J.~M. \& {Lazio}, T.~J.~W. 2002, ArXiv Astrophysics e-prints, arXiv:astro-ph/0207156

\bibitem[{{Corsi} \& {Piro}(2006)}]{cp06}
{Corsi}, A. \& {Piro}, L. 2006, \aap, 458, 741

\bibitem[{{Dai} {et~al.}(2007){Dai}, {Garnavich}, {Prieto}, {Stanek},
  {Kochanek}, {Bechtold}, {Bouche}, {Buschkamp}, {Diolaiti}, {Fan},
  {Giallongo}, {Gredel}, {Hill}, {Jiang}, {McClellend}, {Milne}, {Pedichini},
  {Pogge}, {Ragazzoni}, {Rhoads}, {Smareglia}, {Thompson}, \&
  {Wagner}}]{dgp+07}
{Dai}, X., {Garnavich}, P.~M., {Prieto}, J.~L., {Stanek}, K.~Z., {Kochanek},
  C.~S., {Bechtold}, J., {Bouche}, N., {Buschkamp}, P., {Diolaiti}, E., {Fan},
  X., {Giallongo}, E., {Gredel}, R., {Hill}, J.~M., {Jiang}, L., {McClellend},
  C., {Milne}, P., {Pedichini}, F., {Pogge}, R.~W., {Ragazzoni}, R., {Rhoads},
  J., {Smareglia}, R., {Thompson}, D., \& {Wagner}, R.~M. 2007, ArXiv e-prints,
  712

\bibitem[{{Dickey} \& {Lockman}(1990)}]{dl90}
{Dickey}, J.~M. \& {Lockman}, F.~J. 1990, \araa, 28, 215

\bibitem[{{Evans} {et~al.}(2007){Evans}, {Beardmore}, {Page}, {Tyler},
  {Osborne}, {Goad}, {O'Brien}, {Vetere}, {Racusin}, {Morris}, {Burrows},
  {Capalbi}, {Perri}, {Gehrels}, \& {Romano}}]{ebp+07}
{Evans}, P.~A., {Beardmore}, A.~P., {Page}, K.~L., {Tyler}, L.~G., {Osborne},
  J.~P., {Goad}, M.~R., {O'Brien}, P.~T., {Vetere}, L., {Racusin}, J.,
  {Morris}, D., {Burrows}, D.~N., {Capalbi}, M., {Perri}, M., {Gehrels}, N., \&
  {Romano}, P. 2007, \aap, 469, 379

\bibitem[{{Frail} {et~al.}(2006){Frail}, {Cameron}, {Kasliwal}, {Nakar},
  {Price}, {Berger}, {Gal-Yam}, {Kulkarni}, {Fox}, {Soderberg}, {Schmidt},
  {Ofek}, \& {Cenko}}]{fck+06}
{Frail}, D.~A., {Cameron}, P.~B., {Kasliwal}, M., {Nakar}, E., {Price}, P.~A.,
  {Berger}, E., {Gal-Yam}, A., {Kulkarni}, S.~R., {Fox}, D.~B., {Soderberg},
  A.~M., {Schmidt}, B.~P., {Ofek}, E., \& {Cenko}, S.~B. 2006, \apjl, 646, L99

\bibitem[{{Frail} {et~al.}(2001){Frail}, {Kulkarni}, {Sari}, {Djorgovski},
  {Bloom}, {Galama}, {Reichart}, {Berger}, {Harrison}, {Price}, {Yost},
  {Diercks}, {Goodrich}, \& {Chaffee}}]{fks+01}
{Frail}, D.~A., {Kulkarni}, S.~R., {Sari}, R., {Djorgovski}, S.~G., {Bloom},
  J.~S., {Galama}, T.~J., {Reichart}, D.~E., {Berger}, E., {Harrison}, F.~A.,
  {Price}, P.~A., {Yost}, S.~A., {Diercks}, A., {Goodrich}, R.~W., \&
  {Chaffee}, F. 2001, \apjl, 562, L55

\bibitem[{{Frail} {et~al.}(2005){Frail}, {Soderberg}, {Kulkarni}, {Berger},
  {Yost}, {Fox}, \& {Harrison}}]{fsk+05}
{Frail}, D.~A., {Soderberg}, A.~M., {Kulkarni}, S.~R., {Berger}, E., {Yost},
  S., {Fox}, D.~W., \& {Harrison}, F.~A. 2005, \apj, 619, 994

\bibitem[{{Frail} {et~al.}(2000){Frail}, {Waxman}, \& {Kulkarni}}]{fwk00}
{Frail}, D.~A., {Waxman}, E., \& {Kulkarni}, S.~R. 2000, \apj, 537, 191

\bibitem[{{Frail} {et~al.}(2003){Frail}, {Yost}, {Berger}, {Harrison}, {Sari},
  {Kulkarni}, {Taylor}, {Bloom}, {Fox}, {Moriarty-Schieven}, \&
  {Price}}]{fyb+03}
{Frail}, D.~A., {Yost}, S.~A., {Berger}, E., {Harrison}, F.~A., {Sari}, R.,
  {Kulkarni}, S.~R., {Taylor}, G.~B., {Bloom}, J.~S., {Fox}, D.~W.,
  {Moriarty-Schieven}, G.~H., \& {Price}, P.~A. 2003, \apj, 590, 992

\bibitem[{{Freedman} \& {Waxman}(2001)}]{fw01}
{Freedman}, D.~L. \& {Waxman}, E. 2001, 547, 922

\bibitem[{{Fruchter} {et~al.}(1999){Fruchter}, {Thorsett}, {Metzger}, {Sahu},
  {Petro}, {Livio}, {Ferguson}, {Pian}, {Hogg}, {Galama}, {Gull},
  {Kouveliotou}, {Macchetto}, {Van Paradijs}, {Pedersen}, \& {Smette}}]{ftm+99}
{Fruchter}, A.~S., {Thorsett}, S.~E., {Metzger}, M.~R., {Sahu}, K.~C., {Petro},
  L., {Livio}, M., {Ferguson}, H., {Pian}, E., {Hogg}, D.~W., {Galama}, T.,
  {Gull}, T.~R., {Kouveliotou}, C., {Macchetto}, D., {Van Paradijs}, J.,
  {Pedersen}, H., \& {Smette}, A. 1999, 519, L13

\bibitem[{{Fukugita} {et~al.}(1995){Fukugita}, {Shimasaku}, \&
  {Ichikawa}}]{fsi95}
{Fukugita}, M., {Shimasaku}, K., \& {Ichikawa}, T. 1995, \pasp, 107, 945

\bibitem[{{Garmire} {et~al.}(2003){Garmire}, {Bautz}, {Ford}, {Nousek}, \&
  {Ricker}}]{gbf+03}
{Garmire}, G.~P., {Bautz}, M.~W., {Ford}, P.~G., {Nousek}, J.~A., \& {Ricker},
  Jr., G.~R. 2003, in Presented at the Society of Photo-Optical Instrumentation
  Engineers (SPIE) Conference, Vol. 4851, X-Ray and Gamma-Ray Telescopes and
  Instruments for Astronomy. Edited by Joachim E. Truemper, Harvey D.
  Tananbaum. Proceedings of the SPIE, Volume 4851, pp. 28-44 (2003)., ed. J.~E.
  {Truemper} \& H.~D. {Tananbaum}, 28--44

\bibitem[{{Garnavich} {et~al.}(2007){Garnavich}, {Fan}, {Jiang}, {Dai}, {Kuhn},
  {Bouche}, {Buschkamp}, {Smith}, {Milne}, {Bechtold}, {Stanek}, {Prieto},
  {Wagner}, {Rhoads}, {Hill}, {Baruffolo}, {Desantis}, {Diolaiti}, {Dipaola},
  {Farinato}, {Fontana}, {Gallozzi}, {Gasparo}, {Giallongo}, {Grazian},
  {Pasian}, {Pedichini}, {Ragazzoni}, {Smareglia}, {Speziali}, {Testa}, \&
  {Vernet}}]{gfj+07}
{Garnavich}, P., {Fan}, X., {Jiang}, L., {Dai}, X., {Kuhn}, O., {Bouche}, N.,
  {Buschkamp}, P., {Smith}, P., {Milne}, P., {Bechtold}, J., {Stanek}, K.~Z.,
  {Prieto}, J., {Wagner}, R.~M., {Rhoads}, J., {Hill}, J., {Baruffolo}, A.,
  {Desantis}, C., {Diolaiti}, E., {Dipaola}, A., {Farinato}, J., {Fontana}, A.,
  {Gallozzi}, S., {Gasparo}, F., {Giallongo}, E., {Grazian}, A., {Pasian}, F.,
  {Pedichini}, F., {Ragazzoni}, R., {Smareglia}, R., {Speziali}, R., {Testa},
  V., \& {Vernet}, E. 2007, GRB Coordinates Network, 6165, 1

\bibitem[{{Golenetskii} {et~al.}(2007){Golenetskii}, {Aptekar}, {Mazets},
  {Pal'Shin}, {Frederiks}, \& {Cline}}]{gam+07}
{Golenetskii}, S., {Aptekar}, R., {Mazets}, E., {Pal'Shin}, V., {Frederiks},
  D., \& {Cline}, T. 2007, GRB Coordinates Network, 6049, 1

\bibitem[{{Goodman}(1997)}]{goo97}
{Goodman}, J. 1997, New Astronomy, 2, 449

\bibitem[{{Harrison} {et~al.}(1999){Harrison}, {Bloom}, {Frail}, {Sari},
  {Kulkarni}, {Djorgovski}, {Axelrod}, {Mould}, {Schmidt}, {Wieringa}, {Wark},
  {Subrahmanyan}, {McConnell}, {McCarthy}, {Schaefer}, {McMahon}, {Markze},
  {Firth}, {Soffitta}, \& {Amati}}]{hbf+99}
{Harrison}, F.~A., {Bloom}, J.~S., {Frail}, D.~A., {Sari}, R., {Kulkarni},
  S.~R., {Djorgovski}, S.~G., {Axelrod}, T., {Mould}, J., {Schmidt}, B.~P.,
  {Wieringa}, M.~H., {Wark}, R.~M., {Subrahmanyan}, R., {McConnell}, D.,
  {McCarthy}, P.~J., {Schaefer}, B.~E., {McMahon}, R.~G., {Markze}, R.~O.,
  {Firth}, E., {Soffitta}, P., \& {Amati}, L. 1999, 523, L121

\bibitem[{{Harrison} {et~al.}(2001){Harrison}, {Yost}, {Sari}, {Berger},
  {Galama}, {Holtzman}, {Axelrod}, {Bloom}, {Chevalier}, {Costa}, {Diercks},
  {Djorgovski}, {Frail}, {Frontera}, {Hurley}, {Kulkarni}, {McCarthy}, {Piro},
  {Pooley}, {Price}, {Reichart}, {Ricker}, {Shepherd}, {Schmidt}, {Walter}, \&
  {Wheeler}}]{hys+01}
{Harrison}, F.~A., {Yost}, S.~A., {Sari}, R., {Berger}, E., {Galama}, T.~J.,
  {Holtzman}, J., {Axelrod}, T., {Bloom}, J.~S., {Chevalier}, R., {Costa}, E.,
  {Diercks}, A., {Djorgovski}, S.~G., {Frail}, D.~A., {Frontera}, F., {Hurley},
  K., {Kulkarni}, S.~R., {McCarthy}, P., {Piro}, L., {Pooley}, G.~G., {Price},
  P.~A., {Reichart}, D., {Ricker}, G.~R., {Shepherd}, D., {Schmidt}, B.,
  {Walter}, F., \& {Wheeler}, C. 2001, \apj, 559, 123

\bibitem[{{Hook} {et~al.}(2003){Hook}, {Allington-Smith}, {Beard}, {Crampton},
  {Davies}, {Dickson}, {Ebbers}, {Fletcher}, {Jorgensen}, {Jean}, {Juneau},
  {Murowinski}, {Nolan}, {Laidlaw}, {Leckie}, {Marshall}, {Purkins},
  {Richardson}, {Roberts}, {Simons}, {Smith}, {Stilburn}, {Szeto}, {Tierney},
  {Wolff}, \& {Wooff}}]{hab+03}
{Hook}, I., {Allington-Smith}, J.~R., {Beard}, S.~M., {Crampton}, D., {Davies},
  R.~L., {Dickson}, C.~G., {Ebbers}, A.~W., {Fletcher}, J.~M., {Jorgensen}, I.,
  {Jean}, I., {Juneau}, S., {Murowinski}, R.~G., {Nolan}, R., {Laidlaw}, K.,
  {Leckie}, B., {Marshall}, G.~E., {Purkins}, T., {Richardson}, I.~M.,
  {Roberts}, S.~C., {Simons}, D.~A., {Smith}, M.~J., {Stilburn}, J.~R.,
  {Szeto}, K., {Tierney}, C., {Wolff}, R.~J., \& {Wooff}, R. 2003, in
  Instrument Design and Performance for Optical/Infrared Ground-based
  Telescopes. Edited by Iye, Masanori; Moorwood, Alan F. M. Proceedings of the
  SPIE, Volume 4841, pp. 1645-1656 (2003)., ed. M.~{Iye} \& A.~F.~M.
  {Moorwood}, 1645--1656

\bibitem[{{Hurley} {et~al.}(2007){Hurley}, {Cline}, {Mitrofanov}, {Kozyrev},
  {Litvak}, {Sanin}, {Tret'yakov}, {Parshukov}, {Boynton}, {Fellows},
  {Harshman}, {Shinohara}, {Starr}, {Yamaoka}, {Ohno}, {Fukazawa}, {Takahashi},
  {Tashiro}, {Terada}, {Murakami}, {Makishima}, {Smith}, {Lin}, {McTiernan},
  {Schwartz}, {Wigger}, {Hajdas}, {Zehnder}, {von Kienlin}, {Lichti}, {Rau},
  {Cummings}, {Krimm}, {Barthelmy}, \& {Gehrels}}]{hcm+07}
{Hurley}, K., {Cline}, T., {Mitrofanov}, I., {Kozyrev}, A., {Litvak}, M.,
  {Sanin}, A., {Tret'yakov}, V., {Parshukov}, A., {Boynton}, W., {Fellows}, C.,
  {Harshman}, K., {Shinohara}, C., {Starr}, R., {Yamaoka}, K., {Ohno}, M.,
  {Fukazawa}, Y., {Takahashi}, T., {Tashiro}, M., {Terada}, Y., {Murakami}, T.,
  {Makishima}, K., {Smith}, D.~M., {Lin}, R.~P., {McTiernan}, J., {Schwartz},
  R., {Wigger}, C., {Hajdas}, W., {Zehnder}, A., {von Kienlin}, A., {Lichti},
  G., {Rau}, A., {Cummings}, J., {Krimm}, H., {Barthelmy}, S., \& {Gehrels}, N.
  2007, GRB Coordinates Network, 6024, 1

\bibitem[{{Jarrett} {et~al.}(2006){Jarrett}, {Polletta}, {Fournon}, {Stacey},
  {Xu}, {Siana}, {Farrah}, {Berta}, {Hatziminaoglou}, {Rodighiero}, {Surace},
  {Domingue}, {Shupe}, {Fang}, {Lonsdale}, {Oliver}, {Rowan-Robinson}, {Smith},
  {Babbedge}, {Gonzalez-Solares}, {Masci}, {Franceschini}, \&
  {Padgett}}]{jpf+06}
{Jarrett}, T.~H., {Polletta}, M., {Fournon}, I.~P., {Stacey}, G., {Xu}, K.,
  {Siana}, B., {Farrah}, D., {Berta}, S., {Hatziminaoglou}, E., {Rodighiero},
  G., {Surace}, J., {Domingue}, D., {Shupe}, D., {Fang}, F., {Lonsdale}, C.,
  {Oliver}, S., {Rowan-Robinson}, M., {Smith}, G., {Babbedge}, T.,
  {Gonzalez-Solares}, E., {Masci}, F., {Franceschini}, A., \& {Padgett}, D.
  2006, \aj, 131, 261

\bibitem[{{Jordi} {et~al.}(2006){Jordi}, {Grebel}, \& {Ammon}}]{jga06}
{Jordi}, K., {Grebel}, E.~K., \& {Ammon}, K. 2006, \aap, 460, 339

\bibitem[{{Kocevski} \& {Butler}(2007)}]{kb07}
{Kocevski}, D. \& {Butler}, N. 2007, astro-ph/0707.4478

\bibitem[{{Kulkarni} {et~al.}(1999){Kulkarni}, {Djorgovski}, {Odewahn},
  {Bloom}, {Gal}, {Koresko}, {Harrison}, {Lubin}, {Armus}, {Sari},
  {Illingworth}, {Kelson}, {Magee}, {Dokkum}, {Frail}, {Mulchaey}, {Malkan},
  {McClean}, {Teplitz}, {Koerner}, {Kirkpatrick}, {Kobayashi}, {Yadigaroglu},
  {Halpern}, {Piran}, {Goodrich}, {Chaffee}, {Feroci}, \& {Costa}}]{kdo+99}
{Kulkarni}, S.~R., {Djorgovski}, S.~G., {Odewahn}, S.~C., {Bloom}, J.~S.,
  {Gal}, R.~R., {Koresko}, C.~D., {Harrison}, F.~A., {Lubin}, L.~M., {Armus},
  L., {Sari}, R., {Illingworth}, G.~D., {Kelson}, D.~D., {Magee}, D.~K.,
  {Dokkum}, P.~G.~V., {Frail}, D.~A., {Mulchaey}, J.~S., {Malkan}, M.~A.,
  {McClean}, I.~S., {Teplitz}, H.~I., {Koerner}, D., {Kirkpatrick}, D.,
  {Kobayashi}, N., {Yadigaroglu}, I.-A., {Halpern}, J., {Piran}, T.,
  {Goodrich}, R.~W., {Chaffee}, F.~H., {Feroci}, M., \& {Costa}, E. 1999, \nat,
  398, 389

\bibitem[{{Kumar}(2000)}]{kumar00}
{Kumar}, P. 2000, 538, L125

\bibitem[{{Li} \& {Chevalier}(2003)}]{lc03}
{Li}, Z.-Y. \& {Chevalier}, R.~A. 2003, in Lecture Notes in Physics, Berlin
  Springer Verlag, Vol. 598, Supernovae and Gamma-Ray Bursters, ed.
  K.~{Weiler}, 419--444

\bibitem[{{Meszaros}(2006)}]{mes06}
{Meszaros}, P. 2006, Reports of Progress in Physics, 69, 2259

\bibitem[{{Nakar} \& {Piran}(2005)}]{np05}
{Nakar}, E. \& {Piran}, T. 2005, \apjl, 619, L147

\bibitem[{{Oke} {et~al.}(1995){Oke}, {Cohen}, {Carr}, {Cromer}, {Dingizian},
  {Harris}, {Labrecque}, {Lucinio}, {Schaal}, {Epps}, \& {Miller}}]{occ+95}
{Oke}, J.~B., {Cohen}, J.~G., {Carr}, M., {Cromer}, J., {Dingizian}, A.,
  {Harris}, F.~H., {Labrecque}, S., {Lucinio}, R., {Schaal}, W., {Epps}, H., \&
  {Miller}, J. 1995, \pasp, 107, 375

\bibitem[{{Panaitescu}(2008)}]{pan08}
{Panaitescu}, A. 2008, 
\mnras, 383, 1143 

\bibitem[{{Panaitescu} \& {Kumar}(2001)}]{pk01}
{Panaitescu}, A. \& {Kumar}, P. 2001, 560, L49

\bibitem[{{Pei}(1992)}]{pei92}
{Pei}, Y.~C. 1992, \apj, 395, 130

\bibitem[{{Piran}(1999)}]{pir99}
{Piran}, T. 1999, \physrep, 314, 575

\bibitem[{{Piran}(2005)}]{pir05}
---. 2005, Reviews of Modern Physics, 76, 1143

\bibitem[{{Piran}(2005{\natexlab{b}})}]{p05}
---. 2005{\natexlab{b}}, Reviews of Modern Physics, 76, 1143

\bibitem[{{Prochaska} {et~al.}(2007){Prochaska}, {Chen}, {Dessauges-Zavadsky},
  \& {Bloom}}]{pcd+07}
{Prochaska}, J.~X., {Chen}, H.-W., {Dessauges-Zavadsky}, M., \& {Bloom}, J.~S.
  2007, \apj, 666, 267 

\bibitem[{{Racusin} \& {Vetere}(2007)}]{GCN.6030}
{Racusin}, J. \& {Vetere}, L. 2007, GRB Coordinates Network, 6030, 1

\bibitem[Racusin et al.(2007)]{rcm07} Racusin, J.~L., 
Cummings, J., Marshall, F.~E., Burrows, D.~N., Krimm, H., 
\& Sato, G.\ 2007, GCNR, 28, 3 (2007), 28, 3 

\bibitem[{{Rhoads}(1999)}]{rho99}
{Rhoads}, J.~E. 1999, \apj, 525, 737

\bibitem[{{Sari}(1997)}]{sar97}
{Sari}, R. 1997, \apjl, 489, L37+

\bibitem[{{Sari} \& {Esin}(2001)}]{se01}
{Sari}, R. \& {Esin}, A.~A. 2001, \apj, 548, 787

\bibitem[{{Sari} {et~al.}(1999){Sari}, {Piran}, \& {Halpern}}]{sph99}
{Sari}, R., {Piran}, T., \& {Halpern}, J.~P. 1999, 519, L17

\bibitem[{{Sari} {et~al.}(1998){Sari}, {Piran}, \& {Narayan}}]{spn98}
{Sari}, R., {Piran}, T., \& {Narayan}, R. 1998, 497, L17

\bibitem[{{Sault} {et~al.}(1995){Sault}, {Teuben}, \& {Wright}}]{stw95}
{Sault}, R.~J., {Teuben}, P.~J., \& {Wright}, M.~C.~H. 1995, in Astronomical
  Society of the Pacific Conference Series, Vol.~77, Astronomical Data Analysis
  Software and Systems IV, ed. R.~A. {Shaw}, H.~E. {Payne}, \& J.~J.~E.
  {Hayes}, 433--+

\bibitem[{{Schlegel} {et~al.}(1998){Schlegel}, {Finkbeiner}, \&
  {Davis}}]{sfd98}
{Schlegel}, D.~J., {Finkbeiner}, D.~P., \& {Davis}, M. 1998, \apj, 500, 525

\bibitem[{{Stanek} {et~al.}(1999){Stanek}, {Garnavich}, {Kaluzny}, {Pych}, \&
  {Thompson}}]{sgk+99b}
{Stanek}, K.~Z., {Garnavich}, P.~M., {Kaluzny}, J., {Pych}, W., \& {Thompson},
  I. 1999, 522, L39

\bibitem[Updike et al.(2008)]{uhn+08} Updike, A.~C., et al.\ 
2008, ArXiv e-prints, 805, arXiv:0805.1094 

\bibitem[{{Usov}(1992)}]{u92}
{Usov}, V.~V. 1992, \nat, 357, 472

\bibitem[{{van der Horst}(2007)}]{van07}
{van der Horst}, A.~J. 2007, GRB Coordinates Network, 6042, 1

\bibitem[{{Vreeswijk} {et~al.}(2007){Vreeswijk}, {Ledoux}, {Smette}, {Ellison},
  {Jaunsen}, {Andersen}, {Fruchter}, {Fynbo}, {Hjorth}, {Kaufer}, {M{\o}ller},
  {Petitjean}, {Savaglio}, \& {Wijers}}]{vls+07}
{Vreeswijk}, P.~M., {Ledoux}, C., {Smette}, A., {Ellison}, S.~L., {Jaunsen},
  A.~O., {Andersen}, M.~I., {Fruchter}, A.~S., {Fynbo}, J.~P.~U., {Hjorth}, J.,
  {Kaufer}, A., {M{\o}ller}, P., {Petitjean}, P., {Savaglio}, S., \& {Wijers},
  R.~A.~M.~J. 2007, \aap, 468, 83

\bibitem[{{Walker}(1998)}]{wal98}
{Walker}, M.~A. 1998, \mnras, 294, 307

\bibitem[{{Walker}(2001)}]{wal01}
---. 2001, \mnras, 321, 176

\bibitem[{{Wei} \& {Lu}(1998)}]{wl98}
{Wei}, D.~M. \& {Lu}, T. 1998, \apj, 505, 252

\bibitem[{{Wei} \& {Lu}(2000)}]{wl00}
---. 2000, \aap, 360, L13

\bibitem[Willingale et al.(2007)]{wog+07} Willingale, R., 
O'Brien, P.~T., Goad, M.~R., Osborne, J.~P., Page, K.~L., 
\& Tanvir, N.~R.\ 2007, ArXiv e-prints, 710, arXiv:0710.3727 

\bibitem[{{Woosley}(1993)}]{w93}
{Woosley}, S.~E. 1993, \apj, 405, 273

\bibitem[{{Wu} {et~al.}(2005){Wu}, {Dai}, {Huang}, \& {Lu}}]{wdhl05}
{Wu}, X.~F., {Dai}, Z.~G., {Huang}, Y.~F., \& {Lu}, T. 2005, \apj, 619, 968

\bibitem[{{Yost}(2004)}]{2004PhDT.......429Y}
{Yost}, S.~A. 2004, PhD thesis, AA(CALIFORNIA INSTITUTE OF TECHNOLOGY)

\bibitem[{{Yost} {et~al.}(2003){Yost}, {Harrison}, {Sari}, \&
  {Frail}}]{yhsf03}
{Yost}, S.~A., {Harrison}, F.~A., {Sari}, R., \& {Frail}, D.~A.
  2003, \apj, 597, 459


\bibitem[{{Zhang} {et~al.}(2006){Zhang}, {Fan}, {Dyks}, {Kobayashi},
  {M{\'e}sz{\'a}ros}, {Burrows}, {Nousek}, \& {Gehrels}}]{zfd+06}
{Zhang}, B., {Fan}, Y.~Z., {Dyks}, J., {Kobayashi}, S., {M{\'e}sz{\'a}ros}, P.,
  {Burrows}, D.~N., {Nousek}, J.~A., \& {Gehrels}, N. 2006, \apj, 642, 354

\end{thebibliography}

\clearpage

\begin{deluxetable}{llll}
\tabletypesize{\footnotesize}
\tablecaption{X-ray observations of the GRB 070125 with \Swift\
at $3.594 \times 10^{17}$ Hz
\label{tab:xray}}
\tablewidth{0pt}
\tablehead{
\colhead{Days since} & \colhead{Counts/s (0.3-10.0 keV)} & \colhead{Flux
(0.3-10.0 keV)}& \colhead{Flux Density}\\
\colhead{Explosion} & \colhead{} & \colhead{erg s$^{-1}$ cm$^{-2}$} &
\colhead{$\mu$Jy}
}
\startdata
0.54203 & $0.064175 \pm0.016806$ & $(3.00 \pm 0.79)\times 10^{-12} $ & $0.240\pm 0.063$ \\
0.54468 & $0.096087 \pm0.025044$ & $(4.49 \pm 1.17)\times 10^{-12} $ & $0.360\pm 0.094$ \\
0.54630 & $0.104163 \pm0.027022$ & $(4.86 \pm 1.26)\times 10^{-12} $ & $0.390\pm 0.100$ \\
0.54868 & $0.060351 \pm0.015879$ & $(2.82 \pm 0.74)\times 10^{-12} $ & $0.226\pm 0.059$ \\
0.55235 & $0.065485 \pm0.017149$ & $(3.06 \pm 0.80)\times 10^{-12} $ & $0.245\pm 0.064$ \\
0.55447 & $0.106140 \pm0.023902$ & $(4.96 \pm 1.12)\times 10^{-12} $ & $0.398\pm 0.089$ \\
0.55787 & $0.098574 \pm0.017867$ & $(4.60 \pm 0.83)\times 10^{-12} $ & $0.369\pm 0.067$ \\
0.60723 & $0.083110 \pm0.021560$ & $(3.88 \pm 1.01)\times 10^{-12} $ & $0.311\pm 0.081$ \\
0.60991 & $0.062137 \pm0.016195$ & $(2.90 \pm 0.76)\times 10^{-12} $ & $0.233\pm 0.061$ \\
0.61257 & $0.084761 \pm0.022197$ & $(3.96 \pm 1.04)\times 10^{-12} $ & $0.317\pm 0.083$ \\
0.61587 & $0.056149 \pm0.014844$ & $(2.62 \pm 0.69)\times 10^{-12} $ & $0.210\pm 0.056$ \\
0.61921 & $0.065275 \pm0.017013$ & $(3.05 \pm 0.79)\times 10^{-12} $ & $0.245\pm 0.063$ \\
0.62327 & $0.051077 \pm0.010302$ & $(2.39 \pm 0.48)\times 10^{-12} $ & $0.191\pm 0.038$ \\
0.67575 & $0.077789 \pm0.020564$ & $(3.63 \pm 0.96)\times 10^{-12} $ & $0.291\pm 0.077$ \\
0.67845 & $0.057316 \pm0.015010$ & $(2.68 \pm 0.70)\times 10^{-12} $ & $0.215\pm 0.056$ \\
0.68217 & $0.062115 \pm0.016114$ & $(2.90 \pm 0.75)\times 10^{-12} $ & $0.233\pm 0.060$ \\
0.68565 & $0.055467 \pm0.014663$ & $(2.59 \pm 0.69)\times 10^{-12} $ & $0.208\pm 0.055$ \\
0.68898 & $0.062736 \pm0.016507$ & $(2.93 \pm 0.77)\times 10^{-12} $ & $0.235\pm 0.062$ \\
0.69205 & $0.067433 \pm0.017235$ & $(3.15 \pm 0.81)\times 10^{-12} $ & $0.253\pm 0.064$ \\
0.73354 & $0.081260 \pm0.021280$ & $(3.79 \pm 0.99)\times 10^{-12} $ & $0.304\pm 0.079$ \\
0.73787 & $0.071125 \pm0.013240$ & $(3.32 \pm 0.62)\times 10^{-12} $ & $0.266\pm 0.049$ \\
1.34534 & $0.032554 \pm0.007425$ & $(1.52 \pm 0.35)\times 10^{-12} $ & $0.122\pm 0.028$ \\
1.35747 & $0.041983 \pm0.008103$ & $(1.96 \pm 0.38)\times 10^{-12} $ & $0.157\pm 0.030$ \\
1.41052 & $0.027726 \pm0.006362$ & $(1.29 \pm 0.30)\times 10^{-12} $ & $0.104\pm 0.023$ \\
1.42675 & $0.039339 \pm0.009101$ & $(1.84 \pm 0.43)\times 10^{-12} $ & $0.147\pm 0.034$ \\
1.47713 & $0.032748 \pm0.007624$ & $(1.53 \pm 0.36)\times 10^{-12} $ & $0.123\pm 0.028$ \\
1.48861 & $0.038279 \pm0.007576$ & $(1.79 \pm 0.35)\times 10^{-12} $ & $0.143\pm 0.028$ \\
1.78359 & $0.014033 \pm0.002725$ & $(6.55 \pm 1.27)\times 10^{-13} $ & $0.053 \pm 0.010$ \\
1.89040 & $0.013734 \pm0.003671$ & $(6.41 \pm 1.71)\times 10^{-13} $ & $0.051 \pm 0.014$ \\
1.94946 & $0.015608 \pm0.004296$ & $(7.29 \pm 2.01)\times 10^{-13} $ & $0.058 \pm 0.016$ \\
2.05462 & $0.012729 \pm0.002489$ & $(5.94 \pm 1.16)\times 10^{-13} $ & $0.048 \pm 0.009$ \\
2.18650 & $0.009233 \pm0.002318$ & $(4.31 \pm 1.08)\times 10^{-13} $ & $0.035 \pm 0.009$ \\
2.30888 & $0.010806 \pm0.002868$ & $(5.04 \pm 1.34)\times 10^{-13} $ & $0.040 \pm 0.011$ \\
2.47372 & $0.015085 \pm0.003520$ & $(7.04 \pm 1.64)\times 10^{-13} $ & $0.057 \pm 0.013$ \\
2.68019 & $0.006752 \pm0.001515$ & $(3.15 \pm 0.71)\times 10^{-13} $ & $0.025 \pm 0.006$ \\
2.83975 & $0.010154 \pm0.002670$ & $(4.74 \pm 1.25)\times 10^{-13} $ & $0.038 \pm 0.010$ \\
3.07971 & $0.007542 \pm0.001640$ & $(3.52 \pm 0.77)\times 10^{-13} $ & $0.028 \pm 0.006$ \\
3.38860 & $0.005554 \pm0.001293$ & $(2.59 \pm 0.60)\times 10^{-13} $ & $0.021 \pm 0.005$ \\
3.59126 & $0.006728 \pm0.001769$ & $(3.14 \pm 0.83)\times 10^{-13} $ & $0.025 \pm 0.007$ \\
4.04549 & $0.004496 \pm0.001265$ & $(2.10 \pm 0.59)\times 10^{-13} $ & $0.017 \pm 0.006$ \\
5.97272 & $0.001798 \pm0.000493$ & $(8.40 \pm 2.30)\times 10^{-14} $ & $0.007 \pm 0.002$ \\
9.08277 & $0.001052 \pm0.000293$ & $(4.91 \pm 1.36)\times 10^{-14} $ & $0.004 \pm 0.001$ \\
10.5438 & $0.000785 \pm0.000239$ & $(3.67 \pm 1.11)\times 10^{-14} $ & $0.003 \pm 0.001$ \\
14.5620 & $<0.000277$ & $<1.3\times 10^{-14}$ & $<0.001 $ \\
39.6190 &$<0.000167$    & $< 2.0 \times 10^{-15}$           &  $<0.0002$\tablenotemark{a}          \\
\enddata
\tablenotetext{a}{{\it Chandra} observations, GCN 6186}
\end{deluxetable}

\clearpage

\begin{deluxetable}{cccclccll}
\tabletypesize{\scriptsize}
\tablecaption{Optical/IR/UV observations of the GRB 070125
\label{tab:opt}}
\tablewidth{0pt}
\tablehead{
\colhead{Days since} & \colhead{Filter} & \colhead{$\lambda$} &
\colhead{$F_0$} &
\colhead{Magnitude} & \colhead{$A_\lambda$} & \colhead{Corrected} & \colhead{Flux
density} & \colhead{Reference\tablenotemark{a}}\\
\colhead{Explosion} &  \colhead{} & \colhead{in $\mu$m} &
\colhead{Jy} & \colhead{} & \colhead{mag} &
\colhead{mag} & \colhead{$\mu$Jy} & \colhead{}
}
\startdata
0.54034 &    V & 0.55 & 3590 & $ 18.54 \pm 0.06 $& 0.17& 18.37 & $161.1\pm8.8 $&
 GCN6041, 6036 \\
0.55035 &    B & 0.45 & 4020 & $ 18.92 \pm 0.03 $& 0.23& 18.69 & $134.3\pm3.8 $&
 GCN6036, 6041 \\
0.67441 &    V & 0.55 & 3590 & $ 18.74 \pm 0.07 $& 0.17& 18.57 & $134.0\pm8.5 $&
 GCN6036, 6041 \\
0.68433 &    B & 0.45 & 4020 & $ 19.03 \pm 0.06 $& 0.23& 18.80 & $121.4\pm6.9 $&
 GCN6036, 6041 \\
0.79149 &    R & 0.66 & 3020 & $ 18.59 \pm 0.03$ & 0.14& 18.45 & $125.9 \pm 3.9$
 & P60\\
0.79379 &    R & 0.66 & 3020 & $ 18.51 \pm 0.03$ & 0.14& 18.37 & $135.6 \pm 4.3$
 & P60\\
0.79609 &    R & 0.66 & 3020 & $ 18.57 \pm 0.03$ & 0.14& 18.43 & $128.2 \pm 4.1$
 & P60\\
0.79848 &    R & 0.66 & 3020 & $ 18.61 \pm 0.03$ & 0.14& 18.47 & $123.7 \pm 3.7$
 & P60\\
0.80087 &    R & 0.66 & 3020 & $ 18.53 \pm 0.05$ & 0.14& 18.39 & $133.1 \pm 6.6$
 & P60\\
0.80229 &    R & 0.66 & 3020 & $ 18.60\pm  -   $ & 0.14& 18.46 & $124.8\pm-$&
GCN6028 \\
0.80326 &    R & 0.66 & 3020 & $ 18.55 \pm 0.03$ & 0.14& 18.41 & $130.7 \pm 4.2$
 & P60\\
0.89965 &    r'& 0.63 & 3631 & $ 19.03 \pm 0.09$ & 0.14& 18.89 & $100.9\pm 1.2$
& GMOS\\
0.90833 &    R & 0.66 & 3020 & $ 18.80\pm  -   $ & 0.14& 18.66 & $103.7\pm-$&
GCN6044 \\
1.08990 &    R & 0.66 & 3020 & $ 18.70 \pm 0.06$ & 0.14& 18.56 & $113.8 \pm 6.5$
 & P60\\
1.09277 &    R & 0.66 & 3020 & $ 18.63 \pm 0.07$ & 0.14& 18.49 & $121.3 \pm 7.4$
 & P60\\
1.09564 &    R & 0.66 & 3020 & $ 18.69 \pm 0.03$ & 0.14& 18.55 & $114.8 \pm 3.5$
 & P60\\
1.10146 &    R & 0.66 & 3020 & $ 18.66 \pm 0.04$ & 0.14& 18.52 & $118.1 \pm 4.0$
 & P60\\
1.10450 &    i'& 0.77 & 3631 & $ 18.61 \pm 0.03$ & 0.10& 18.51 & $143.2\pm 4.5$
& P60 \\
1.10742 &    i'& 0.77 & 3631 & $ 18.61 \pm 0.03$ & 0.10& 18.51 & $143.2\pm 4.3$
& P60 \\
1.11337 &    i'& 0.77 & 3631 & $ 18.53 \pm 0.03$ & 0.10& 18.43 & $154.2 \pm 4.8$ & P60 \\
1.11630 &    i'& 0.77 & 3631 & $ 18.59 \pm 0.03$ & 0.10& 18.49 & $146.0 \pm 4.6$ & P60 \\
1.11929 &    R & 0.66 & 3020 & $ 18.60 \pm 0.03$ & 0.14& 18.46 & $124.8 \pm 3.6$ & P60\\
1.12222 &    R & 0.66 & 3020 & $ 18.60 \pm 0.03$ & 0.14& 18.46 & $124.8 \pm 4.0$ & P60\\
1.12516 &    R & 0.66 & 3020 & $ 18.64 \pm 0.04$ & 0.14& 18.50 & $120.2 \pm 4.1$ & P60\\
1.12811 &    R & 0.66 & 3020 & $ 18.60 \pm 0.03$ & 0.14& 18.46 & $124.8 \pm 3.6$ & P60\\
1.13108 &    R & 0.66 & 3020 & $ 18.62 \pm 0.03$ & 0.14& 18.48 & $122.4 \pm 3.6$ & P60\\
1.13405 &    i'& 0.77 & 3631 & $ 18.64 \pm 0.04$ & 0.10& 18.54 & $139.3\pm 4.8$& P60 \\
1.13698 &    i'& 0.77 & 3631 & $ 18.62 \pm 0.04$ & 0.10& 18.52 & $141.9\pm 5.1$& P60 \\
1.14743 &    R & 0.66 & 3020 & $ 18.73 \pm 0.15$ & 0.14& 18.59 & $110.6\pm14.6$&   GCN6035 \\
1.15466 &    R & 0.66 & 3020 & $ 18.63 \pm 0.04$ & 0.14& 18.49 & $121.3 \pm 4.6$ & P60\\
1.15705 &    R & 0.66 & 3020 & $ 18.64 \pm 0.03$ & 0.14& 18.50 & $120.2 \pm 3.6$ & P60\\
1.16182 &    R & 0.66 & 3020 & $ 18.63 \pm 0.03$ & 0.14& 18.49 & $121.3 \pm 3.9$ & P60\\
1.16421 &    R & 0.66 & 3020 & $ 18.66 \pm 0.03$ & 0.14& 18.52 & $118.1 \pm 3.8$ & P60\\
1.16661 &    i'& 0.77 & 3631 & $ 18.65 \pm 0.04$ & 0.10& 18.55 & $138.0\pm 4.9$& P60 \\
1.16900 &    i'& 0.77 & 3631 & $ 18.64 \pm 0.03$ & 0.10& 18.54 & $139.3\pm 4.3$& P60 \\
1.17378 &    i'& 0.77 & 3631 & $ 18.66 \pm 0.04$ & 0.10& 18.56 & $136.9\pm 4.8$& P60 \\
1.17617 &    i'& 0.77 & 3631 & $ 18.68 \pm 0.03$ & 0.10& 18.58 & $134.2\pm 4.1$& P60 \\
1.17857 &    R & 0.66 & 3020 & $ 18.65 \pm 0.04$ & 0.14& 18.51 & $119.2 \pm 3.9$ & P60\\
1.18096 &    R & 0.66 & 3020 & $ 18.66 \pm 0.04$ & 0.14& 18.52 & $118.1 \pm 3.9$ & P60\\
1.18336 &    R & 0.66 & 3020 & $ 18.64 \pm 0.04$ & 0.14& 18.50 & $120.2 \pm 3.9$ & P60\\
1.18576 &    R & 0.66 & 3020 & $ 18.72 \pm 0.04$ & 0.14& 18.58 & $111.7 \pm 3.9$ & P60\\
1.18815 &    R & 0.66 & 3020 & $ 18.66 \pm 0.05$ & 0.14& 18.52 & $118.1 \pm 5.5$ & P60\\
1.19060 &    i'& 0.77 & 3631 & $ 18.71 \pm 0.04$ & 0.10& 18.61 & $130.6\pm 4.6$& P60 \\
1.19300 &    i'& 0.77 & 3631 & $ 18.69 \pm 0.04$ & 0.10& 18.59 & $133.1\pm 4.5$& P60 \\
1.19539 &    i'& 0.77 & 3631 & $ 18.67 \pm 0.04$ & 0.10& 18.57 & $135.6\pm 4.8$& P60 \\
1.20019 &    i'& 0.77 & 3631 & $ 18.70 \pm 0.04$ & 0.10& 18.60 & $131.8\pm 4.9$& P60 \\
1.20261 &    R & 0.66 & 3020 & $ 18.67 \pm 0.05$ & 0.14& 18.53 & $116.9 \pm 5.3$ & P60\\
1.20500 &    R & 0.66 & 3020 & $ 18.65 \pm 0.04$ & 0.14& 18.51 & $117.2 \pm 4.2$ & P60\\
1.20740 &    R & 0.66 & 3020 & $ 18.65 \pm 0.09$ & 0.14& 18.51 & $119.2 \pm 9.7$ & P60\\
1.20980 &    R & 0.66 & 3020 & $ 18.71 \pm 0.04$ & 0.14& 18.57 & $112.8 \pm 4.0$ & P60\\
1.21220 &    R & 0.66 & 3020 & $ 18.72 \pm 0.04$ & 0.14& 18.58 & $111.7 \pm 4.2$ & P60\\
1.21464 &    i'& 0.77 & 3631 & $ 18.71 \pm 0.05$ & 0.10& 18.61 & $130.6\pm 5.5$& P60 \\
1.21704 &    i'& 0.77 & 3631 & $ 18.68 \pm 0.04$ & 0.10& 18.58 & $134.2\pm 5.2$& P60 \\
1.21944 &    i'& 0.77 & 3631 & $ 18.71 \pm 0.04$ & 0.10& 18.61 & $130.6\pm 4.9$& P60 \\
1.22185 &    i'& 0.77 & 3631 & $ 18.74 \pm 0.04$ & 0.10& 18.64 & $127.0\pm 4.8$& P60 \\
1.22425 &    i'& 0.77 & 3631 & $ 18.67 \pm 0.04$ & 0.10& 18.57 & $135.6\pm 4.8$& P60 \\
1.22521 &    Rc& 0.66 & 3020 & $ 18.90\pm  0.2 $ & 0.14& 18.76 & $94.6\pm  23.3$&  GCN6050 \\
1.22910 &    R & 0.66 & 3020 & $ 18.71 \pm 0.05$ & 0.14& 18.57 & $112.8 \pm 5.0$ & P60\\
1.23151 &    R & 0.66 & 3020 & $ 18.70 \pm 0.09$ & 0.14& 18.56 & $113.8 \pm 9.3$ & P60\\
1.23391 &    R & 0.66 & 3020 & $ 18.75 \pm 0.05$ & 0.14& 18.61 & $108.7 \pm 4.9$ & P60\\
1.23632 &    R & 0.66 & 3020 & $ 18.76 \pm 0.04$ & 0.14& 18.62 & $107.7 \pm 4.5$ & P60\\
1.27104 &    Rc& 0.66 & 3020 & $ 19.00 \pm  0.3 $ & 0.14& 18.86 &$86.3\pm21.3$&GCN6050 \\
1.31410 &    Ic& 0.81 & 2380 & $ 18.00 \pm  0.3$ & 0.10& 17.90 & $164.7\pm6.8$& GCN6050 \\
1.31410 &    g'& 0.49 & 3631 & $ 19.60 \pm  0.2 $ & 0.06& 19.54 & $ 55.4\pm8.6$&  GCN6050 \\
1.31410 &    Rc& 0.66 & 3020 & $ 18.80 \pm  0.2 $ & 0.14& 18.66 & $ 103.8\pm17.8$&  GCN6050 \\
1.35377 &    V & 0.55 & 3590 & $ 19.26 \pm 0.27 $& 0.17& 19.09 & $83.0\pm 18.5 $& GCN6041 \\
1.37590 &    Rc& 0.66 & 3020 & $ 18.70 \pm  0.2 $ & 0.14& 18.56 & $113.8\pm19.5$&  GCN6050 \\
1.37605 &    R & 0.66 & 3020 & $ 19.09 \pm 0.05 $& 0.14& 18.95 & $79.6\pm 3.7$& GCN6039 \\
1.40437 &    Rc& 0.66 & 3020 & $ 19.40 \pm  0.4 $ & 0.14& 19.26 & $59.7\pm18.8 $& GCN6050 \\
1.79844 &    R & 0.66 & 3020 & $ 19.43 \pm 0.03$ & 0.14& 19.29 & $58.1 \pm 1.6$& P60\\
1.81062 &    i'& 0.77 & 3631 & $ 19.49 \pm 0.03$ & 0.10& 19.39 & $63.7 \pm 2.0$& P60 \\
1.84702 &    R & 0.66 & 3020 & $ 19.51 \pm 0.03$ & 0.14& 19.37 & $54.0 \pm 1.7$& P60\\
1.85907 &    i'& 0.77 & 3631 & $ 19.52 \pm 0.03$ & 0.10& 19.42 & $61.9 \pm 1.9$& P60 \\
1.90324 &    R & 0.66 & 3020 & $ 19.59 \pm 0.03$ & 0.14& 19.45 & $50.1 \pm 1.3$& P60\\
1.91542 &    i'& 0.77 & 3631 & $ 19.58 \pm 0.03$ & 0.10& 19.48 & $58.6 \pm 1.6$& P60 \\
1.96396 &    Rc& 0.66 & 3020 & $ 19.71 \pm 0.02 $& 0.14& 19.57 & $44.9\pm0.8$& GCN6096 \\
1.97868 &    R & 0.66 & 3020 & $ 19.62 \pm 0.03$ & 0.14& 19.48 & $48.7 \pm 1.4$& P60\\
1.99132 &    i'& 0.77 & 3631 & $ 19.69 \pm 0.03$ & 0.10& 19.59 & $52.9 \pm 1.6$& P60 \\
2.64396 &   Rc & 0.66 & 3020 & $ 20.23 \pm  0.1 $& 0.14& 20.09 & $27.8\pm2.5$ &GCN6047 \\
2.65995 &   Rc & 0.66 & 3020 & $ 20.26 \pm 0.11 $& 0.14& 20.12 & $27.0\pm2.7$ &GCN6047 \\
2.67696 &   Rc & 0.66 & 3020 & $ 20.21 \pm 0.11 $& 0.14& 20.07 & $28.3\pm2.8 $&GCN6047 \\
2.78096 &   Rc & 0.64 & 3020 & $ 20.25 \pm 0.11 $& 0.14& 20.11 & $27.3\pm2.7 $&GCN6047 \\
2.79596 &   Rc & 0.66 & 3020 & $ 20.35 \pm 0.12 $& 0.14& 20.21 & $24.9\pm2.7 $&GCN6047 \\
2.81479 &   Ks & 2.22 & 670  & $ 17.86 \pm 0.25 $& 0.02& 17.84 & $49.0\pm10.1 $& GCN6054 \\
2.81479 &   J &  1.26 & 1600 & $ 18.82 \pm 0.26 $& 0.05& 18.77 & $49.7\pm10.6 $& GCN6054 \\
2.81479 &  H & 1.66 & 1024 & $ 18.33 \pm 0.25 $ & 0.01 & 18.32 & $ 47.7 \pm 9.8$ & GCN6054 \\
2.99296 &   R &  0.66 & 3020 & $ 20.44 \pm 0.03 $& 0.14& 20.30 & $22.9\pm0.6$&GCN6096 \\
3.02006 &   R &  0.66 & 3020 & $ 20.44 \pm 0.04$ & 0.14& 20.30 & $22.9 \pm 0.8$& P60\\
3.03328 &   i' & 0.77 & 3631 & $ 20.47 \pm 0.04$ & 0.10& 20.37 & $25.9 \pm 1.1$& P60 \\
3.62096 &   Rc & 0.66 & 3020 & $ 20.80 \pm  0.2 $ & 0.14& 20.66 & $16.4\pm2.8$& GCN6064 \\
3.91797 &   r' & 0.63 & 3631 & $ 21.22 \pm 0.09$ & 0.14& 21.08 & $13.4 \pm 1.5$& GMOS\\
4.03995 &   R &  0.66 & 3020 & $ 21.07 \pm 0.07$ & 0.14& 20.93 & $12.8\pm0.8 $&GCN6096 \\
4.07106 &   i' & 0.77 & 3631 & $ 21.03 \pm 0.10$ & 0.10& 20.93 & $15.5 \pm 1.0$& P60 \\
4.08226 &   R &  0.66 & 3020 & $>20.44$          & 0.14& $>20.30$ & $<22.5 $ & P60\\
8.85767 &   R &  0.66 & 3020 & $>21.63$          & 0.14& $>21.49$ & $<7.5 $ & P60\\
10.0099 &   i' & 0.77 & 3631 & $ 23.75 \pm 0.14$ & 0.10& 23.65 & $1.3\pm 0.2$ &GMOS \\
11.82110 &  R &  0.66 & 3020 & $>22.57$          & 0.14& $>22.43$& $<3.2 $ & P60\\
12.00096 &  R &  0.66 & 3020 & $>23.80$          & 0.14& $>23.66$ & $<1.0$&  GCN6096 \\
21.99400 &  R &  0.66 & 3020 & $>25.40$          & 0.14& $>25.26$ & $<0.2 $ & LRIS\\
21.99400 &  g' & 0.49 & 3631 & $>26.10$          & 0.06& $>26.04$ & $<0.2 $ & LRIS\\
26.79396 &  r &  0.67 & 3631 & $ 26.30 \pm  0.3 $ & 0.14& 26.16 &$0.13\pm-$&   GCN6165 \\
\enddata
\tablecomments{P60$\equiv$ Palomer 60-inch Telescope observations.}
\tablenotetext{a}{$\bullet$ GCN 6041: \Swift\ UVOT, Marshall, F.E., vanden Berk, D.E.
and Racusin, J. $\bullet$ GCN 6036: \Swift\ UVOT, Marshall, F.E., Racusin, J. $\bullet$ GCN 6028: Palomer 60in, Cenko, S.B. and Fox, D.B. $\bullet$
GCN 6044: 16in PROMPT telescope, Haislip, J., Reichart, D. and 
LaCluyze, A. et al. $\bullet$ GCN 6035: TNT 0.8m telescope, 
Xing, L.P., Zhai, M., Qiu, Y.L. et al. $\bullet$ GCN 6050:
50cm MITSuME Telescope, Yoshida, M., Yanagisawa, K., and Kawai, N.,
$\bullet$ GCN 6039: KANATA 1.5-m telescope, Uemura, M.,
Arai, A., and Uehara, T. $\bullet$ GCN 6096: MDM 2.4m and 1.3m telescopes,
Mirabal, N., and Thorstensen, J.R. $\bullet$
GCN 6047: 152 cm Cassini Telescope, Greco, G., Terra, F.,
and Bartolini, C. et al. $\bullet$ GCN 6054: PAIRITEL 1.3m telescope,
Bloom, J.S., Starr, D., and Blake, C.H. $\bullet$ GCN 6064:
152 cm Loiano telescope, Greco, G., Terra, F., and Bartolini, C. et al.
$\bullet$ GCN 6165: Large Binocular Telescope, Garnavich, P., Fan, X.,
and Jiang, L. et al.
}
\end{deluxetable}

\clearpage

\begin{deluxetable}{cccccc}
\tabletypesize{\scriptsize}
\tablecaption{Radio observations of the GRB 070125
\label{tab:radio}}
\tablewidth{0pt}
\tablehead{
\colhead{Date of} & \colhead{Days since} & \colhead{Telescope} & \colhead{Frequency} & \colhead{Flux
density} & \colhead{Error}\\
\colhead{observation} & \colhead{explosion} & \colhead{} & \colhead{in GHz} & \colhead{$\mu$Jy} & \colhead{$\mu$Jy}
}
\startdata
2007 Jan 26.82 & 1.51  & WSRT& 4.86 &    $ <174$     & 87\\
2007 Jan 29.32 & 4.01  & VLA& 8.46 &    360     &  42\\
2007 Jan 30.24 & 4.93  & VLA& 8.46 &    454     & 38 \\
2007 Jan 30.95 & 5.64  & WSRT& 4.86 &    102            & 26 \\
2007 Jan 31.07 & 5.76  & VLA& 4.86 &    <141             & 58 \\
2007 Jan 31.11 & 5.80  & VLA& 8.46 &    382     & 52 \\
2007 Jan 31.76 & 6.45  & GMRT& 0.61 &  $<300$       &      150 \\
2007 Jan 31.83 & 6.52 & MAMBO2& 250 &  3140   &     590\\
2007 Feb 01.92 & 7.61 & MAMBO2& 250 &  1910   &     720\\
2007 Feb 02.06 & 7.75  & VLA& 8.46 &    563     & 62 \\
2007 Feb 04.92 & 10.61 & MAMBO2& 250 & $<1470$  &   710\\
2007 Feb 05.03 & 10.72 & VLA& 8.46 &    482     &  52\\
2007 Feb 05.04 & 10.73 & VLA& 22.50 &    1594   &   70    \\
2007 Feb 05.07 & 10.76 & VLA& 4.86 &    $<132 $          & 42 \\
2007 Feb 05.29 & 10.98 & CARMA& 95 &   2300  & 700  \\
2007 Feb 06.24 & 11.93 & VLA& 8.46 &    489     &  43\\
2007 Feb 06.26 & 11.95 & VLA& 4.86 &    $<124  $         & 34 \\
2007 Feb 07.11 & 12.80 & VLA& 22.50 &    1603   &   235    \\
2007 Feb 07.12 & 12.81 & VLA& 8.46 &    405     &  20\\
2007 Feb 07.38 & 13.07 &  VLA&14.94 &     1159    &   234   \\
2007 Feb 08.21 & 13.90 & VLA& 14.94 &     917     &   293    \\
2007 Feb 08.23 & 13.92 & VLA& 4.86 &    $<145   $        & 50 \\
2007 Feb 08.24 & 13.93 & VLA& 8.46 &    559     &  26\\
2007 Feb 08.40 & 14.09 & VLA& 22.50 &    1621   &   151    \\
2007 Feb 09.21 & 14.90 & VLA& 4.86 &    $<108   $    & 49 \\
2007 Feb 09.25 & 14.94 & VLA& 8.46 &    399     & 69 \\
2007 Feb 09.27 & 14.96 & VLA& 22.50 &    1343   &   162   \\
2007 Feb 09.27 & 14.96 & CARMA& 95  &  2300  & 800 \\
2007 Feb 09.28 & 14.97 & VLA& 14.94 &     891     &   129    \\
2007 Feb 10.37 & 16.06 & VLA& 14.94 &     1410    &   137    \\
2007 Feb 10.79 & 16.48 & MAMBO2& 250 &  2670    &     930\\
2007 Feb 11.08 & 16.77 & VLA& 8.46 &    385     &  42\\
2007 Feb 11.12 & 16.81 & VLA& 22.50 &    1367   &   104    \\
2007 Feb 11.16 & 16.85 & VLA& 4.86 &    $<196 $     & 71 \\
2007 Feb 12.08 & 17.77 & VLA& 1.46 &    $<920 $     &       460 \\
2007 Feb 12.16 & 17.85 & VLA& 8.46 &    596     & 109\\
2007 Feb 12.20 & 17.89 & VLA& 14.94 &     1226    &   155    \\
2007 Feb 12.21 & 17.90 & CARMA& 95  &  2100  & 700  \\
2007 Feb 12.26 & 17.95 & VLA& 22.50 &    1222   &   167    \\
2007 Feb 12.79 & 18.48 & MAMBO2& 250 & $<1270$  &   930\\
2007 Feb 13.07 & 18.76 &  VLA&1.46 &    $<940$     &       600 \\
2007 Feb 13.11 & 18.80 & VLA& 4.86 &    $<150  $     & 52 \\
2007 Feb 13.17 & 18.86 & VLA& 8.46 &    660     &  39\\
2007 Feb 13.21 & 18.90 & VLA& 14.94 &     1217    &   197    \\
2007 Feb 13.25 & 18.94 & VLA& 22.50 &    1248   &   83     \\
2007 Feb 14.18 & 19.87 & VLA& 8.46 &    581     &  14\\
2007 Feb 16.17 & 21.86 & VLA& 1.46 &    $<1068$     &       540  \\
2007 Feb 16.19 & 21.88 & VLA& 4.86 &    $<159  $     &  47 \\
2007 Feb 17.02 & 22.71 & VLA& 1.46 &    $<1152$     &       766  \\
2007 Feb 17.04 & 22.73 & VLA& 4.86 &    $<262$         &  64 \\
2007 Feb 18.16 & 23.85 & VLA& 22.50 &    1168   &   118    \\
2007 Feb 18.18 & 23.87 & VLA& 8.46 &    303     &  63\\
2007 Feb 18.21 & 23.90 & VLA& 14.94 &     798     &   182    \\
2007 Feb 18.27 & 23.96 & CARMA& 95  &  2400  & 700  \\
2007 Feb 21.05 & 26.74 & VLA& 14.94 &     1101    &   148    \\
2007 Feb 22.21 & 27.90 & VLA& 4.86 &    $308$         &  78 \\
2007 Feb 23.07 & 28.76 & VLA& 1.46 &    $<984$     &       804  \\
2007 Feb 25.16 & 30.85 & VLA& 22.50 &    839    &   73     \\
2007 Feb 27.20 & 32.89 & VLA& 1.46 &    $<978$     &       639  \\
2007 Mar 01.22 & 34.91 & VLA& 4.86 &    $<322$         &  95 \\
2007 Mar 01.24 & 34.93 & VLA& 14.94 &     432     &   149    \\
2007 Mar 01.28 & 34.97 & VLA& 8.46 &    414     &  66\\
2007 Mar 01.30 & 34.99 & VLA& 22.50 &    788    &   74     \\
2007 Mar 13.18 & 46.87 & VLA& 1.46 &    $<1078$     &       739 \\
2007 Mar 15.20 & 48.89 & VLA& 4.86 &    229         &  49 \\
2007 Mar 15.22 & 48.91 & VLA& 8.46 &    443     & 59 \\
2007 Mar 21.22 & 54.91 & VLA& 4.86 &    262         &  52 \\
2007 Mar 21.26 & 54.95 & VLA& 8.46 &    473     &  44\\
2007 Mar 23.08 & 56.77 & VLA& 22.50 &    559    &   161   \\
2007 Mar 25.15 & 58.84 & VLA& 1.46 &    $<580$     &       280 \\
2007 Apr 02.08 & 66.77 & VLA& 4.86 &    226         &  57 \\
2007 Apr 02.12 & 66.81 & VLA& 8.46 &    345     & 48 \\
2007 Apr 02.16 & 66.85 & VLA& 1.46 &    $<680$     &       480 \\
2007 Apr 02.20 & 66.89 & VLA& 14.94 &     530     &   137 \\
2007 Apr 02.24 & 66.93 & VLA& 22.50 &    568    &   137    \\
2007 Apr 21.13 & 85.82 & VLA& 1.46 &    $<1200$     &       570 \\
2007 Apr 21.18 & 85.87 & VLA& 4.86 &    302         &  55 \\
2007 Apr 22.18 & 86.87 & VLA& 8.46 &    403     & 56 \\
2007 Apr 22.99 & 87.68 & VLA& 22.50 &    615    &   167   \\
2007 Apr 23.02 & 87.71 & VLA& 14.94 &     $<450$  &         226\\
2007 May 15.03 & 109.72& VLA& 8.46 &    267         & 42\\
2007 May 17.00 & 111.69& VLA& 1.46 &    $<556$     &       278\\
2007 May 18.81 & 113.50& VLA& 4.86 &    $<290$      & 145\\
2007 May 19.91 & 114.60& VLA& 14.94 &     $<1440$ & 720\\
2007 May 20.04 & 114.73& VLA& 22.50 &  $<176$   &   88\\
2007 Jul 04.76 & 160.45&  VLA&8.46 & 145 & 33\\
2007 Jul 04.80 & 160.49&  VLA&4.86 & 174 & 45\\
2007 Jul 04.85 & 160.54& VLA& 1.46 & $<162 $ & 160\\
2007 Aug 03.87 & 190.56& VLA& 22.50 & $<446$ & 223\\
2007 Aug 11.68 & 198.37& VLA& 4.86 & $<100$  & 42\\
2007 Aug 11.72 & 198.41& VLA& 8.46 & 166 & 46\\
2007 Aug 13.71 & 200.40& VLA& 14.94 & $<350$ & 169\\
2007 Sep 08.64 & 226.33& VLA & 4.86 & $<141$ & 57\\
2007 Oct 04.52 & 252.21& VLA & 8.46 & 148 & 60 \\
2007 Oct 04.60 & 252.29& VLA & 4.86 & 203 & 34\\
2007 Nov 18.39 & 297.08& VLA & 8.46 & 111 & 19\\
2007 Nov 18.48 & 297.17& VLA & 4.86 & 161 & 23\\
2008 Jan 02.27 & 341.96& VLA & 8.46 & 64 & 18 \\
2008 Jan 02.35 & 342.04& VLA & 4.86 & 133 & 21\\
\enddata
\end{deluxetable}

\clearpage

\begin{deluxetable}{lcc|cc}
  \tabletypesize{\scriptsize}
  \tablecaption{Optical/X-ray Afterglow Model Fits}
  \tablecolumns{5}
  \tablewidth{0pc}
  \tablehead{
      \colhead{Parameters} & \multicolumn{4}{c}{Models}\\
     \cline{2-5}\\
\colhead{} & \multicolumn{2}{c}{Single powerlaw} & 
      \multicolumn{2}{c}{Broken powerlaw}\\
      \colhead{} & \colhead{same slope}&\colhead{independent slope}
      & \colhead{same slope} & \colhead{independent slope}
  }
  \startdata
  \colhead{$\alpha_{1}$} & 1.80\tablenotemark{a} &$\alpha_{1}^X(1.85)$, $\alpha_{1}^O(1.80)$ & $1.73\pm0.02$& $\alpha_{1}^X(1.76\pm0.30)$, $\alpha_{1}^O(1.73\pm0.02)$\\ 
  \colhead{$\alpha_{2}$} & \ldots &\ldots & $2.49 ^{+0.85}_{-0.18}$&
  $2.49 ^{+0.86}_{-0.18}$\\
  \colhead{$t_{\mathrm{b}}$(days)} &\ldots &\ldots &$3.8 ^{+2.6}_{-0.4}$
  & $3.8 ^{+2.6}_{-0.4}$\\
  \colhead{$\chi^{2}_{\mathrm{r}}(\mathrm{Total})$} &2.23 (99 d.o.f.) &
  2.25 (98 d.o.f.)& 1.30 (97 d.o.f.)& 1.31 (96 d.o.f.) \\
  \colhead{$\chi^{2}_{\mathrm{r}}(\mathrm{X-ray})$} & 0.88 (22 d.o.f.)&
  0.87 (22 d.o.f.)& 1.32 (20 d.o.f.)& 1.31 (20 d.o.f.)\\
  \colhead{$\chi^{2}_{\mathrm{r}}(R\mathrm{-band})$} &2.22 (52 d.o.f.) &
  2.21 (52 d.o.f.)& 1.54 (52 d.o.f.)& 1.54 (52 d.o.f.)\\
  \colhead{$\chi^{2}_{\mathrm{r}}(i^{\prime}\mathrm{-band})$}& 
  3.43 (25 d.o.f.)& 3.59 (24 d.o.f.) &0.77 (25 d.o.f.) & 0.81 (24 d.o.f.) \\
  \enddata
  \tablecomments{All fits were performed jointly to constrain the relevant
    decay indices and break time to be 
    the same in all bandpasses.  All errors
    quoted are 90\% confidence intervals.}
  \tablenotetext{a}{The poor quality of the overall 
  fit precludes meaningful
    estimates of the 90\% confidence intervals.}
\label{tab:jet}
\end{deluxetable}

\clearpage

\begin{deluxetable}{l|ll|ll|ll}
\scriptsize
\tablecaption{Comparison between Wind and ISM model
\label{tab:models}}
\tablewidth{0pt}
\tablehead{
\colhead{parameters} & \multicolumn{2}{c}{ISM} & \multicolumn{2}{c}{Wind}
&  \multicolumn{2}{c}{Wind-fixed $\epsilon_e$}\\
\colhead{} & \colhead{with IC} & \colhead{no IC} & \colhead{with IC} & \colhead{no IC} & \colhead{with IC} & \colhead{no IC}\\
}
\startdata
$E_{K,\rm iso}$($10^{52}$erg)\tablenotemark{a} &
$6.45_{-0.24}^{+1.03}$  & 
$14.57_{-4.01}^{+3.20}$  &
$0.29_{-0.07}^{+0.43}$
& $0.15_{-0.08}^{+90.0}$ & $1.25^{+0.02}_{-0.17}$ &
$0.41^{+0.29}_{-0.01}$\\
$\theta_j$ (rad)&  
$0.23_{-0.01}^{+0.01}$  
& $0.12_{-0.01}^{+0.02}$
& $0.36_{-0.02}^{+0.02}$ &
$0.33_{-0.04}^{+90.0}$ & $0.24^{+0.01}_{0.02}$ & $0.22^{+0.01}_{-0.01}$\\
 $p$ & 
$2.45_{-0.02}^{+0.01}$  &
$2.11_{-0.01}^{+0.02}$
& $2.17_{-0.01}^{+0.01}$ 
& $2.14_{-0.01}^{+0.84}$ & $2.25^{+0.01}_{-0.01}$ &
$2.18^{+0.01}_{-0.03}$\\
$\epsilon_e$&
$0.27_{-0.01}^{+0.03}$
& $0.13_{-0.02}^{+0.03}$
& $0.99_{-0.13}^{+0.01}$
& $0.77_{-0.74}^{+0.02}$ & 0.4 & 0.4\\
$\epsilon_B$(\%)
& $2.77_{-0.75}^{+0.44}$
&$99.99_{-0.0}^{+0.1}$
& $6.78_{-7.28}^{+4.11}$ &
$100_{-98.0}^{+0.00}$ & $3.73^{+0.87}_{-0.03}$ & $100^{+0.00}_{-9.90}$\\
$n$ or $A_\star$\tablenotemark{b} &
$42.07_{-3.86}^{+1.59}$
& $4.43_{-0.20}^{+0.62}$ &
$2.81_{-0.15}^{+0.55}$
& $0.99_{-0.30}^{+0.03}$ & $3.52^{+0.02}_{-0.43}$ &
$1.00^{+0.12}_{-0.02}$\\
$t_{jet}$ (days)&
$3.69_{-0.07}^{+0.03}$  &
$1.61_{-0.08}^{+0.04}$&
$7.29_{-1.09}^{+0.07}$ &
$6.87_{-3.12}^{+1.73}$ & $3.34^{+0.10}_{-0.01}$
& $3.47^{+0.03}_{-0.08}$\\
$t_{nonrel}$ (days) &
$53.45_{-0.88}^{+0.52}$ &
$91.74_{-4.02}^{+2.30}$ &
$27.45_{-2.08}^{+1.40}$ &
$30.88_{-9.20}^{+7.80}$ & $30.03^{+1.17}_{-0.20}$ &
$36.68^{+0.02}_{-1.07}$\\
$t_{cool}$(days)&
$8.49_{-1.72}^{+0.62}$
& $4.59_{-0.11}^{+0.17}$
& $28.40_{-3.70}^{+0.73}$
& $30.89_{-18.10}^{+0.20}$ & $18.13^{+2.24}_{-1.13}$ &
$31.37^{+1.04}_{-1.80}$\\
$A_V$ &
$7.4 \times 10^{-8}$ &
0.06 &
$0.08$ &
0.11 & $4.0 \times 10^{-7}$ & 0.08\\
Fit statistic & 592.02 & 633.02 & 556.26 & 575.36 & 584.85 & 588.96\\
$\chi^2/{d.o.f.}$ &
$\frac{326.14}{186}$ &
$\frac{411.42}{186}$
& $\frac{254.71}{186}$ &
$\frac{298.52}{186}$
& $\frac{292.66}{187}$& $\frac{335.44}{186}$\\
\enddata
\tablenotetext{a}{Isotropic kinetic energy at the time when $\nu_c=\nu_m$.}
\tablenotetext{b}{Density $n$ for ISM model in cm$^{-3}$
and $A_\star$ for wind model.}
\end{deluxetable}

\clearpage

\clearpage

\clearpage

\clearpage

\clearpage

\clearpage

\clearpage

\clearpage

\clearpage

\clearpage

\end{document}